\pdfoutput=1
\newif\ifarxiv
\arxivtrue
\def\paperversiondraft{draft}

\ifarxiv
  \documentclass[sigplan,screen,nonacm]{acmart}
\else
  \documentclass[sigplan,screen]{acmart}
\fi

\usepackage[utf8]{inputenc}
\usepackage[T1]{fontenc}
\usepackage{microtype}
\usepackage{makecell}
\usepackage{listings}

\ifarxiv
  \usepackage[frozencache]{minted}
\else
  \usepackage[finalizecache]{minted}
\fi
\usepackage{svg}

\makeatletter
\patchcmd{\@addmarginpar}{\ifodd\c@page}{\ifodd\c@page\@tempcnta\m@ne}{}{}
\makeatother
\ifx\paperversion\paperversiondraft
  \makeatletter
    \paperwidth=\dimexpr \paperwidth + 6cm\relax
    \oddsidemargin=\dimexpr\oddsidemargin + 3cm\relax
    \evensidemargin=\dimexpr\evensidemargin + 3cm\relax
    \marginparwidth=\dimexpr \marginparwidth + 3cm\relax
  \makeatother
\fi

\usepackage[textsize=tiny]{todonotes}
\usepackage[normalem]{ulem}

\makeatletter
\font\uwavefont=lasyb10 scaled 652
\newcommand\colorwave[1][blue]{\bgroup\markoverwith{\lower3\p@\hbox{\uwavefont\textcolor{#1}{\char58}}}\ULon}

\newcommand\highlight[2]{{{\colorwave[#1]{#2}}}}
\makeatother

\makeatletter
\newcommand\InFloat[2]{\ifnum\@floatpenalty<0\relax#1\else#2\fi}
\makeatother

\ifx\paperversion\paperversiondraft
\newcommand\createtodoauthor[2]{
  \def\tmpdefault{emptystring}
  \expandafter\newcommand\csname #1\endcsname[2][\tmpdefault]{
    \def\tmp{##1}
    \InFloat{
        \smash{
	  \marginnote{
	    \todo[inline,linecolor=#2,backgroundcolor=#2,bordercolor=#2]
	      {\textbf{#1 (Figure):} ##2}
          }
        }
    }{\ifthenelse{\equal{\tmp}{\tmpdefault}} 
      {\todo[linecolor=#2,backgroundcolor=#2,bordercolor=#2]{\textbf{#1:} ##2}\ignorespaces}
      {\ifthenelse{\equal{##2}{}} 
        {\highlight{#2}{##1}}
        {\highlight{#2}{##1}\todo[linecolor=#2,backgroundcolor=#2,bordercolor=#2]
	  {\textbf{#1:} ##2}
	}
      }
    }
  }
}
\else
\newcommand\createtodoauthor[2]{%
\expandafter\newcommand\csname #1\endcsname[2][]{##1}%
}%
\fi

\usepackage{etoolbox}

\makeatletter
\ifcsdef{minted@optlistcl@quote}
{
\ifwindows
  \renewcommand{\minted@optlistcl@quote}[2]{%
    \ifstrempty{#2}{\detokenize{#1}}{\detokenize{#1="#2"}}}
\else
  \renewcommand{\minted@optlistcl@quote}[2]{%
    \ifstrempty{#2}{\detokenize{#1}}{\detokenize{#1='#2'}}}
\fi

\newcommand{\minted@def@optcl@novalue}[2]{%
  \define@booleankey{minted@opt@g}{#1}%
    {\minted@addto@optlistcl{\minted@optlistcl@g}{#2}{}%
     \@namedef{minted@opt@g:#1}{true}}
    {\@namedef{minted@opt@g:#1}{false}}
  \define@booleankey{minted@opt@g@i}{#1}%
    {\minted@addto@optlistcl{\minted@optlistcl@g@i}{#2}{}%
     \@namedef{minted@opt@g@i:#1}{true}}
    {\@namedef{minted@opt@g@i:#1}{false}}
  \define@booleankey{minted@opt@lang}{#1}%
    {\minted@addto@optlistcl@lang{minted@optlistcl@lang\minted@lang}{#2}{}%
     \@namedef{minted@opt@lang\minted@lang:#1}{true}}
    {\@namedef{minted@opt@lang\minted@lang:#1}{false}}
  \define@booleankey{minted@opt@lang@i}{#1}%
    {\minted@addto@optlistcl@lang{minted@optlistcl@lang\minted@lang @i}{#2}{}%
     \@namedef{minted@opt@lang\minted@lang @i:#1}{true}}
    {\@namedef{minted@opt@lang\minted@lang @i:#1}{false}}
  \define@booleankey{minted@opt@cmd}{#1}%
      {\minted@addto@optlistcl{\minted@optlistcl@cmd}{#2}{}%
        \@namedef{minted@opt@cmd:#1}{true}}
      {\@namedef{minted@opt@cmd:#1}{false}}
}
\minted@def@optcl@novalue{customlexer}{-x}
}
{
}
\makeatother

\makeatletter
\ifcsdef{minted@optlistcl@quote}
{
  \newminted[mlir]{tools/MLIRLexer.py:MLIRLexerOnlyOps}{customlexer, mathescape, fontsize=\footnotesize, linenos}
  \newmintinline[mlirinline]{tools/MLIRLexer.py:MLIRLexerOnlyOps}{customlexer, mathescape, escapeinside=||, linenos}
}
{
  \newminted[mlir]{tools/MLIRLexer.py:MLIRLexerOnlyOps -x}{mathescape, fontsize=\footnotesize, linenos}
  \newmintinline[mlirinline]{tools/MLIRLexer.py:MLIRLexerOnlyOps -x}{mathescape, escapeinside=||, linenos}
}
\makeatother

\newminted[ccode]{c}{mathescape, linenos, fontsize=\footnotesize, style=bw, frame=lines, breaklines=True}
\newminted[python]{python}{mathescape, fontsize=\footnotesize, linenos, style=vs}

\usepackage{enumitem}
\usepackage{xargs}
\usepackage{lipsum}
\usepackage{xparse}
\usepackage{xifthen, xstring}
\usepackage{xspace}

\usepackage{subcaption}
\usepackage{graphicx}

\usepackage{multirow}
\definecolor{pairedNegOneLightGray}{HTML}{cacaca}
\definecolor{pairedNegTwoDarkGray}{HTML}{827b7b}
\definecolor{pairedOneLightBlue}{HTML}{a6cee3}
\definecolor{pairedTwoDarkBlue}{HTML}{1f78b4}
\definecolor{pairedThreeLightGreen}{HTML}{b2df8a}
\definecolor{pairedFourDarkGreen}{HTML}{33a02c}
\definecolor{pairedFiveLightRed}{HTML}{fb9a99}
\definecolor{pairedSixDarkRed}{HTML}{e31a1c}

\graphicspath{{./images/}}

\usepackage{booktabs}

\usepackage{color}
\definecolor{deepblue}{rgb}{0,0,0.5}
\definecolor{figureblue}{RGB}{0,102,186}
 \definecolor{figureyellow}{RGB}{239,180,0}
\definecolor{deepred}{rgb}{0.6,0,0}
\definecolor{deepgreen}{rgb}{0,0.5,0}

\copyrightyear{2024}
\acmYear{2024}
\ifarxiv
  \setcopyright{cc}
\else
  \setcopyright{rightsretained}
\fi
\acmConference[ASPLOS '24]{29th ACM International Conference on Architectural Support for Programming Languages and Operating Systems, Volume 3}{April 27-May 1, 2024}{La Jolla, CA, USA}
\acmBooktitle{29th ACM International Conference on Architectural Support for Programming Languages and Operating Systems, Volume 3 (ASPLOS '24), April 27-May 1, 2024, La Jolla, CA, USA}
\acmDOI{10.1145/3620666.3651344}
\acmISBN{979-8-4007-0386-7/24/04}

\title{A shared compilation stack for distributed-memory parallelism in stencil DSLs}

\usepackage{cleveref}
\author{George Bisbas}
\authornote{The first four authors contributed equally.}
\affiliation{
   \institution{Imperial College London}
   \country{United Kingdom}
}
\email{g.bisbas18@imperial.ac.uk}
\author{Anton Lydike}
\authornotemark[1]
\affiliation{
   \institution{The University of Edinburgh}
   \country{United Kingdom}
}
\email{anton.lydike@ed.ac.uk}
\author{Emilien Bauer}
\authornotemark[1]
\affiliation{
   \institution{The University of Edinburgh}
   \country{United Kingdom}
}
\email{emilien.bauer@ed.ac.uk}
\author{Nick Brown}
\authornotemark[1]
\affiliation{
   \institution{The University of Edinburgh}
   \country{United Kingdom}
}
\email{nick.brown@ed.ac.uk}

\author{Mathieu Fehr}
 \affiliation{
   \institution{The University of Edinburgh}
   \country{United Kingdom}
}
\email{mathieu.fehr@ed.ac.uk}
\author{Lawrence Mitchell}
 \affiliation{
   \institution{}
   \country{United Kingdom}
}
\email{lawrence@wence.uk}
 
\author{Gabriel Rodriguez-Canal}
 \affiliation{
   \institution{The University of Edinburgh}
   \country{United Kingdom}
}
\email{gabriel.rodcanal@ed.ac.uk}
\author{Maurice Jamieson}
\affiliation{
   \institution{The University of Edinburgh}
   \country{United Kingdom}
}
\email{maurice.jamieson@ed.ac.uk}
\author{Paul H.~J.~Kelly}
\affiliation{
   \institution{Imperial College London}
   \country{United Kingdom}
}
\email{p.kelly@imperial.ac.uk}
\author{Michel Steuwer}
\affiliation{
   \institution{Technische Universität Berlin}
   \country{Germany}
}
\email{michel.steuwer@tu-berline.de}
\author{Tobias Grosser}
\affiliation{
   \institution{University of Cambridge}
   \country{United Kingdom}
}
\email{tobias.grosser@cst.cam.ac.uk}

\begin{document}
\begin{abstract}
Domain Specific Languages (DSLs) increase programmer productivity and provide high
performance. Their targeted abstractions allow scientists to express
problems at a high level, providing rich details that optimizing compilers can
exploit to target current- and next-generation supercomputers.
The convenience and performance of DSLs come with significant development and
maintenance costs. The siloed design of DSL compilers and the resulting
inability to benefit from shared infrastructure cause uncertainties around longevity and
the adoption of DSLs at scale. By tailoring the broadly-adopted MLIR
compiler framework to HPC, we bring the same synergies that the
machine learning community already exploits across their DSLs (e.g.\
Tensorflow, PyTorch) to the finite-difference stencil HPC community. We introduce new
HPC-specific abstractions for message passing targeting distributed stencil
computations. We demonstrate the sharing of common components across three
distinct HPC stencil-DSL compilers: Devito, PSyclone, and the Open Earth Compiler,
showing that our framework generates high-performance executables
based upon a shared compiler ecosystem.
\end{abstract}

\ifarxiv
\fi

\maketitle

\section{Introduction}

%
%
\begin{figure*}[tp]
  \centering
  \begin{subfigure}[t]{0.49\textwidth}
    \includegraphics[width=\columnwidth]{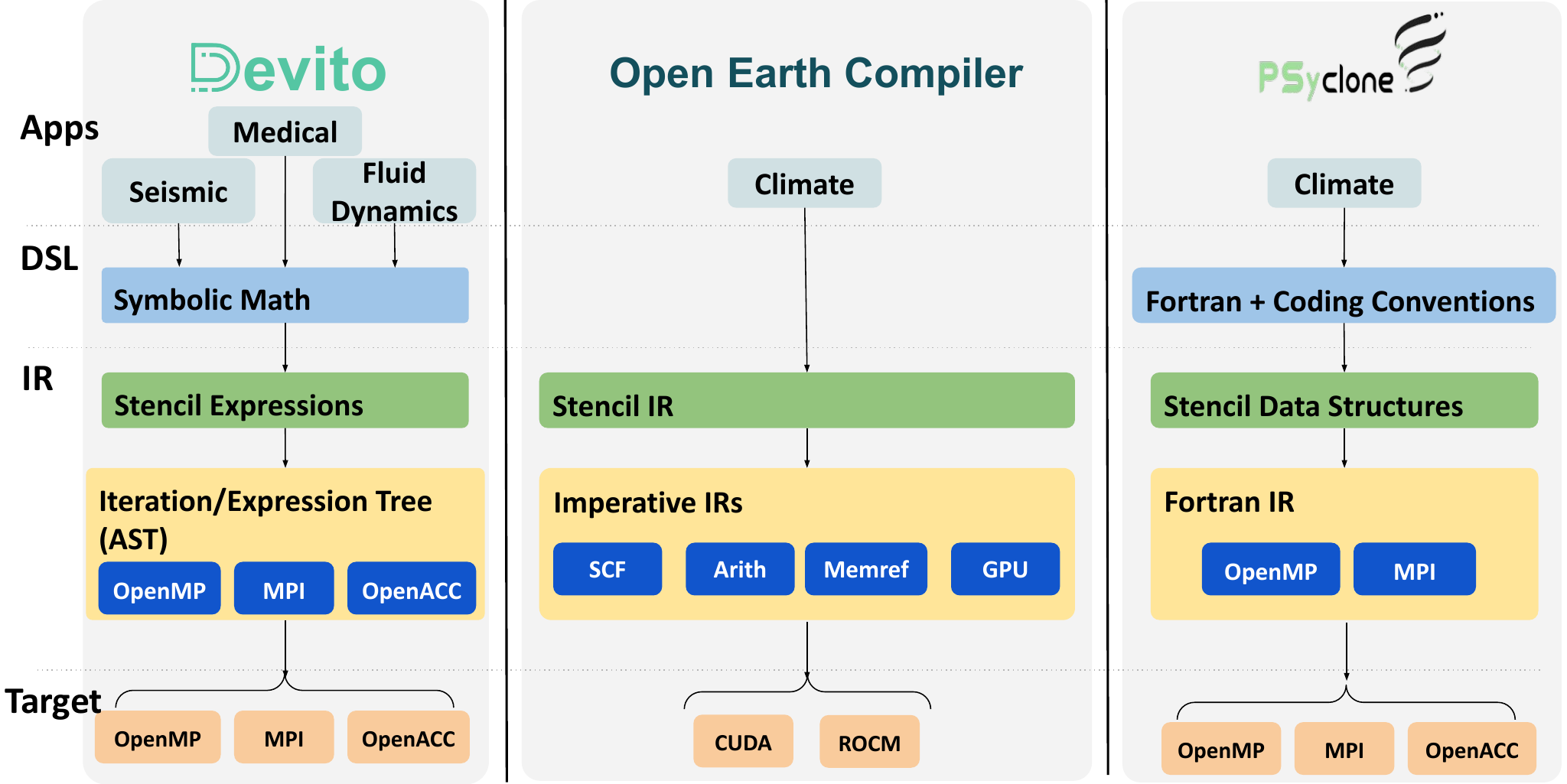}
	  \caption{Devito, the Open Earth Compiler (OEC), and PSyclone independently
	  maintain abstractions for stencils and use similar imperative
	  constructs. However, HPC features such as parallelism
    with MPI (in OEC) and GPUs (in PSyclone) 
    are not universal.\label{fig:today}}
  \end{subfigure}  \hfill
  \begin{subfigure}[t]{0.49\textwidth}
      \centering
      \includegraphics[width=\columnwidth]{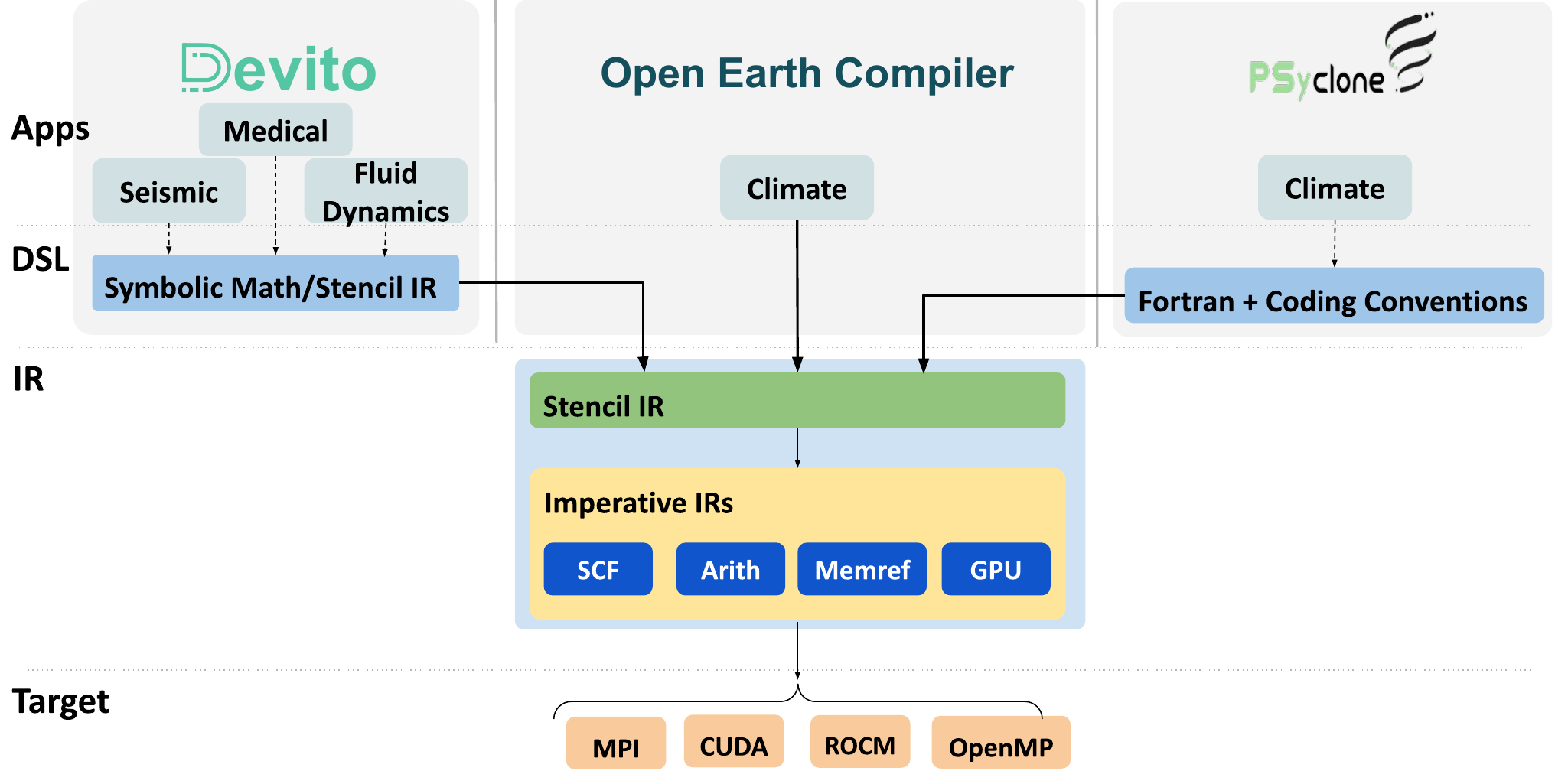}
      \caption{We combine the optimization and code generation pipelines of Devito,
      the Open Earth Compiler, and PSyclone. As a result, optimization passes can be
      shared and advanced HPC features are available to all tools.\label{fig:vision}}
  \end{subfigure}
  \caption{Our work enables reuse of HPC and target-specific abstractions across DSL and
  compiler frameworks and consequently offers synergies across DSL communities,
  while maintaining the community-tailored interfaces of each DSL compiler.\label{fig:today-vision}}
\end{figure*}

Most DSLs are purposefully designed for specific problem domains,
providing focused and specialized language constructs tailored to those domains.
They capture concepts for a specific domain, yet a large portion of their code base
is dedicated to reasoning about generic concepts in HPC.\@ These include
the generation of directives for shared-memory parallelism, message-passing
communications for distributed-memory parallelism, vectorization,
arithmetic (factorization, sub-expression elimination), and loop optimizations
(blocking, fusion, fission). These general-purpose optimizations are often combined
with domain-specific ones. After lowering away domain-specific
knowledge, compiling an \emph{intermediate representation} (IR)
to executable code often requires the DSL library developers to implement
standard compiler passes. Sharing the developed technology between distinct
projects is challenging due to integration issues with different programming languages, lack of standardization, and maintenance support.

MLIR \citep{mlir} has partially solved this interoperability problem,
especially for deep-learning libraries, by providing an extensible and
community-driven unified IR.\@ However, efforts are still needed to bridge
the gap between HPC concepts and MLIR.\@
This work aims to address this gap by introducing key HPC abstractions designed
for interoperability with existing MLIR dialects. We utilize xDSL \citep{xdsl},
a Python-native clone of the MLIR toolchain to achieve this goal.
In this work we focus on explicit finite-difference (FD) stencil computations on structured grids. While FD stencils serve as a meaningful
and illustrative example to showcase the applicability of the proposed techniques,
the intention is to contribute to the advancement of HPC practices more broadly,
with the FD stencil computation acting as a representative case study within a
larger ecosystem of potential applications.
This paper makes the following contributions:
\begin{itemize}
\item A set of HPC-specific compilation components comprising SSA dialects
      and lowerings between them:
  \begin{itemize}
    \item An \emph{SSA dialect to facilitate automated domain decomposition}
          for distributed-memory execution of stencil kernels via message-passing
          (\cref{sec:dmp}).
    \item An \emph{SSA dialect for message passing} as a set of modular 
          operations in a standardized SSA-based IR
          (\cref{sec:mpi});
  \end{itemize}
\item A prototype implementation of abstraction-sharing compilation stack
      for two HPC stencil-DSL compilers, PSyclone and Devito,
      based on the concepts of \emph{SSA} and \emph{Region},
      utilizing the MLIR and xDSL compiler frameworks and extending concepts
      from the Open Earth Compiler (\cref{sec:analysis_dsls}).
\item A performance evaluation demonstrating that our approach offers
      competitive performance for a range of FD stencil computations, 
      compared to the existing domain-specific compiler stacks,
      for CPU shared- and distributed-memory parallelism, GPUs and FPGAs
      running at scale on the ARCHER2 and Cirrus supercomputers (\cref{sec:eval}).
\end{itemize}
The paper is organized as follows:
\cref{sec:hpc_dsls} discusses our compiler-centric approach,
\cref{sec:ssa_region} introduces MLIR for abstraction sharing,
\cref{sec:ssa_abstractions_hpc} presents our SSA dialect abstractions that are the main
contributions of this work, \cref{sec:analysis_dsls} presents a prototype implementation
for the stencil DSLs used in this work to validate our contributions, and \cref{sec:eval}
evaluates the performance of our final software stack.\@ \Cref{sec:related,sec:conclusions} refer to related work and conclude with future directions.

\section{Stencil DSLs with Shared Abstractions}\label{sec:hpc_dsls}

We propose a compiler-centric approach where stencil DSLs maximize the sharing of their
implementation (e.g.\ common abstractions from their application domains or
commonly used HPC abstractions) while maintaining the community-tailored
interfaces and workflows that have proven critical to communicate effectively
with their domain experts. By compiler-centric we mean that we build and contribute
compiler infrastructure and leverage capabilities and optimizations provided by compilation frameworks,
such as transformation passes, abstractions, performance optimizations, portability, and productivity.

We focus on three DSLs targetting explicit FD stencils on structured meshes,
Devito \citep{devitoTOMS2020}, the Open Earth Compiler \citep{openearth2022}, and PSyclone \citep{adams2019lfric}.

\begin{figure}[htpb]
  \includegraphics[width=0.6\columnwidth]{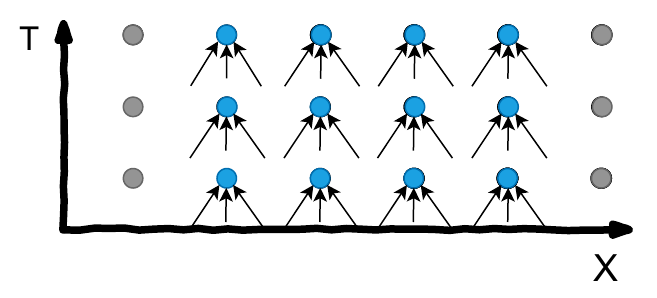}
  \caption{A 1D-3pt Jacobi stencil. Point updates depend on neighbouring values of the previous timestep}\label{fig:stencil_intro}
\end{figure}

FD-stencils computation pattern is an iterative computation update of an element in an $n$-dimensional spatial
grid at time $t$ as a function of neighboring grid elements (space dependencies) at
previous timesteps {$t-1, \dots, t-k$} (time dependencies).
\Cref{fig:stencil_intro} illustrates a 1D-3pt Jaocobi stencil
and its flow dependences. Each point is updated using values from
the previous timestep and the right and left neighbors.
Arrows illustrate the flow dependencies.
Halo points (grey) are read-only and extend the
computational domain by the size of the stencil radius.

All the prementioned DSLs and their compiler implementations are developed as siloed, standalone frameworks
with zero code sharing (\cref{fig:today}).
However, despite this lack of code sharing, all three stencil compilers use similar concepts and transformations for domain-specific and general-purpose HPC optimizations.
To address this lack of code sharing, we propose a novel shared compilation stack for these three frameworks (\cref{fig:vision}), where domain-specific optimizations and HPC concepts are shared.
User interfaces differ, but the underlying
computational abstractions reason about common concepts.

Devito offers a DSL and compiler framework to automate the FD method and is primarily used in seismic inversion and medical imaging.
The Open Earth Compiler was built for weather and climate models
using stencils.
Finally, PSyclone is designed for weather, climate, and ocean models.
Devito is an embedded DSL and compiler framework written in Python.
The Open Earth compiler is a domain-specific
compiler offering a frontend at the stencil specification level,
and PSyclone uses Fortran, augmented with specific coding conventions
to connect with the HPC community.

Nevertheless, all three frameworks
eventually arrive at the domain-specific concept of stencil computations.
Stencil kernels (and sequences thereof) are optimized and translated into imperative
code that runs on HPC machines. Hence, both the domain-specific
optimization at the stencil level and the code generation for an HPC
machine are conceptually similar.

We show that these similarities can be exploited to reuse
and share the lower-level compilation flow across the three frameworks.
The key to enabling such reuse is to share abstractions across the stencil
DSL compilers. The compiler community has embraced the concepts
of static single assignment (SSA) and nested control flow regions
as recommended program expression and transformation approaches.
The MLIR project \citep{mlir} has recently demonstrated that
these concepts can be instantiated for different domains (see
\cref{sec:ssa_region}) and used to provide a principled way to exchange
information through different stages of the compilation pipeline.
We embrace these concepts to design a shared compilation flow
(\cref{fig:vision}) that connects these three stencil DSL compilers.
Specifically, we demonstrate that all three compilers can use a
common stencil abstraction to optimize and generate code for their stencil kernels.
Starting from this stencil-level abstraction
(which is generated from disparate, domain-specific inputs),
the shared infrastructure translates the common stencil abstractions
to stencil kernels and leverages established SSA-based compiler IRs for loops, arithmetic, and memory operations.
For building a stencil-level abstraction we adopt the \emph{stencil} and \emph{GPU}
IRs in the form of MLIR dialects\footnote{https://mlir.llvm.org/docs/Dialects/}.
These two \emph{dialects}, introduced in \citet{openearth2022}, are
discussed in detail in \cref{sec:ssa_abstractions_hpc} where we expand their features to better support the stencil DSLs studied here. This was selected as a least-friction way to lower stencil-related concepts to MLIR and LLVM.\@
The Open Earth Compiler \citep{openearth2022} has already successfully targeted the lowering of stencil concepts
to the MLIR framework. We reuse and extend its infrastructure, and more specifically, its Stencil IR (\cref{fig:today}), used as the basis for the \emph{stencil} dialect in this work (refer to \cref{sec:stencil} for more details).

This lowering process involves a series of passes across SSA-based compiler IRs.
Additionally, we introduce new IRs based on the SSA+Regions abstraction (\cref{sec:ssa_region}) for lowering the \emph{stencil} dialect IR to distributed memory parallelism via the Message Passing Interface (MPI) standard.
The result is a comprehensive lowering stack that automatically
generates shared- and distributed-memory and GPU-accelerated
computations for Devito, the Open Earth Compiler, and PSyclone.

\section{Sharing Abstractions through IRs\label{sec:ssa_region}}

The abstractions transformed by compilers must concretely be realized
in code as some IR.\@ IRs that maintain SSA form \citep{cytron1991efficiently}
have proven particularly useful as their direct encoding of a program's
data flow allows transformations without requiring complex analyses.
Recently, the MLIR project \citep{mlir} demonstrated that adding
hierarchical structure to SSA-based IRs via nested regions facilitates the
modeling of higher-level domain-specific constructs, enabling
development of compilers in deep learning \citep{pytorch2019nips, tensorflow, tensorflow2015-whitepaper},
fully-homomorphic encryption \citep{SyFER2020}, and hardware
design \citep{eldridge2021mlir, majumder2021hir}. This \emph{SSA+Regions} concept
provides a principled way to interface distinct domain-specific IRs, which we adopt as a
foundational principle in our compiler stack.

In our SSA-based IRs, the primary constructs are \emph{operations}, chained 
by the \emph{values} they define and use. Each operation in an SSA-based IR has a
name, a list of values it uses called \emph{operands}, and may
define zero or more values called \emph{results}. All values have a name,
and following the SSA property, each name can be assigned at most once at
any program location. We model SSA-based IRs using MLIR syntax, both
for familiarity and to simplify interoperability with the existing
MLIR ecosystem.
For example, an addition operation (\texttt{arith.add}) naturally takes
two operands and returns one value. Operations may also carry \emph{attributes}
that encode static information on the operation directly. For instance,
the \texttt{arith.constant} operation carries a \texttt{value} attribute,
corresponding to the constant value it returns.
Here is an example in MLIR syntax (producing the integer value 84 in \mlirinline{|\%|res}):
\begin{mlir}
\end{mlir}

Each value has an associated \emph{type}, such as
the 32-bit integer type \texttt{i32} used in the example. For readability,
we sometimes shorten the syntax of operations by omitting the parentheses,
the attribute names, or the redundant types.
This results in the shortened syntax \mlirinline{|\%|0 = arith.constant 42: i32} for the constant operation.

To represent control flow and to model higher-level abstractions, operations can be
nested in \emph{regions}, which are themselves attached to operations. A region contains a
control-flow graph of \emph{blocks}, each containing both a list of
values (the \emph{block arguments}) and a list of operations.
Since the abstractions we introduce in this paper only use regions with a single block,
we will omit the notion of \emph{blocks} for the rest of the paper, and will
use the term \emph{region arguments} to refer to the region arguments
of the first (single) block in a region. As an example of the structure that regions capture,
a \texttt{for} loop can be expressed as
\begin{mlir}
scf.for(
{
  ^(
    ...
}
\end{mlir}
This loop iterates from \mlirinline{|\%|lo} to \mlirinline{|\%|hi} with a stride of \mlirinline{|\%|stride}. Its body is represented as a nested region that
takes the loop induction variable \mlirinline{|\%|i} as a \emph{region argument} and then contains a list of operations in the block's body.

Sets of operations, types, and attributes related to a common abstraction are organized into
distinct units called \emph{dialects}.
For example, the \mlirinline{add} and \mlirinline{constant} operations
are part of the \emph{arith} dialect that covers arithmetic operations.
On the other hand, integer attributes (\mlirinline{42 : i32}) and integer
types (\mlirinline{i32}) are part of the \emph{builtin} dialect.

While various compiler frameworks use SSA-based IRs, the introduction
and use of regions is more recent and central in MLIR and xDSL.\@
MLIR is an LLVM subproject that supports SSA and regions in a performance-focused
C++ compiler framework. It comes with pre-defined
abstractions for arithmetic, structured control flow, memory references, GPUs,
and many more. The xDSL project is a Python-native clone of MLIR that provides
an embedding of SSA and regions into Python facilitating interoperability with Python-based
DSLs. Similar to MLIR, xDSL is open-source and publicly available on GitHub\footnote{https://github.com/xdslproject/xdsl}.
Both frameworks are built on the previously introduced concepts of \emph{SSA} and \emph{Region}
and share the same textual representation to share infrastructure without tight coupling
of code. Users may introduce their own abstractions by defining new
dialects specific to their application domain. Since most DSLs we are targeting are written in Python,
we start our compilation pipeline from xDSL and define our custom
abstractions there. We develop dialect-specific lowerings from xDSL to
existing MLIR dialects and then hand off to the MLIR toolchain where
we apply further lowerings to target particular hardware.

\section{SSA Compiler Abstractions for HPC\label{sec:ssa_abstractions_hpc}}

The memory requirements of computational science problems in HPC often exceed the capacity of DRAM on a single platform.
Fortunately, many HPC problems lend themselves to spatial domain
decomposition, allowing for the distribution of computational tasks
across multiple compute nodes within a cluster.
Thus, domain decomposition and message-passing abstractions are considered
integral parts of HPC.\@

While MLIR provides several dialects that also benefit HPC users,
such as \emph{omp} for OpenMP shared-memory parallelism (SMP),
\emph{gpu} for generic GPU programming across NVIDIA and AMD GPUs,
and \emph{vector} for vectorizing computations, it does not provide
a means to control distributed memory parallelism (DMP).

We extend the capabilities of the \emph{stencil} dialect introduced in \citet{openearth2022}
to target multi-node HPC clusters through a novel, extended version of this dialect (\cref{sec:stencil}).
This dialect captures the geometric decomposition pattern of where the domain is decomposed into rank-local domains.
The rank-local domains communicate at predefined points in the computation.
One of the uses of the \emph{dmp} dialect is in the developed lowerings from the existing \emph{stencil} dialect.
While the \emph{stencil} dialect contains the mathematical abstract representation of the problem, the \emph{dmp} carries a more concrete realization.
Although still at a fairly high level from a parallelism perspective, we consequently lower to the \emph{mpi} dialect,
an even lower-level realization of the parallelism and decomposition.

Frameworks can represent their stencil computations using the \emph{stencil} dialect,
benefiting from common IR and semantics.
The \emph{dmp} dialect is an abstraction for distributed-memory parallel (DMP) operations,
such as data exchanges through message passing (\cref{sec:dmp}).
Through the \emph{dmp} dialect, frameworks that have expressed their IR in \emph{stencil}
benefit from automatically decomposing and distributing the computational domain to the MPI ranks that are used for the computation.
We refer to the data owned by an MPI rank as ``rank-local'' data.
Finally, we introduce an MPI-based dialect (\cref{sec:mpi}) that mirrors MPI's
point-to-point and collective communications and serves as a lowering target for the \emph{dmp} dialect.
Communications are expressed using this novel dialect, facilitating the development of codes leveraging message passing.
The implementation of \emph{stencil}, \emph{dmp}, and \emph{mpi} dialects is open-source and available online.

\subsection{A Stencil Dialect\label{sec:stencil}}

The initial implementation of the \emph{stencil} dialect was introduced as part of the Open Earth compiler \citep{openearth2022}, to represent stencil computations and their sequences by capturing essential information, including iteration space bounds and stencil update patterns. This dialect acts as an intermediate step, helping to close the semantic gap between the scientist's problem description and the concrete implementation on CPUs and GPUs. 

A major benefit of the \emph{stencil} dialect is that it is problem, domain, and hardware independent, in contrast, for instance, to other approaches such as \citet{essadki2023} and \citet{li2024}. For example, \citet{essadki2023} introduced a \emph{cfd} MLIR dialect for stencils associated with Computational Fluid Dynamics (CFD) problems, with operations and transformations focused on this specific problem. Consequently, their dialect is tied to one class of DSL rather than providing the ability to generalize across domains. By contrast, the work described in \citet{li2024} targets the porting of stencil applications to the Sunway SW26010 heterogeneous many-core processor, with the dialect containing operations influenced by the target machine architecture. Instead, when using the \emph{stencil} dialect from \citet{openearth2022}, a programmer's code is lowered to a general-purpose description of a stencil computation independent of the hardware architectures. A major contribution of \citet{openearth2022} was to demonstrate that this general description of a stencil computation contains enough information for subsequent transformations to effectively operate upon when generating optimized HPC target executables.

In this work, we integrated the \emph{stencil} dialect into the xDSL framework and extended the dialect to enable integration with our DMP and MPI abstractions, facilitating support for various lowering targets, including CPUs, GPUs, and FPGAs. Due to its general purpose nature, the \emph{stencil} dialect contains 11 operations in total, compared, for instance, to \citet{essadki2023}, which only contains two. The following list briefly highlights the most important operations and types in the \emph{stencil} dialect.
\begin{itemize}
    \item \mlirinline{stencil.field} is a type that represents the memory buffer from which stencil input values will be loaded, or to which stencil output values will be stored
    \item \mlirinline{stencil.temp} is a type which represents stencil values and which the \mlirinline{stencil.apply} operation operates over
    \item \mlirinline{stencil.access} is an operation that accesses a value from a \mlirinline{stencil.temp} given the specified offset relative to the current position.
    \item \mlirinline{stencil.store} is an operation that writes values to a field on a user-defined range.
    \item \mlirinline{stencil.load} is an operation that takes a field and returns its values.
    \item \mlirinline{stencil.apply} is an operation that accepts a stencil function plus parameters and applies this stencil function to those parameters, generating an output \mlirinline{stencil.temp} field.
    \item \mlirinline{stencil.return} is an operation that terminates the \mlirinline{stencil.apply} stencil function and writes the results of the stencil operator to the temporary values returned by the stencil.apply operation.
\end{itemize}
\Cref{lst:stencil_example} provides an example of the \emph{stencil} dialect to compute a 1-dimensional 3-point Jacobi calculation (see \cref{fig:stencil_intro}).
\begin{listing}[bp]
  \begin{mlir}
                               -> !temp<?xf64>
                    -> !temp<?xf64> {
    // 
    stencil.return 
  }
  
  stencil.store 
       : !temp<?xf64> to !field<[0,128]xf64>
  \end{mlir}
  \caption{Example MLIR for 1-dimensional 3-point Jacobi stencil.\label{lst:stencil_example}}
\end{listing}
The \mlirinline{stencil.load} operation loads the values held in a buffer (the \mlirinline{!field} type) into a \mlirinline{!temp} type, which can be operated upon, whilst \mlirinline{stencil.store} stores the resulting values back to a buffer. It was found in \citet{openearth2022} that working with value semantics enables powerful optimization across stencil operators, and whilst those are not demonstrated in this work, we purposefully keep this design to ease their integration in further work.
The \mlirinline{stencil.apply} operation defines the Jacobi stencil computation, which is applied to values passed as operands. Its region represents the computation of individual outputs and runs across the entire field, where \mlirinline{stencil.access} accesses input values with a relative offset to the current index. For instance, in \cref{lst:stencil_example}, the direct grid neighbors and the value itself are accessed. The \mlirinline{stencil.return} operation then returns the result for the current grid cell.

Working with authors of the \emph{stencil} dialect, we have made various enhancements.
For instance, in the original dialect, bounds information was encoded as attributes of MLIR operations.
However, this required additional analysis and transformations in lowering from the logical coordinates of the stencil representation to zero-based memory operation indexing. To avoid these complexities, we modified the dialect to associate domain bounds information with the values themselves through their types. This allows any operation using stencil-related types to access this information directly through their operands. It not only simplifies the existing stencil transformations but enables independent lowering to the DMP abstraction, decoupling this from stencil-specific operations that define logical coordinates.

Previously, the \emph{stencil} dialect was only capable of representing 3D stencils and we have also enhanced the dialect so that it can be used with stencils of any number of dimensions. Not only does this enable us to run simpler, 1D and 2D problems, but furthermore provides flexibility for more complicated future domains. The original dialect was specifically tailored to target GPUs and-so we have enhanced the stencil transformations by providing an additional lowering pipeline which is better suited for shared memory parallelism by leveraging loop tiling to improve data locality.

By incorporating these enhancements, we improved the \emph{stencil} dialect's compatibility with other MLIR dialects, and specifically based upon requirements from other contributed dialects in this section. For instance, it is possible for subsequent transforms to determine the minimal halo shape and size that is required for distributed memory by scanning the \mlirinline{stencil.access} offsets which operate used on inputs of a \mlirinline{stencil.apply}. Such information can then be used as a basis for automatically distributing the computation on multiple nodes.

\paragraph{Limitations} The main functional limitation of the \emph{stencil} dialect is the requirement for compile-time known bounds. Whilst this requires recompilation of code when the problem sizes changes, the performance benefits delivered by having this information statically available during compilation are significant. For example, known bounds enable constant-folding of most of the memory access address computations and thus reduce register pressure which is critical to increasing parallelism on GPUs. Due to the nature of DSL compilers, propagating bounds from the call site to the stencil expression is usually not difficult in practice, while the existence of compile-time known bounds greatly aids in implementing more complex compiler analysis and optimization passes.

Another limitation is the absence of a high-level representation for modeling boundary conditions, and this feature is scheduled for future work. The current version of the dialect is flexible enough to manually encode boundary conditions as conditionals over the memory accesses. However, expressing these at a higher level could lead to more opportunities for optimizing code generation and enhancing the supported use cases for the users.

\subsection{\emph{dmp} dialect: An IR for Domain Decomposition\label{sec:dmp}}
This section introduces a novel dialect for distributed memory parallelism (DMP), the \emph{dmp} dialect.
\emph{dmp} is used to express parallel communication patterns as modular building blocks that lower computational dialects, such as \emph{stencil} to an IR for DMP.\@
\emph{dmp} offers a mechanism for describing the exchange of rectangular subsections of data among nodes.
Additionally, we introduce a pass to convert a stencil kernel into an IR for distributed computations.
This pass allows for optimizing intra- and inter-node performance gains using the \emph{dmp} dialect for targeting large-scale runs.

\begin{figure}[htbp]
  \includegraphics[width=0.96\columnwidth]{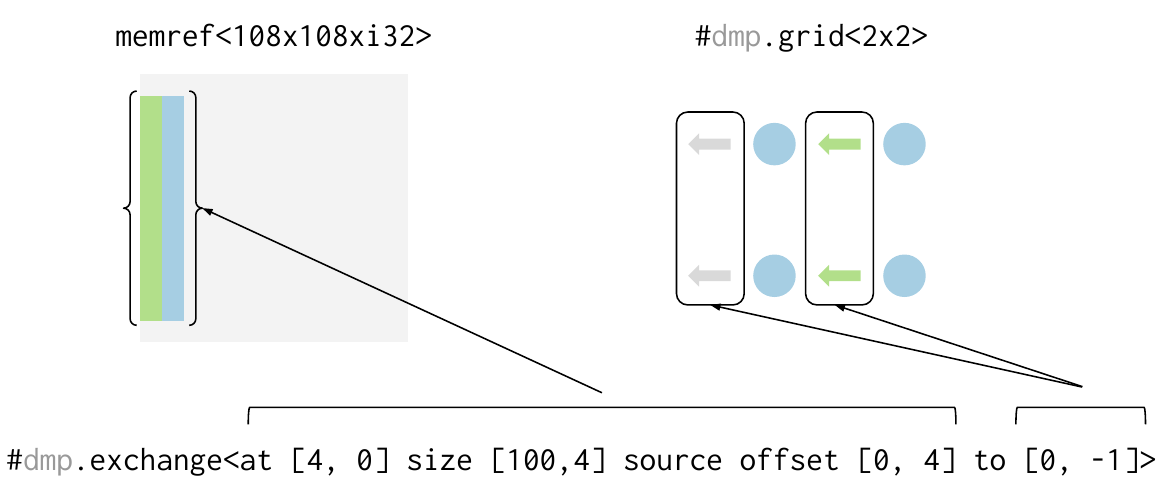}
  \caption{The exchange declaration defines a rectangular region of size 100 by 4, starting at (4, 0) inside a
  \mlirinline{memref} that needs to be updated with data from the neighbor at the relative position (0, -1).
  In exchange for receiving this data, a rectangular region of the same size, but offset by (0, 4) will be sent to
  the neighbor. This allows us to effectively model halo exchanges in a declarative way.\label{fig:exchange_decl}}
\end{figure}

The contributed \emph{dmp} dialect expresses this communication pattern using the \mlirinline{dmp.swap} operation (\cref{lst:halo-swap-op}), which takes
a \mlirinline{memref} as an argument and has attributes that specify which subsections should be exchanged with which ranks.
\begin{listing}[htbp]
\begin{mlir}
dmp.swap(
{
  "grid" = #dmp.grid<2x2>, 
  "swaps" = [
    #dmp.exchange<at [4, 0] size [100, 4]
                  source offset [0, 4] to [0, -1]>, 
    #dmp.exchange<at [4, 104] size [100, 4]
                  source offset [0, -4]  to [0, 1]>
  ]
} : (memref<108x108xf32>) -> ()
\end{mlir}

\caption{A high-level declarative expression of a data subsection exchange from some buffer.}\label{lst:halo-swap-op}
\end{listing}
The swap is configured using two parameters, the number and the cartesian topology of the ranks participating
in the swap (using the \mlirinline{dmp.grid} attribute) and a list of exchanges that are to be carried out (using the 
\mlirinline{dmp.exchange} attribute). 
As illustrated in \cref{fig:exchange_decl}, each exchange marks two rectangular subsections of the memory region to
exchange (one to send from, one to receive into) and the relative offset of the rank with which these regions are to be
exchanged.

\begin{figure*}[tp]
  \includegraphics[width=\textwidth]{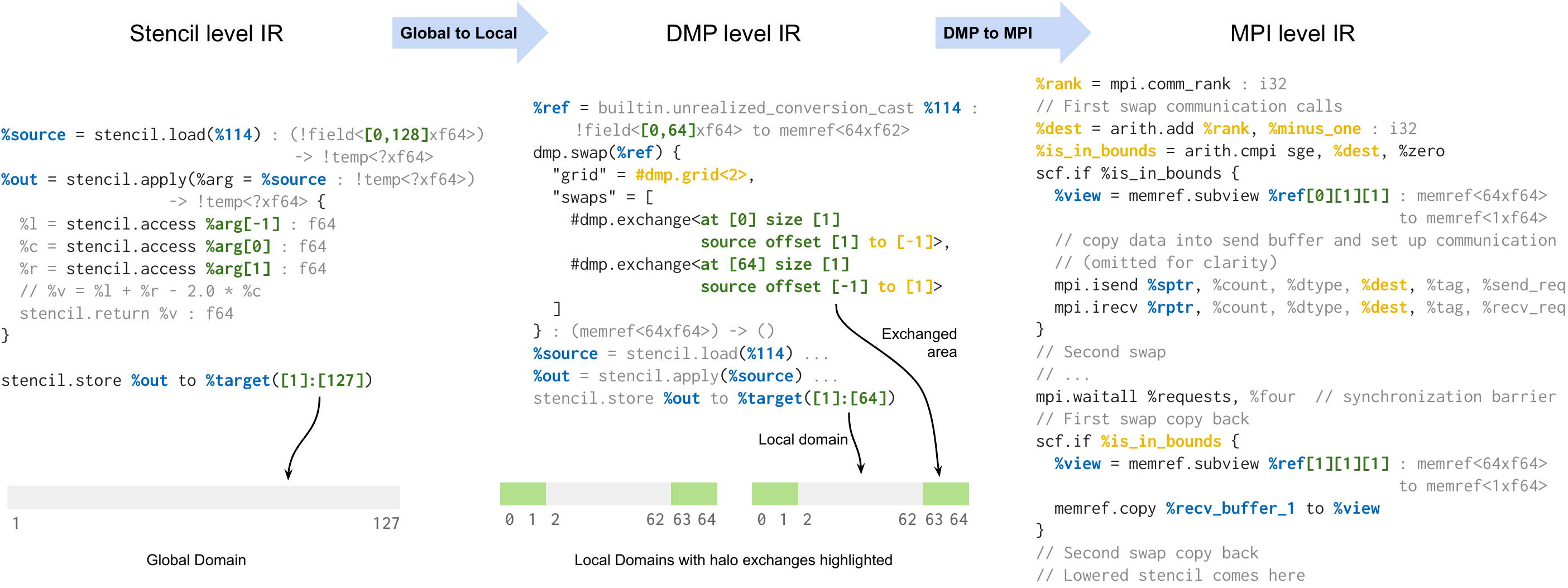}
  \caption{A global stencil being transformed to a local \emph{stencil} + \emph{dmp}, and then lowered to MPI.\@
  We highlight {\color{figureblue}the data being operated on},
  {\color{deepgreen}shape and halo information}, and {\color{figureyellow}communication-related information}.
  This showcases how we can enrich the IR with relevant information to perform efficient rewrites
  at every level of abstraction.}\label{fig:stencil_dmp_mpi_lowering}
\end{figure*}

While \emph{dmp} operations could be inserted by the frontend framework as desired, we offer a shared pass that automatically prepares \emph{stencil} programs for distributed execution. 
This pass is parameterized by information on the topology of MPI ranks in the computation, along with a \emph{decomposition strategy}.
This strategy describes how the data should be distributed among nodes.
Given this information, we equally decompose the domain represented in \emph{stencil} to a ``local'' data domain,
to the available ranks.
The \emph{stencil} dialect is also responsible for adding the necessary halos to local domains.
Subsequently, \mlirinline{dmp.swap} operations are inserted before each load, ensuring that neighboring ranks hold the updated data before proceeding to the following stencil computation.
While this may generate redundant data exchanges, a subsequent pass eliminates them via a further
pass analyzing the SSA data flow.
\Cref{fig:stencil_dmp_mpi_lowering} illustrates the lowering of the IR and the resulting stencil domain.

Internally, a \emph{decomposition strategy} is represented by a class that exposes an 
interface that allows a rewrite pass to calculate the local domain from the global domain.
It also provides the rank layout (the \mlirinline{dmp.grid} attribute) and generates the halo exchange
declarations (the \mlirinline{dmp.exchange} attributes) from the stencil access patterns. 
This extensible design allows adopters to supplement our default slicing strategy
with their own optimized versions if they require more exotic exchange patterns.
We provide a standard slicing strategy that supports 1D, 2D, and 3D decomposition.

The design of the \emph{dmp} dialect is directly influenced by design choices made in the \emph{stencil} dialect.
The compile-time known bounds of the stencil expression greatly aided in reducing the complexity of the \emph{dmp} operations and transformations,
demonstrating that the simplicity of the \emph{stencil} dialect directly benefits the available compiler optimizations.

\paragraph{Limitations} The \emph{dmp} dialect is a prototype implementation of domain decompositions for stencil computations.
It is designed to operate on contiguous, hyper-rectangular subsections of the domain and, therefore, does not support operations on sparse or non-rectangular-shaped domains. Extending the dialect to cover more data layouts is projected for future work.

Another limitation arises due to the fact that one exchange operation corresponds to one halo exchange, making the implementation of optimizations such as communication-computation overlap and exchanging multiple halos simultaneously more complicated than it needs to be. Finding a suitable extension to the \emph{dmp} dialect for more application domains and more efficient rewrites is also projected as an interesting direction for future work.

\subsection{\emph{mpi} dialect: An IR for Message Passing\label{sec:mpi}}

Our final contribution to the MLIR ecosystem is the \emph{MPI} dialect, designed to
mirror MPI's point-to-point and collective communications.
The \emph{mpi} dialect has already been upstreamed to the MLIR compiler toolchain\footnote{https://mlir.llvm.org/docs/Dialects/MPI/}.
This dialect serves as a target for our \emph{dmp} dialect and is lowered to MPI library function calls.

Our work here comprises the dialect itself, containing the abstract operations and types.
The operations correspond to the MPI calls, while the types represent MPI types such as request handles, communicators, and data types.
Additionally, we have introduced operations to reduce the friction between the MPI and the MLIR ecosystems.
These operations enable the expression of common MPI concepts outside of the library calls, such as request object lists and interactions with memory references.
Because MLIR and the LLVM backends lack an intrinsic notion of MPI, we have also developed
a lowering that will convert the MPI dialect into the underlying function calls
and associated memory allocations using the low-level \emph{llvm} and \emph{func} MLIR dialects.
The contributed \emph{mpi} dialect currently supports the following subset of the MPI 1.0 standard operations:
\begin{itemize}
  \raggedright
  \item Blocking and Non-blocking Point-to-Point communications:
  \texttt{MPI\_Send}, \texttt{MPI\_Recv}, \texttt{MPI\_Isend}, \texttt{MPI\_Irecv}
  \item Nonblocking request operations: \texttt{MPI\_Test}, \texttt{MPI\_Wait} 
  and  \texttt{MPI\_Waitall}
  \item Blocking reductions: \texttt{MPI\_Reduce} and \texttt{MPI\_Allreduce}
  \item Broadcast and Gather collective: \texttt{MPI\_Bcast}, \texttt{MPI\_Gather}
  \item Process Management: \texttt{MPI\_Init}, \texttt{MPI\_Finalize},
  \texttt{MPI\_Comm\_rank} and \texttt{MPI\_Comm\_size}
\end{itemize}
While this subset of MPI operations provides the necessary building blocks
for the current needs of our MPI applications, the design allows for extensions.
Additionally, we provide operations that simplify targeting \emph{mpi} from higher levels of abstraction.
Our compiler transformations and lowerings leverage these operations to implement DMP.\@

One of the lowerings we provide transforms \emph{dmp} to \emph{mpi} operations.
Note that there is nothing MPI-specific in the \emph{dmp} dialect, allowing targeting
other communication libraries if desired.
Lowering to \emph{mpi} involves several steps, including
allocating temporary buffers, building the MPI exchange mapping,
packing/unpacking data to/from buffers, and issuing non-blocking send/receive calls.
\Cref{fig:stencil_dmp_mpi_lowering} shows the resulting transformation,
with some details omitted for readability, such as setting skipped request
objects to the null request and temporary send/receive buffer allocations and de-allocations.
Since MPI communication often happens inside loops, any loop invariant calls are hoisted as part of this transformation.

As LLVM has no concept of MPI, we lower these operations to regular function 
calls using the \emph{func} dialect.
Motivated by the distribution on our target platform (ARCHER2), we support the \emph{mpich} implementation of the MPI standard. To make use of MPI, it is usually required
to include the implementations C header file, a notion not supported by MLIR.\@
Instead, we extract magic values from our library's header file and substitute them
for e.g.\ datatype constants during the lowering process. This makes our provided
MPI lowering specific to the \emph{mpich} library, but this strategy can be adapted to
other MPI libraries like OpenMPI.\@ Futhermore, there is an ongoing effort to provide
one standard MPI ABI, greatly simplifying this problem \citep{hammond2023mpi}.

\Cref{lst:mpi_mpi_send} illustrates the operation \mlirinline{mpi.send}
for sending a \mlirinline{memref} containing 128 double-precision floats.
In line 1, an operation in the MPI dialect is employed to unwrap a memref into an LLVM pointer to
the underlying buffer \mlirinline{|\%|buff}, the number of elements in the buffer \mlirinline{|\%|count},
and the corresponding MPI datatype.
IR similar to the one in \cref{lst:mpi_mpi_send} is generated by our lowering pass from
\emph{dmp} to \emph{mpi}, except using non-blocking communication.

As described, our \emph{mpi} dialect must be lowered to function calls that can be effectively
compiled by LLVM.\@ The lowered version of \cref{lst:mpi_mpi_send} is sketched in \cref{lst:lowered_mpi_send}.
In lines 1--6 of \cref{lst:lowered_mpi_send}, it can be seen that the \mlirinline{mpi.unwrap_memref} operation
has been expanded to extract a pointer to a contiguous memory area.
\begin{listing}[htbp]
\begin{mlir}
  : (memref<64x2xf64>) -> (!llvm.ptr, i32, !mpi.datatype)
mpi.send(
  : (!llvm.ptr, i32, !mpi.datatype, i32, i32) -> ()
\end{mlir}
\caption{The \emph{mpi} dialect makes it easy to target distributed memory 
systems from higher levels of abstraction by providing interfaces to work either 
with raw pointers or memrefs. Here we show a basic \texttt{MPI\_Send} call of an 
MLIR \mlirinline{memref}.}\label{lst:mpi_mpi_send}
\end{listing}
\begin{listing}[htbp]
\begin{mlir}
  : (memref<64x2xf64>) -> index
func.call @MPI_Send(
// ...
func.func @MPI_Send(!llvm.ptr, i32, i32, i32, i32, i32)
\end{mlir}
\caption{The \emph{mpi} dialect provides a uniform layer for higher-level dialects to target,
which are then lowered to implementation-specific function calls and constants.}\label{lst:lowered_mpi_send}
\end{listing}
Furthermore, since the size of the \mlirinline{memref} is compile-time known, the \mlirinline{|\%|count} can be statically calculated as well.
The MPI datatype constant here is specific to the MPI library and was extracted from a C header file.
It is worth noting that switching to a different underlying MPI implementation would require a modification to this part of the lowering process.
The \mlirinline{mpi.send} is replaced with \mlirinline{func.call}, an operation in the standard \emph{func} dialect that issues a function call.
In line 11, an external function definition of the \texttt{MPI\_Send} function, is appended to the end of the MLIR module.

\section{Lowering stencil DSLs to MLIR\label{sec:synergies}\label{sec:analysis_dsls}}

This section briefly introduces the frameworks used and then details the methodology to
leverage the synergies exploited from the contributed shared infrastructure.
\Cref{fig:vision} schematically presented the shared infrastructure contributed by this work.
Devito and PSyclone lower their IRs to the \emph{stencil}, \emph{dmp} and \emph{mpi} dialects,
benefiting from the advantages presented in \cref{sec:ssa_abstractions_hpc}.
Additionally, they leverage ``out-of-the-box'' lowerings to other dialects optimized for HPC,
such as \emph{omp}, \emph{gpu} designed for parallel execution on CPUs and GPUs.

\subsection{Devito\label{sec:devito}}

Devito \citep{devitoTOMS2020, devito2019} is an open-source Python DSL and compiler framework
widely used in academia and industry, aiming to ease the development of HPC finite-difference PDE solvers.
Devito's symbolic DSL is based on Sympy \citep{SymPy}, facilitating expressing and solving PDE simulations.
Devito's compiler framework automates the conversion into optimized FD kernels for efficient execution on CPUs and GPUs.
An example of the Devito DSL for modeling a 2D heat diffusion problem is shown in \cref{lst:devito-dsl}.
\begin{listing}[htbp]
\begin{python}
# Model the problem and automatically generate code
grid = Grid(shape=(126,))
u = TimeFunction(name='u', grid=grid, space_order=2)
eqn = Eq(u.dt, 0.5 * u.laplace)
op = Operator(Eq(u.forward, solve(eqn, u.forward)))
# JIT-compile the code and run
op(t=timesteps)
\end{python}
\caption{The high-level symbolic code to model 1-dimensional heat diffusion in Devito DSL.
         Users can only focus on modelling, all machinery for optimized HPC code generation is abstracted away.}\label{lst:devito-dsl}
\end{listing}

Devito's separation of concerns is benefiting interdisciplinary research and development by reducing time-consuming and error-prone
practices widely encountered in stencil codes.
Devito uses a high-level symbolic mathematic-textbook-like abstraction as input that is very close to what
mathematicians, physicists and geoscientists are familiar with.
It assists interdisciplinary scientists with the rapid development of simulation codes for wave propagators or other PDE-dominated problems,
as it facilitates the development of mathematically sophisticated PDE simulations,
the implementation of stencil kernel adjoints, boundary conditions,
sparse non-aligned-to-the-structured-grid operations \citep{bisbas2021} and others.
All the compiler optimizations are abstracted away from the DSL user: 
arithmetic and loop performance optimizations (e.g., common sub-expression elimination,
loop tiling, factorization), and backend-specific optimizations, such as SIMD vectorization, SMP and DMP, and GPUs.
Devito's compiler automates optimization passes, allowing users to apply further tuning heuristically.
On completion of the compiler passes, high-performing C code for efficient PDE solvers is automatically generated.

\begin{figure*}[t]
  \centering
  \includegraphics[width=0.90\textwidth]{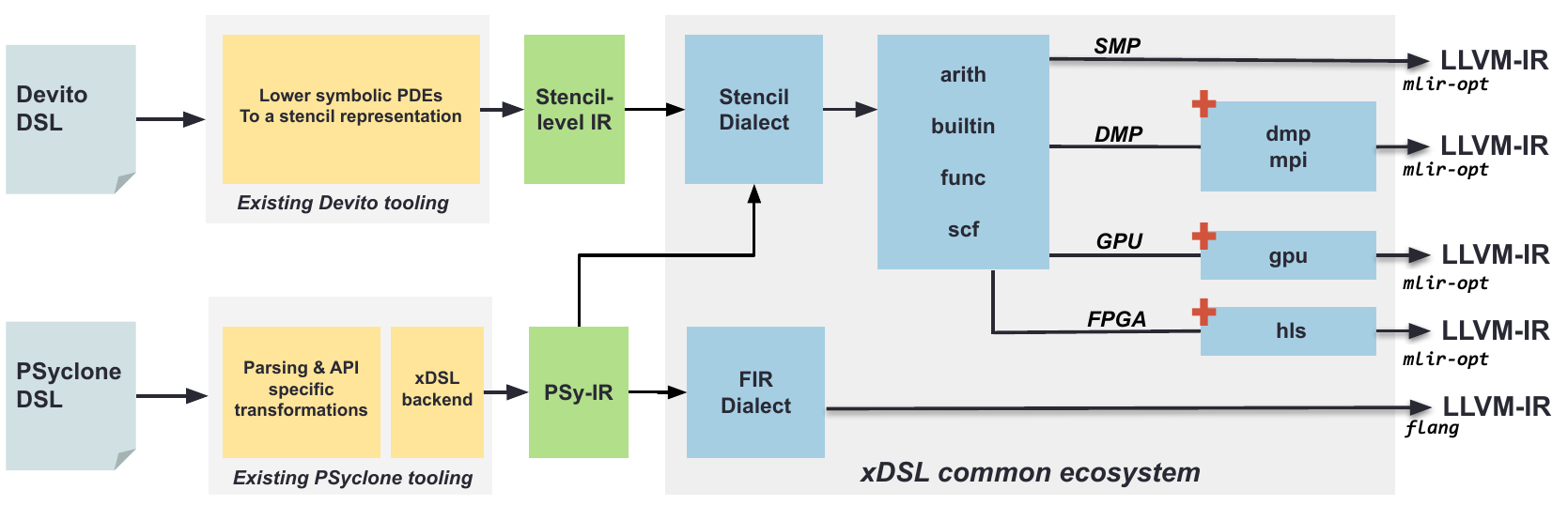}
  \caption{An overview of the Devito and PSyclone xDSL integration.
  The existing Devito IR (green) is lowered to xDSL's \emph{stencil} dialect, whereas an xDSL backend for PSyclone has been written and integrated into the existing tooling, which generates a bespoke PSy-IR matching PSyclone's own IR.\@ Irrespective, both progressively lower
  to xDSL's cloned MLIR dialects, plus the novel-contributed ones (\emph{dmp}, \emph{mpi}, \emph{hls}) following the same shared path.
  Finally, the LLVM-IR code is generated.\label{fig:shared_xdsl_flow}}
\end{figure*}

The shared Devito-PSyclone approach to lowering stencil IRs to xDSL is shown in \cref{fig:shared_xdsl_flow}.
The upper part of the \cref{fig:shared_xdsl_flow} illustrates an overview of lowering Devito's stencil representation
to xDSL and, subsequently, MLIR and LLVM.\@ Initially, Devito's symbolic input undergoes lowering to a representation, which encapsulates the FD-stencil representation regarding iteration spaces, memory accesses, mathematical operations, and many more.

\begin{figure}[htbp]
  \includegraphics[width=\columnwidth]{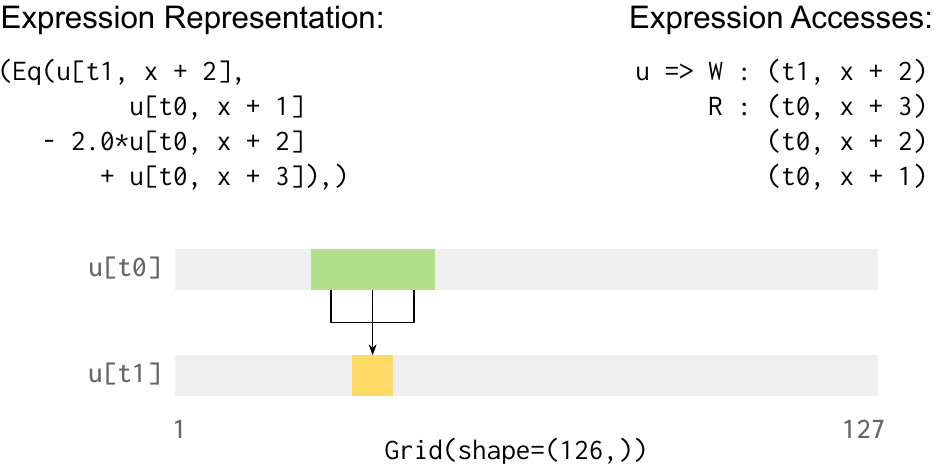}
  \vspace{-1em}
  \caption{Extracting Devito's IR lets us generate stencil dialect expressions.
  Devito provides info on read and write accesses (refer to \cref{fig:stencil_dmp_mpi_lowering}).}\label{fig:devito-semantics}
\end{figure}

The primary Devito object that contains the necessary semantics in Devito is a group of expressions. 
Subsequently, these are parsed and lowered to stencil expressions (\cref{fig:devito-semantics}).
Building upon the example presented in \cref{lst:devito-dsl}, \cref{fig:devito-semantics}
shows some of the IR that Devito is using to represent mathematical equations.
We parse info on read and write accesses from Devito's IR, and use this information to construct expressions
using the \emph{stencil} dialect, which will then be further lowered as described in \cref{fig:stencil_dmp_mpi_lowering}.

While a number of dialects are used when lowering Devito to the xDSL representation (refer to \cref{fig:shared_xdsl_flow}),
some of the main corresponding matches involved while visiting Devito structures are: the translation of mathematical expressions using operations from the \emph{func} and \emph{arith}, translation of Devito's memory accesses to \emph{memref} and \emph{stencil} dialect, translation of the problem dimensions to structured control flow using the \emph{scf} dialect and targetting specific backends using the \emph{mpi} and \emph{gpu} dialects. The integration with Devito happens at the highest possible level of Devito's compiler IRs to evaluate the optimization capabilities MLIR offers compared to the main Devito optimization pipeline.
This approach helps to compare pipelines that do not share any optimizations and start from the same level of abstraction.

The contributed \emph{stencil} dialect facilitates the lowering from Devito to build the FD structured grid, the necessary ghost-cell read-only region, the stencil coefficients, and memory accesses.
Subsequently, we add the temporal and spatial loops, including time-buffering.
We benefit from applying transformation and optimization passes from the shared infrastructure available from the MLIR community\footnote{https://mlir.llvm.org/docs/Passes/} such as \emph{cse, loop-invariant-code-motion, fold-memref-alias-ops} and \emph{convert-scf-to-openmp}.
Finally, HPC code leveraging SMP, DMP for CPUs, and GPUS is generated for several targets and then compiled and executed.
We evaluate these variants against standalone Devito in \cref{sec:eval}.

\subsection{PSyclone\label{sec:psyclone}}

PSyclone \citep{adams2019lfric} is a DSL for Fortran codes, enabling scientists to write their kernels in Fortran, a language that they are already familiar with, and then leverage a translation layer that abstracts the mechanics of the computation and parallelism. PSyclone is popular with weather and climate, for instance, the Met Office's next-generation weather model, LFric \citep{melvin2017lfric}, is using this technology and they are also involved in several other activities around PSyclone such as the NEMO ocean model port \citep{porter2018portable}. The major design feature of PSyclone is to enable a separation of concerns, where scientists develop the \emph{algorithm} and \emph{kernel} layers, both representing the mathematics of their problem and directing the sequence of maths operations that should be run, with the tool then generating the \emph{PSy} layer. This \emph{PSy} layer connects the \emph{algorithm} and \emph{kernel} and layers together, for instance, applying distributed memory parallelism. Furthermore, the PSyclone compiler will analyze the scientist's provided code and undertake optimizations and transformations, such as applying OpenMP or OpenACC at the loop level for threaded parallelism and GPU acceleration respectively.

However, the compilation stack of PSyclone is siloed with a bespoke translation layer that has been written in Python and, after parsing the Fortran source code, builds an internal intermediate representation based in a Directed Acyclical Graph (DAG) form. This IR is then operated upon by transformations bespoke to the PSyclone compiler before the IR is transformed into either Fortran or C. Domain-specific APIs enhance the core functionality of PSyclone, where an API contains a specialized set of transformations that target individual codes or scientific areas. For the results discussed in this work we leverage the NEMO API which has been developed to enable PSyclone to target the NEMO ocean model \citep{porter2018portable}.

\subsubsection{Lowering PSyclone to xDSL/MLIR/LLVM\label{sec:psyclone-lowering}}

\Cref{fig:shared_xdsl_flow} illustrates the shared Devito-PSyclone approach for lowering abstractions to xDSL, then building upon the shared transformations, existing MLIR dialects and new HPC dialects presented in this paper. Consequently, the PSyclone DSL is now a thin abstraction layer a-top a common shared ecosystem which is also used by Devito as described in \cref{sec:devito}. We still use the lexing and parsing of PSyclone to build the PSyclone IR.\@ This is then passed directly to our PSyclone xDSL backend to generate our own \emph{PSy IR}, an xDSL-based IR that closely resembles PSyclone's own IR.\@ Whilst this is still heavily PSyclone dependent, as it is in DAG form, there is a rich structure to the representation that appropriate transformations can exploit and a one-to-one mapping between PSyclone's existing IR and our xDSL \emph{PSy-IR}. 

An example of such a transformation that can be applied at this stage by the PSyclone xDSL backend is the identification of stencils from Fortran loops. Any identified stencils are represented in the \emph{PSy-IR} dialect which is then lowered to SSA form. Once in SSA form the flow is within the common xDSL ecosystem and there are two main dialects that \emph{PSy-IR} is lowered to. The stencil dialect, with the flow exactly as described for Devito in \cref{sec:devito}; and the FIR dialect, used by Flang to represent Fortran. Lowering the non-stencil aspects into this standard Fortran dialect not only means that we preserve the semantics of the surrounding code but also the \emph{escape hatch} of PSyclone which is the ability to handle all Fortran that is not transformed by PSyclone itself.

Whilst a mix of dialects is natural when working with MLIR, this raises a challenge with the LLVM tooling because the \emph{FIR} dialect is separated from other dialects. Flang, which defines the \emph{FIR} dialect, is only capable of working with that dialect and a very small subset of standard dialects such as \emph{arith} and \emph{func}. Likewise, \emph{mlir-opt}, which is used to operate on the standard dialects is unaware of \emph{FIR}. These different standard dialects are crucial; for instance, the \emph{gpu} dialect is required for launching on the GPU and the \emph{memref} dialect for handling memory within stencils, but Flang is not able to process this. 

As shown in \cref{fig:shared_xdsl_flow}, the FIR and stencil aspects are kept separate, with Flang used to build the LLVM-IR from the FIR dialect and mlir-opt the lowered dialects from transformations on the stencil dialect which are wrapped in functions for each stencil. Ultimately, this generated LLVM-IR is passed to the corresponding LLVM backend for the target architecture, and compiled into object files that are linked together to form the target executable where FIR can then call into the stencil dialect components by issuing the appropriate function calls.

\section{Evaluation}\label{sec:eval}

This section presents the experimental evaluation of this paper's
contributions versus the standalone Devito and PSyclone compiler stacks.
The systems used for the experiments were: 
(a) ARCHER2 HPE Cray EX Supercomputer nodes comprising a dual AMD
EPYC 7742 64-core 2.25GHz processor with 128 cores.
Each node has 8 NUMA regions and 16 cores per NUMA region,
32kB of L1, 512kB of L2 cache per core, 16MB L3 cache for every 4 cores
and supports AVX2 vectorization.
It has an HPE Slingshot interconnect with 200 Gb/s bandwidth,
dragonfly topology, and nodes are organized into groups of 128.
CCE's Cray Clang 11.0.4 was used.
(b) Cirrus GPU compute nodes consisting of four NVIDIA Tesla
V100-SXM2-16GB (Volta) GPU accelerators. On Cirrus, we used
version Cray Clang 11.4.100, nvc 22.11 from NVIDIA HPC SDK 22.11, and CUDA 11.6.

We use modified versions of xDSL, PSyclone, and Devito.
Experiments were executed with the Devito code available here
[\href{https://zenodo.org/doi/10.5281/zenodo.10854710}{xdslproject/devito}]
and xDSL code available here [\href{https://zenodo.org/doi/10.5281/zenodo.10854813}{xdslproject/xdsl}].
For the performance evaluation of the stencil kernels we use \emph{GPts/s} (a.k.a \emph{GCells/s})
for throughput.

\subsection{Devito\label{sec:devito-benchmarking}}

We use two benchmarks from CFD and seismic imaging: (i) heat diffusion, a Jacobi-like stencil, and
(ii) the isotropic acoustic wave equation.
We benchmark both problems in 2D and 3D for varying space discretization
orders (SDO) of 2, 4, and 8, resulting in 5-, 9-, 13-pt for 2D and 7-,
13-, 19pt stencils for 3D.
The isotropic acoustic wave equation uses a 2nd order accurate approximation in time, resulting in more points being read at the time dimension.
On ARCHER2, the problem size is $16384^2$ for the 2D and $1024^3$ for the 3D case.
On Cirrus, the problem size is $8192^2$ for the 2D and $512^3$ for the 3D case.
The simulation time in timesteps is $1024$ for the 2D and $512$ for the 3D case.
For the strong scaling experiments, we focus on the 3D problems with a bigger working set.
For all benchmarks, we use a 32-bit single-precision floating point numbers.

\begin{figure}[htbp]
  \centering
  \begin{subfigure}[b]{0.60\columnwidth}
    \includegraphics[width=\textwidth]{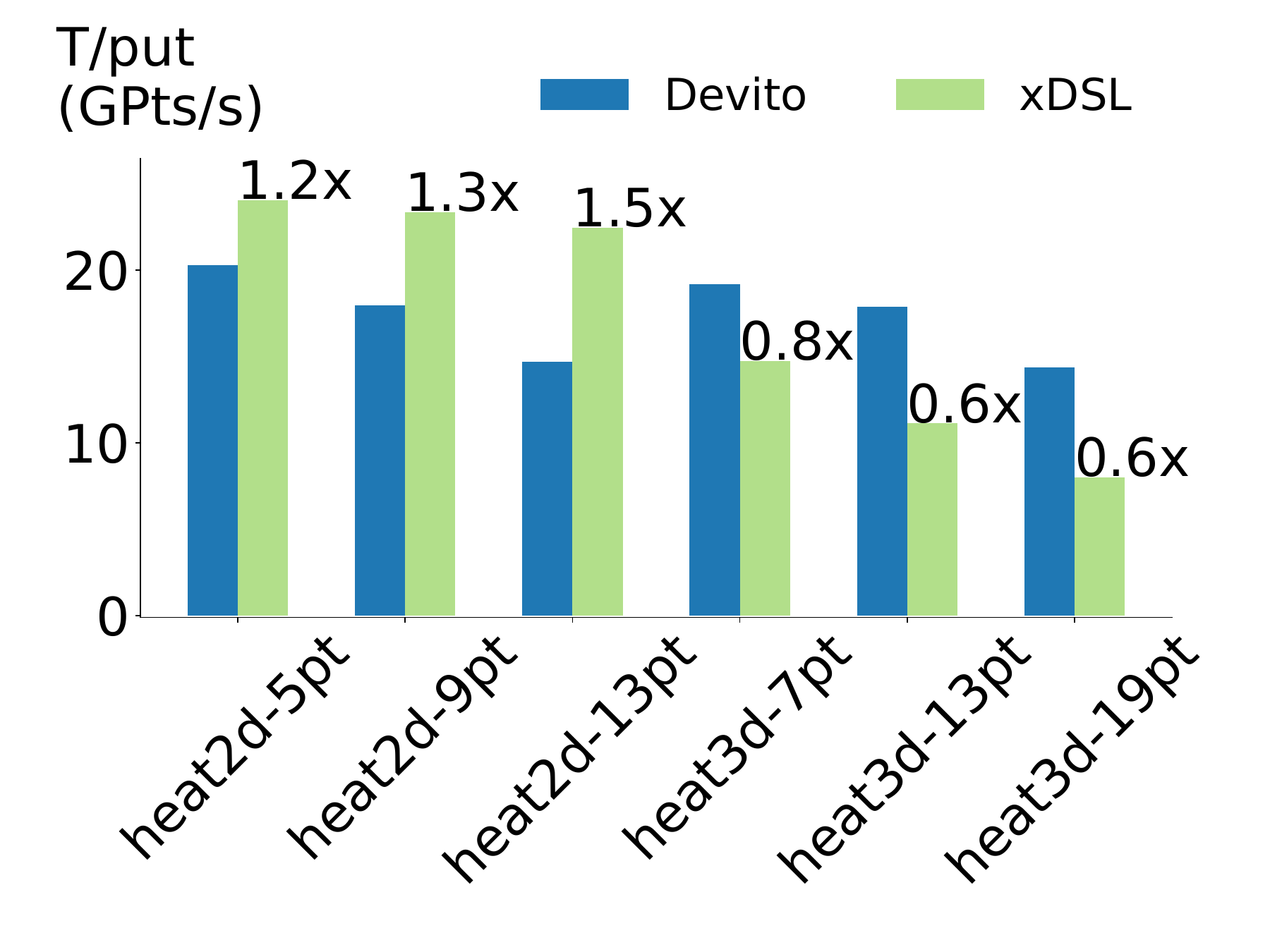}
    \caption{Heat diffusion kernels}\label{fig:singlenode_heat}
  \end{subfigure}
  \begin{subfigure}[b]{0.60\columnwidth}
    \includegraphics[width=\textwidth]{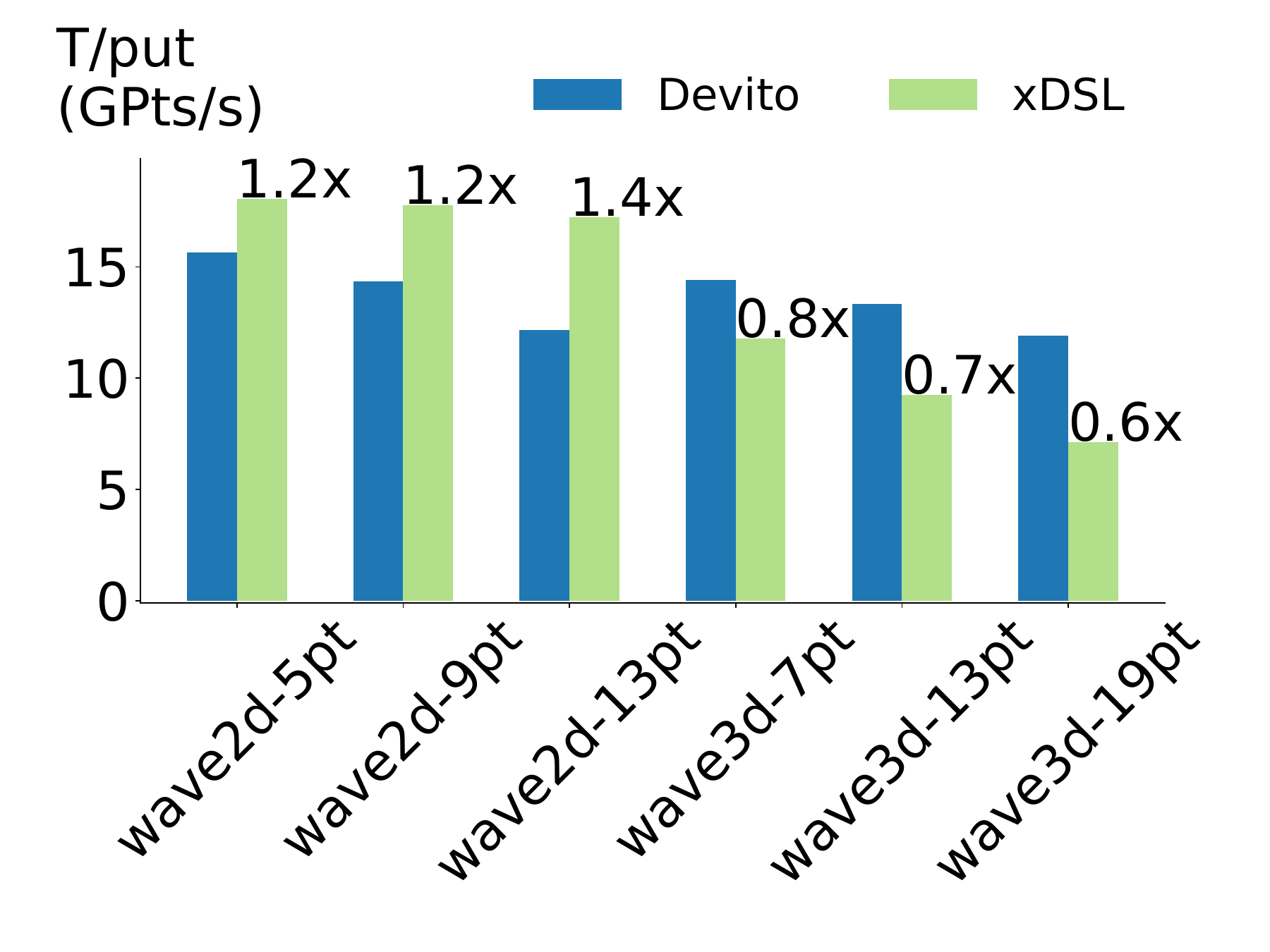}
    \caption{Acoustic wave kernels}\label{fig:singlenode_wave}
  \end{subfigure}
  \caption{xDSL-Devito does well on lower AI kernels, mostly dominated by
  memory bandwidth, but does not reach performance parity with Devito for high-AI problem instances.}\label{fig:singlenode_cpu}
\end{figure}
\Cref{fig:singlenode_cpu} evaluates xDSL-Devito versus the native Devito.
This experiment uses 8 MPI ranks and 16 OpenMP threads per rank on an
AMD EPYC 7742 node of ARCHER2, to better suit the NUMA architecture of the underlying architecture.
The Devito benchmark was optimized using the full suite of flop reduction, DMP, SMP, and SIMD vectorization optimizations \citep{devitoTOMS2020, bisbas2023automated}.
Devito's solvers are highly efficient, as established on roofline plots in
related work \citep{devitoTOMS2020, bisbas2021}, ensuring we compare against a
highly competitive baseline.\@ xDSL-Devito's performance outperforms Devito for low arithmetic
intensity (AI) kernels. However, Devito's flop reduction optimizations pay off for higher AI.\@
One reason for this is the limited vectorization performance of our
current lowered LLVM IR, which is essential for stencil kernels \citep{henretty2011}.

\begin{figure}[htbp]
  \centering
  \begin{subfigure}[b]{0.58\columnwidth}
    \includegraphics[width=\textwidth]{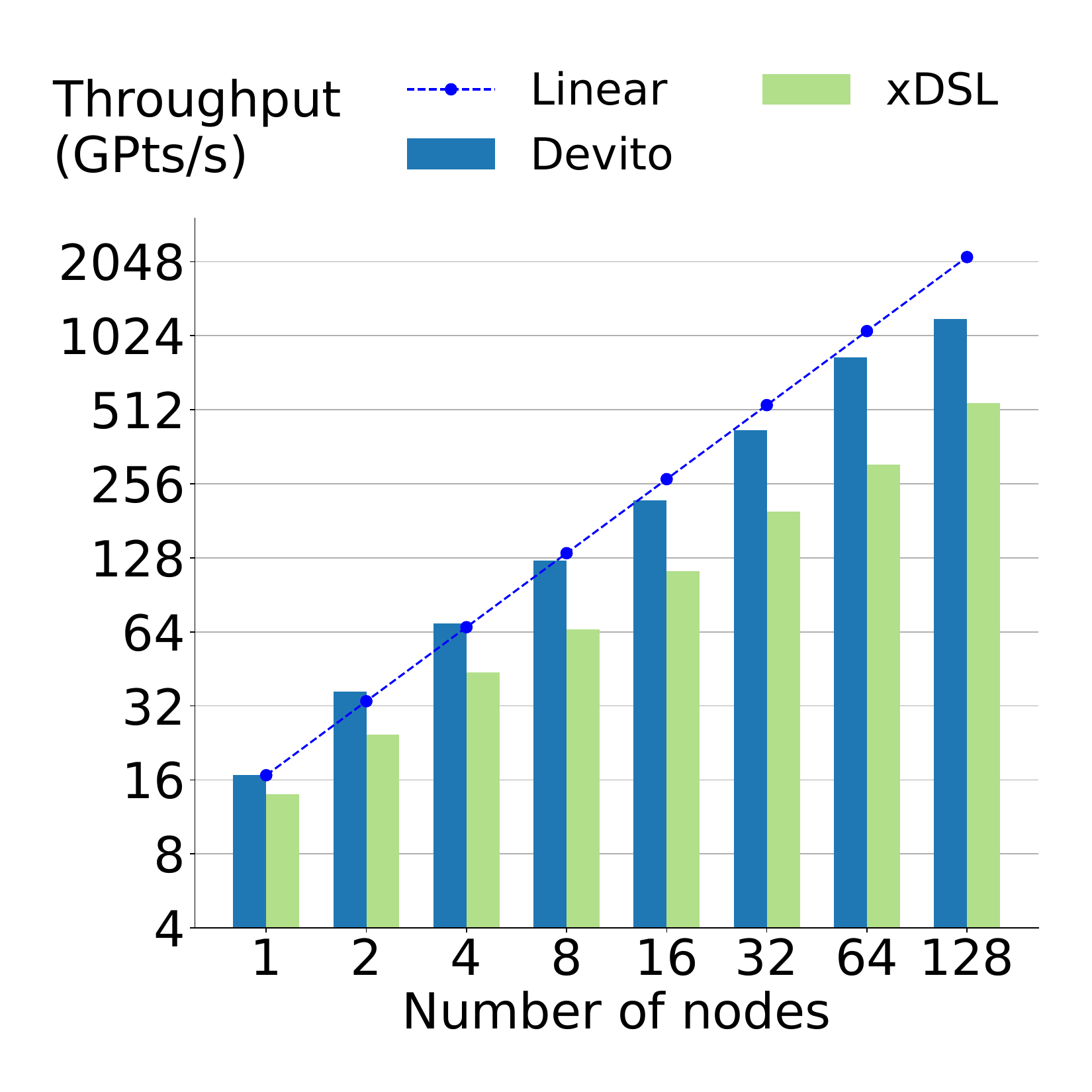}
    \caption{so4 Heat diffusion kernel}\label{fig:heat_multinode_gpts}
  \end{subfigure}
  \begin{subfigure}[b]{0.58\columnwidth}
    \includegraphics[width=\textwidth]{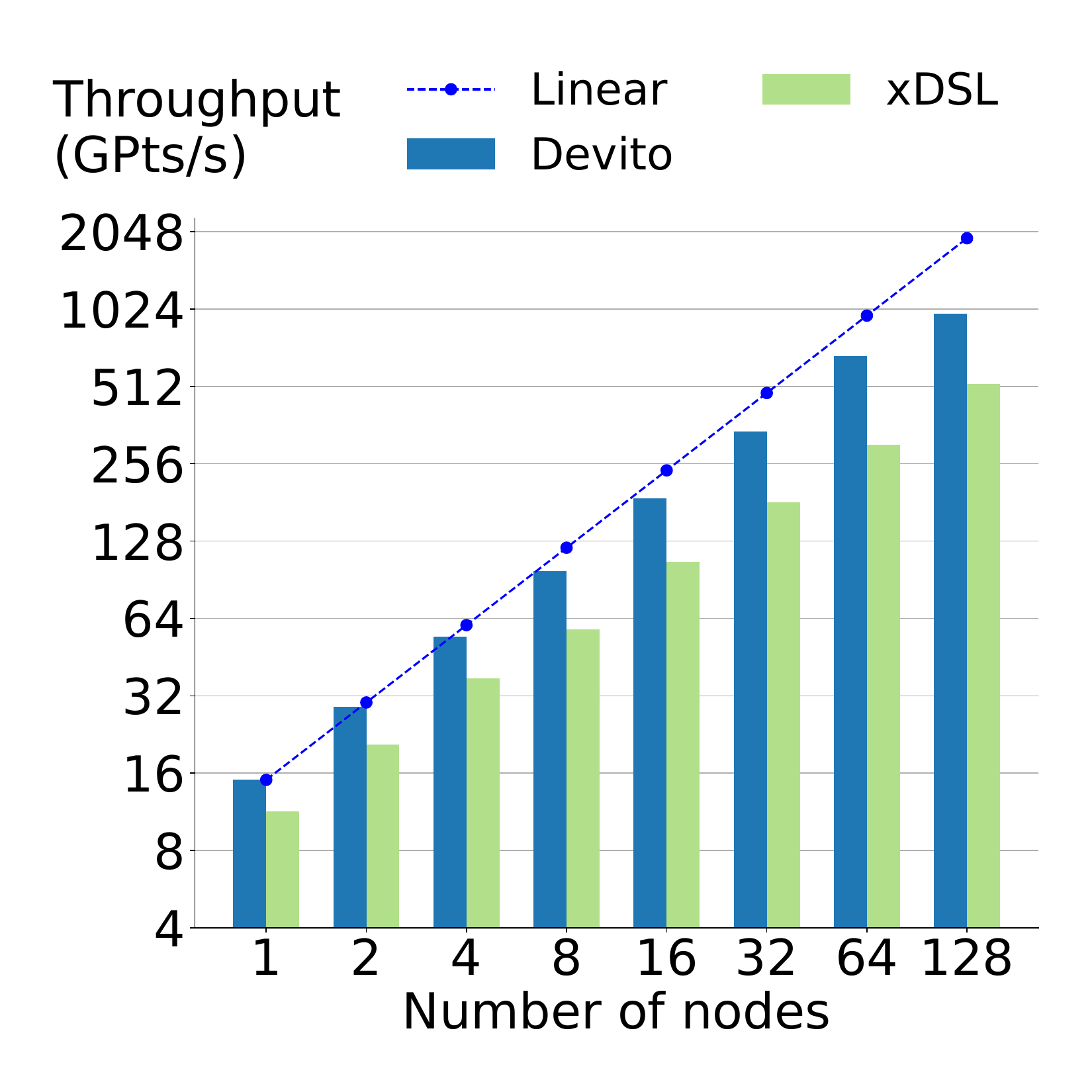}
    \caption{so4 Acoustic wave kernel}\label{fig:wave_multinode_gpts}
  \end{subfigure}
  \caption{xDSL-Devito's strong scaling follows the trend of Devito, but Devito still
           outperforms xDSL with its advanced communication techniques.}\label{fig:strong-scaling}
\end{figure}
\Cref{fig:heat_multinode_gpts,fig:wave_multinode_gpts} show the strong scaling
of the heat diffusion and acoustic wave stencil kernel, respectively.
We benchmark both kernels for a 3D problem and SDO of 4.
We use a whole ARCHER2 group of 128 nodes scaling up to 1024 MPI ranks, totaling 16384 cores.
We observe that xDSL-Devito exhibits strong scaling that may not match Devito's performance but still maintains the scaling trend.
This regression is expected, as Devito supports more advanced communication techniques, consisting of 3D diagonal exchanges leading to more robust and efficient scaling.
More details on the domain decomposition techniques employed by Devito can be found in \citet{bisbas2023automated}.
Extending computation/communication patterns in the \emph{dmp} dialect is planned for future work.
There is definitely space for more improvements on xDSL's end, which is discussed in \cref{sec:conclusions}.

\begin{figure}[htbp]
  \centering
  \begin{subfigure}[b]{0.58\columnwidth}
    \includegraphics[width=\textwidth]{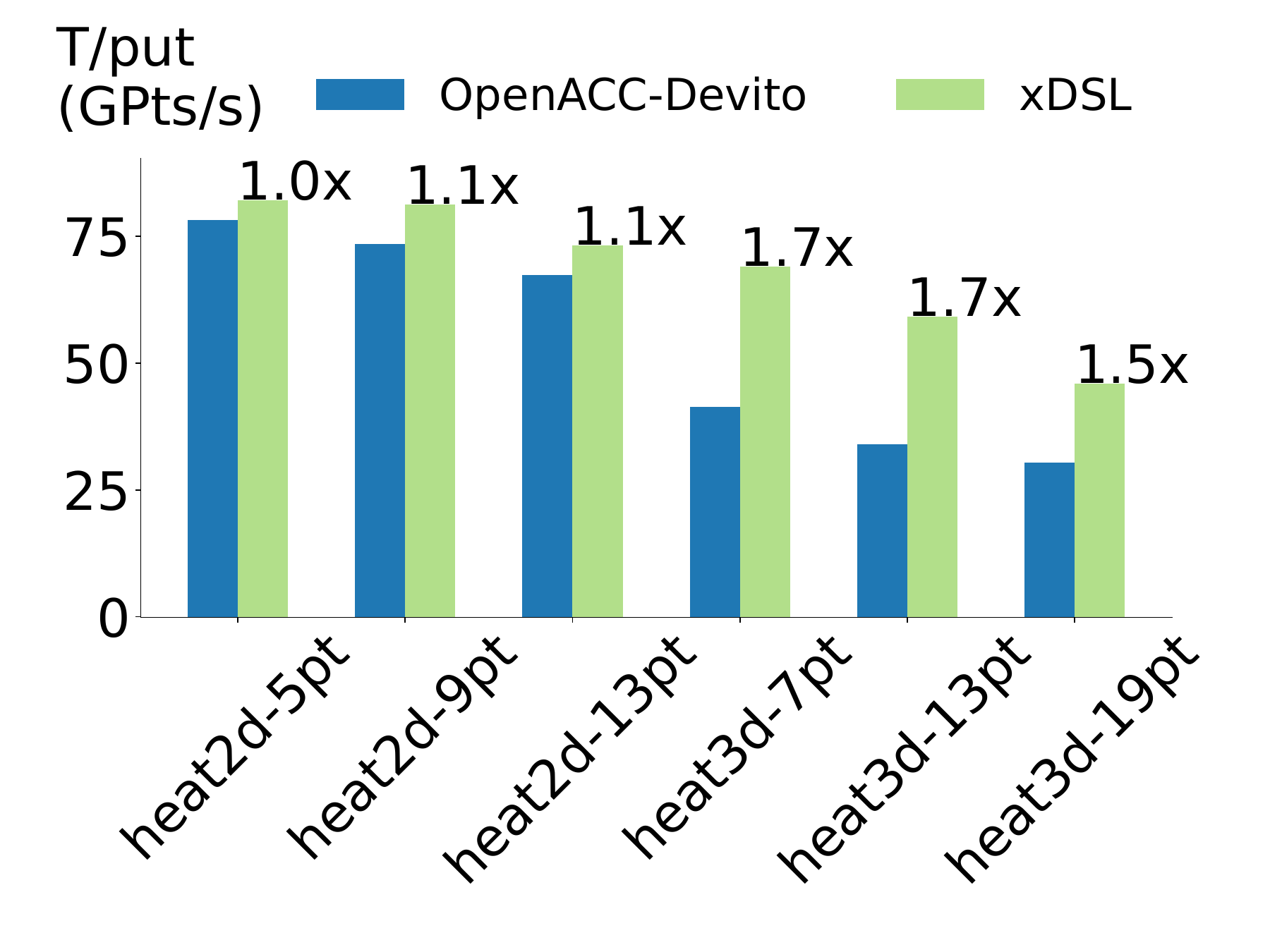}
    \caption{Heat diffusion kernels}\label{fig:singlenode_gpu_heat}
  \end{subfigure}
  \begin{subfigure}[b]{0.58\columnwidth}
    \includegraphics[width=\textwidth]{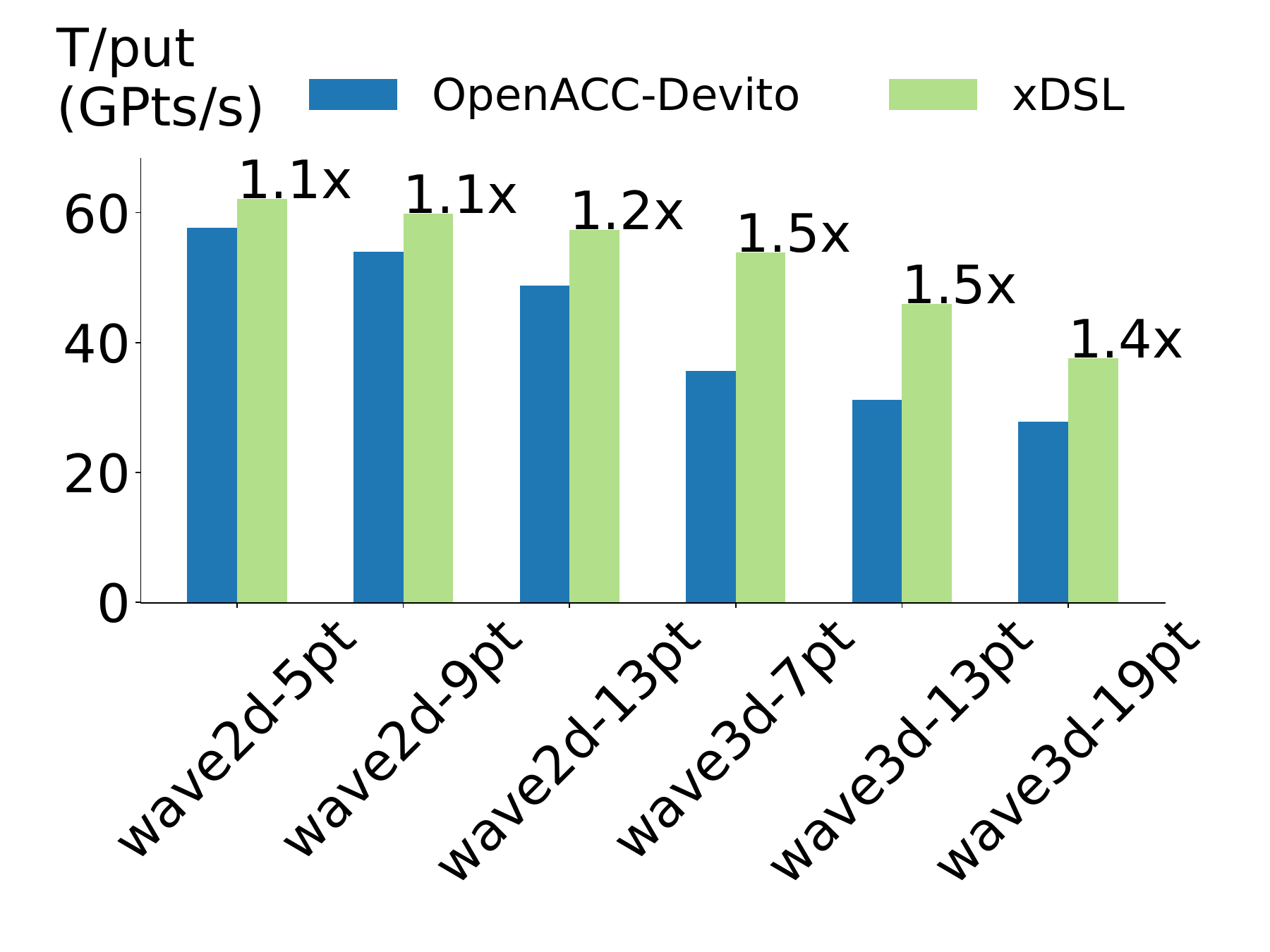}
    \caption{Acoustic wave kernels}\label{fig:singlenode_gpu_wave}
  \end{subfigure}
  \caption{xDSL's lowerings through CUDA outperform Devito's tiled OpenACC kernels for more than 1.5x
  when it comes for 3D kernels on an NVIDIA V100}\label{fig:singlenode_gpu}
\end{figure}
\Cref{fig:singlenode_gpu} presents the GPU evaluation.
We run on a V100, using OpenACC GPU offloading for Devito\footnote{DevitoPRO's license would be needed to compare against CUDA},
reporting the best performance among using
\emph{collapse(2)}, \emph{collapse(3)} or \emph{tile(32,4,8)} for the spatial loops of the stencil.
xDSL-Devito's kernels use MLIR lowerings to CUDA \citep{openearth2022} and use tiled execution.
These lowerings allow us to benefit from an out of the box CUDA backend.
Evaluation shows, as expected, that Devito/xDSL CUDA kernels outperform Devito's OpenACC for the higher-intensity wave
kernels while being mostly on par with the rest of them.
Profiling using \texttt{nsys}\footnote{https://developer.nvidia.com/nsight-systems},
we observe that the main reason is superfluous synchronization overhead on each kernel launch,
which is amortized in the larger kernels by the kernel runtime itself.
The MLIR version used does not yet offer a built-in solution for this.

\subsection{PSyclone}\label{sec:psyclone-benchmarking}
We have selected two benchmarks to explore performance for our approach with PSyclone,
the first is the Piacsek and Williams advection scheme \citep{piacsek1970conservation},
commonly used by Met Office codes such as the MONC high-resolution atmospheric model \citep{brown2020highly}, for calculating the movement of quantities through the atmosphere due to kinetic effects (e.g.\ wind). The second benchmark is the tracer advection kernel from the NEMO ocean model and part of the PSyclone benchmark suite \citep{psyclone_bench}. Both these schemes are extracted from production codes and the benchmarks are different in a couple of crucial ways. Firstly, PW advection contains three separate stencil computations across three fields, whereas tracer advection comprises 24 stencil computations across six fields. Each stencil computation involves many individual calculations, and for tracer advection, there are many more computations than fields due to dependencies between computations generating intermediate results. Furthermore, the tracer advection benchmark contains an outer loop where the entire stencil calculation is executed for 100 iterations.

\begin{figure}[thbp]
  \centering
  \begin{subfigure}[b]{0.58\columnwidth}
    \includegraphics[width=\textwidth]{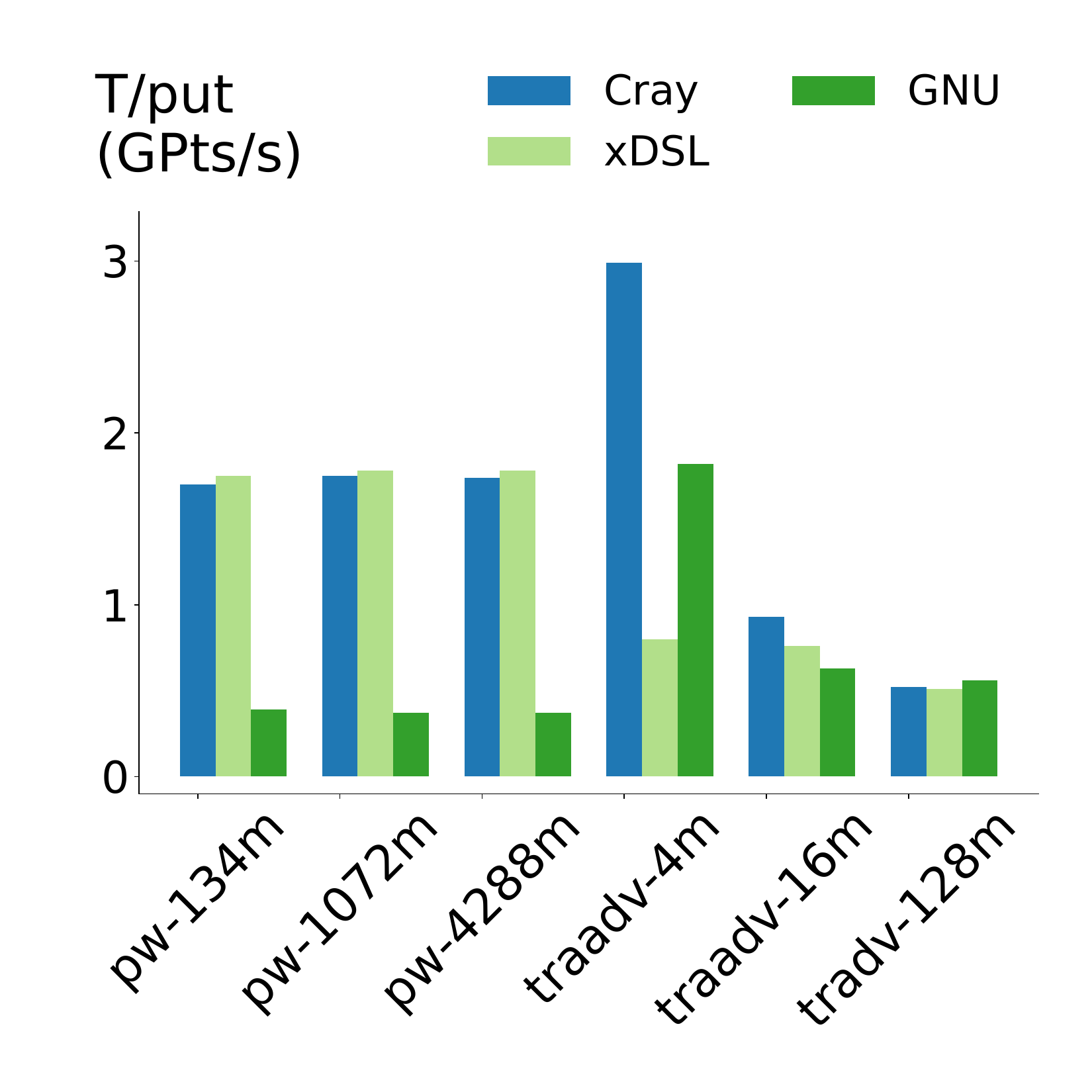}
    \caption{Single CPU node}\label{fig:singlenode_psyclone}
  \end{subfigure}
  \begin{subfigure}[b]{0.58\columnwidth}
    \includegraphics[width=\textwidth]{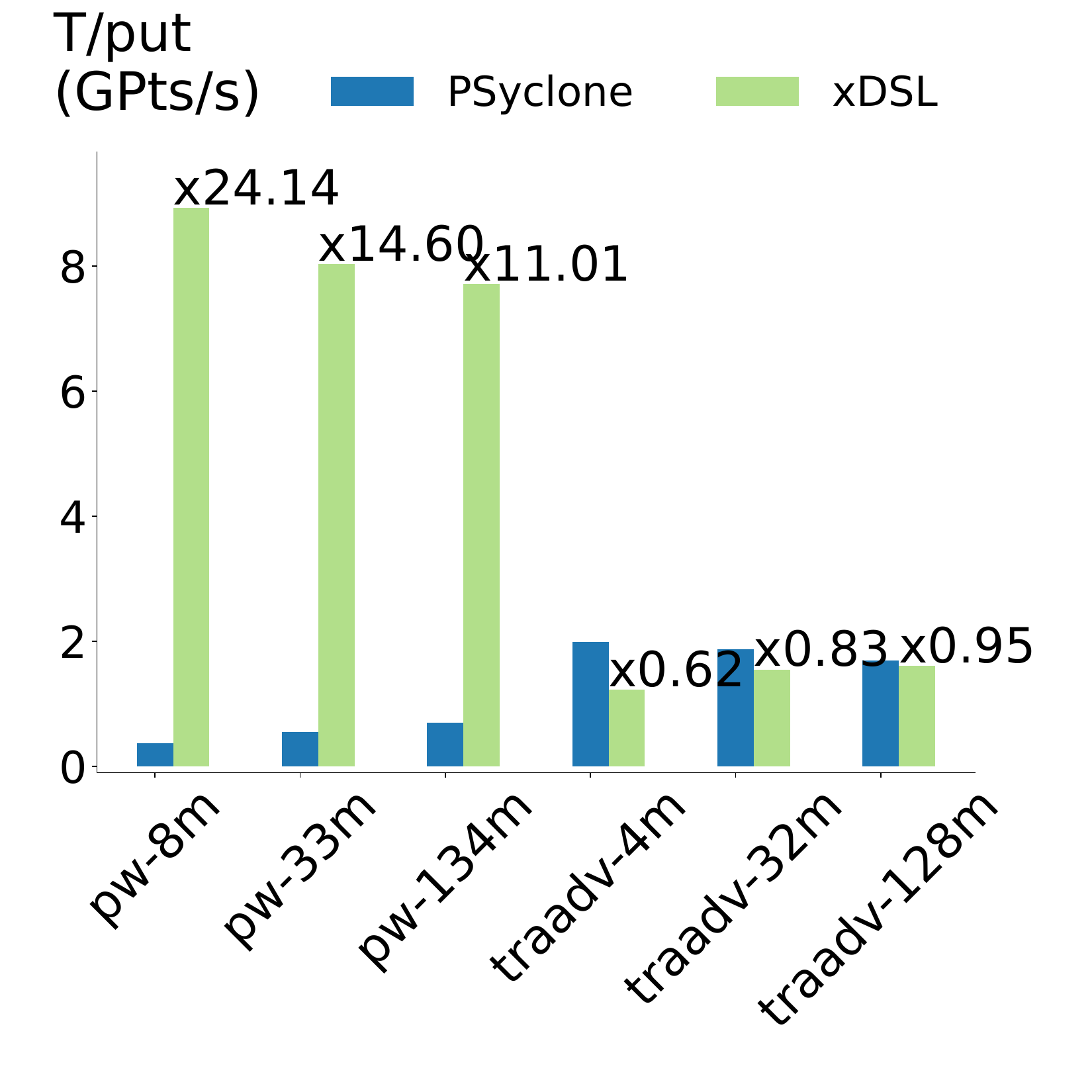}
    \caption{V100 NVIDIA GPU}\label{fig:accelerator_psyclone}
  \end{subfigure}
  \caption{xDSL-PSyclone single node CPU (ARCHER2) and GPU (Cirrus) throughput, where tracer advection benchmark performance is limited by the MLIR scf parallel lowering transformations.}
\end{figure}
\Cref{fig:singlenode_psyclone} provides a performance comparison of using our approach, \emph{xDSL-PSyclone}, against PSyclone where the target code is compiled with the Cray compiler, \emph{Cray-PSyclone} and the GNU compiler \emph{GNU-PSyclone}. We report throughput across our benchmarks (\emph{pw} for PW advection and \emph{traadv} for tracer advection) at different problem sizes. It can be seen that for the PW advection benchmarks the performance delivered by xDSL slightly exceeds that of PSyclone with the Cray compilers, whereas PSyclone with the GNU compiler is performing considerably worse. This demonstrates that the Cray compiler is undertaking numerous HPC optimizations over and above the GNU compiler, and given that xDSL is slightly outperforming the Cray compiler our stack can undertake similar levels of optimizations.

However, in \cref{fig:singlenode_psyclone} it can be seen that the performance is different for tracer advection, where for smaller problem sizes xDSL is considerably slower than PSyclone with both the Cray and GNU compilers, although the performance gap narrows for larger problems. The reason for this is the number of individual stencils, where for the PW advection benchmark the three stencil computations are fused into one single stencil region by xDSL, but with tracer advection there are 18 individual stencil regions due to dependencies. Limitations in the lowering from \emph{scf.parallel} to the OpenMP dialect result in each stencil region being lowered by MLIR into a separate parallel region. Consequently, the \emph{kmp\_wait\_template} function, which spins on a barrier waiting for parallel worker threads to complete, was the most runtime-intensive function when profiling smaller domain sizes. At larger problem sizes this issue, whilst still present, is ameliorated.

We then explored performance on a NVIDIA V100 GPU accelerator which is reported in \cref{fig:accelerator_psyclone}. On the GPU xDSL considerably outperforms PSyclone with the NVIDIA compiler for the PW advection benchmark and this is due to how data allocation is handled. PSyclone uses the managed memory option, whereas the xDSL GPU lowering detects that explicit memory allocation on the device will be more performant and handles this directly. Consequently, when the PSyclone compiled PW advection benchmarks were executed we found that there were a large number of unified memory GPU page faults which do not occur with xDSL.\@

The performance pattern for tracer advection on GPUs demonstrated in \cref{fig:accelerator_psyclone} is similar to that of a single node illustrated in \cref{fig:singlenode_psyclone}, where xDSL lags PSyclone, especially for smaller problem sizes. This is also because of limitations in MLIR lowerings, this time from \emph{scf.parallel} to the GPU dialect, where MLIR invokes a synchronous kernel execution for each parallel loop and thus execution across the CPU and GPU blocks at the end of each kernel. 

\begin{table}[tp]
\centering
\caption{xDSL-PSyclone supports execution on an Alveo U280 FPGA, which is not supported by PSyclone and can undertake auto-tuning of the kernel to a dataflow architecture delivering significant performance improvements compared to the initial version.}\label{tab:fpga-psyclone}
\resizebox{\columnwidth}{!}{ 
    \begin{tabular}{lccc}
      \toprule
      \multirow{2}{*}{\textbf{Benchmark}} & \multicolumn{2}{c}{\textbf{Throughput (GPts/s)}} & \textbf{Performance} \\
                                          & \textbf{Initial} & \textbf{Optimized} & \textbf{improvement} \\
      \midrule
      pw-8m      & $1.0 \times 10^{-3}$ & $1.0 \times 10^{-1}$ & 100x \\
      pw-33      & $8.5 \times 10^{-3}$ & $1.4 \times 10^{-1}$ & 165x \\
      pw-134m    & $8.6 \times 10^{-3}$ & $1.5 \times 10^{-1}$  & 175x \\
      traadv-4m  & $4.5 \times 10^{-4}$ & $5.1 \times 10^{-2}$  & 113x \\
      traadv-32m & $3.6 \times 10^{-4}$ & $7.7 \times 10^{-2}$  & 214x \\
      \bottomrule
    \end{tabular}
  }
\end{table}
\Cref{tab:fpga-psyclone} reports performance for both benchmarks on an Alveo U280 FPGA.\@ We support FPGAs by the development of an MLIR \emph{HLS} dialect and transformations \citep{rodriguez2023stencil} which lower from this to LLVM-IR and others which lower to the HLS dialect from the \emph{stencil} dialect. LLVM-IR is then provided to the AMD Xilinx HLS LLVM backend which synthesizes the corresponding kernel \citep{fortran-fpga}. The \emph{Initial} version represents the algorithm running on the FPGA unchanged from it's Von Neumann based CPU design, whereas the \emph{optimized} version has been transformed by the compiler into a form tuned for dataflow architectures. This transformation has involved breaking the constituent components of the algorithm into separate dataflow regions and the use of a 3D shift buffer \citep{brown2021accelerating} which is a bespoke cache that, when filled, enables all the current grid cell's stencil values to be provided to the calculation each cycle but one value needs to be read from DDR external memory per cycle. Ultimately, this means that calculation loops are able to be pipelined and run without stalling, computing a grid cell each cycle.

These performance numbers demonstrate two important benefits of our flow, firstly that we provide the ability to target FPGAs from a single Fortran source code, adding a new capability to PSyclone. Secondly, as can be seen by the performance improvement between optimized and initial versions in \cref{tab:fpga-psyclone}, working with a high-level stencil description of the problem enables transformations to undertake automatic tuning to a dataflow architecture. Manual FPGA tuning is a very time-consuming step that requires considerable expertise \citep{brown2020exploring}, and so the ability for xDSL to automatically undertake this provides an important capability that has the potential to enhance vendor tools. Whilst the FPGA numbers reported in \cref{tab:fpga-psyclone} fall short of the NVIDIA V100 GPU performance, these are not only in agreement with the performance of tuned PW advection manually optimized for FPGAs \citep{brown2021accelerating}, but it should also be highlighted that the MLIR GPU and CPU pipelines that we leverage are of production quality maturity and optimized by vendor teams.

\begin{figure}[htbp]
  \centering
  \begin{subfigure}[b]{0.58\columnwidth}
    \includegraphics[width=\textwidth]{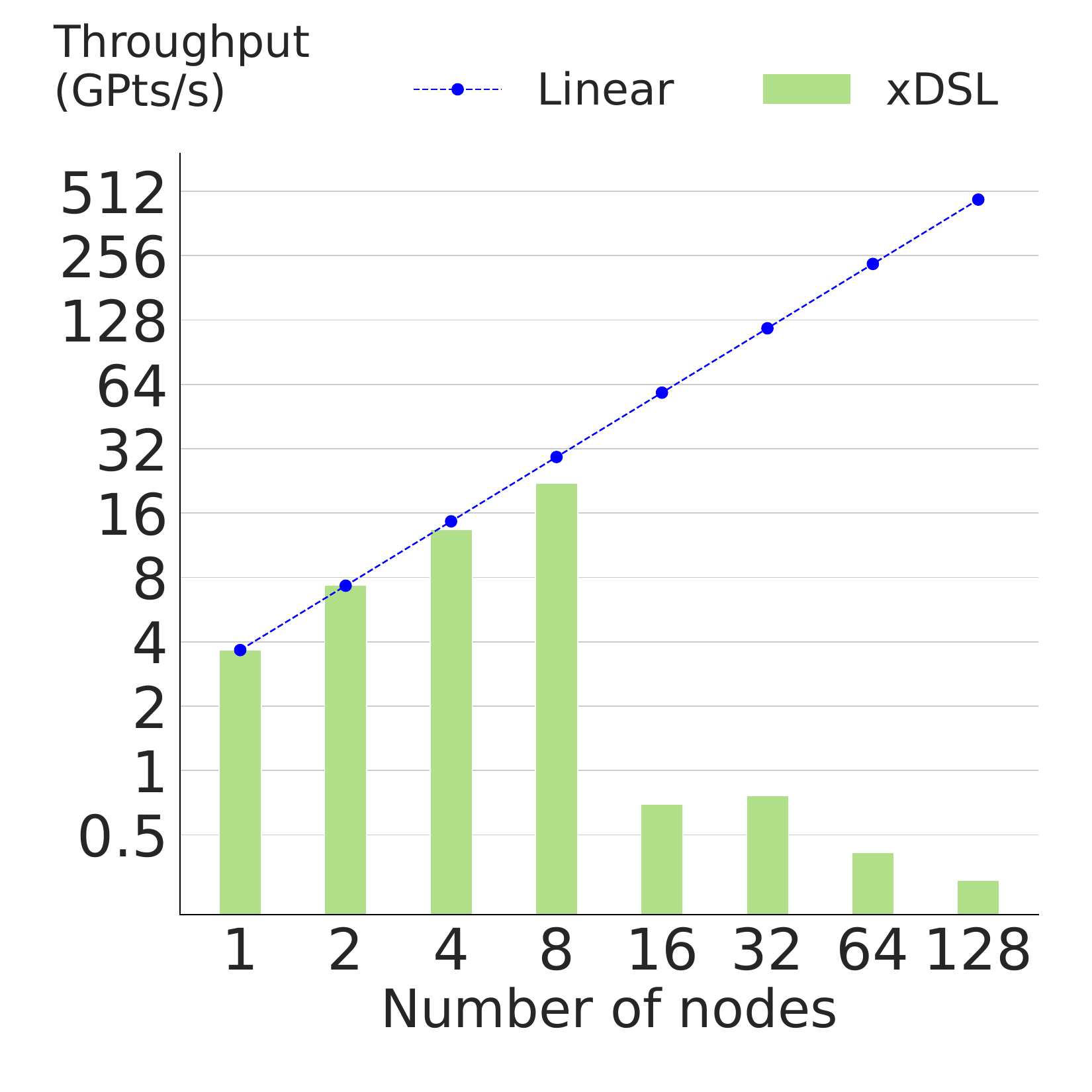}
    \caption{PW advection}\label{fig:pwadvection_dist_throughput}
  \end{subfigure}
  \begin{subfigure}[b]{0.58\columnwidth}
    \includegraphics[width=\textwidth]{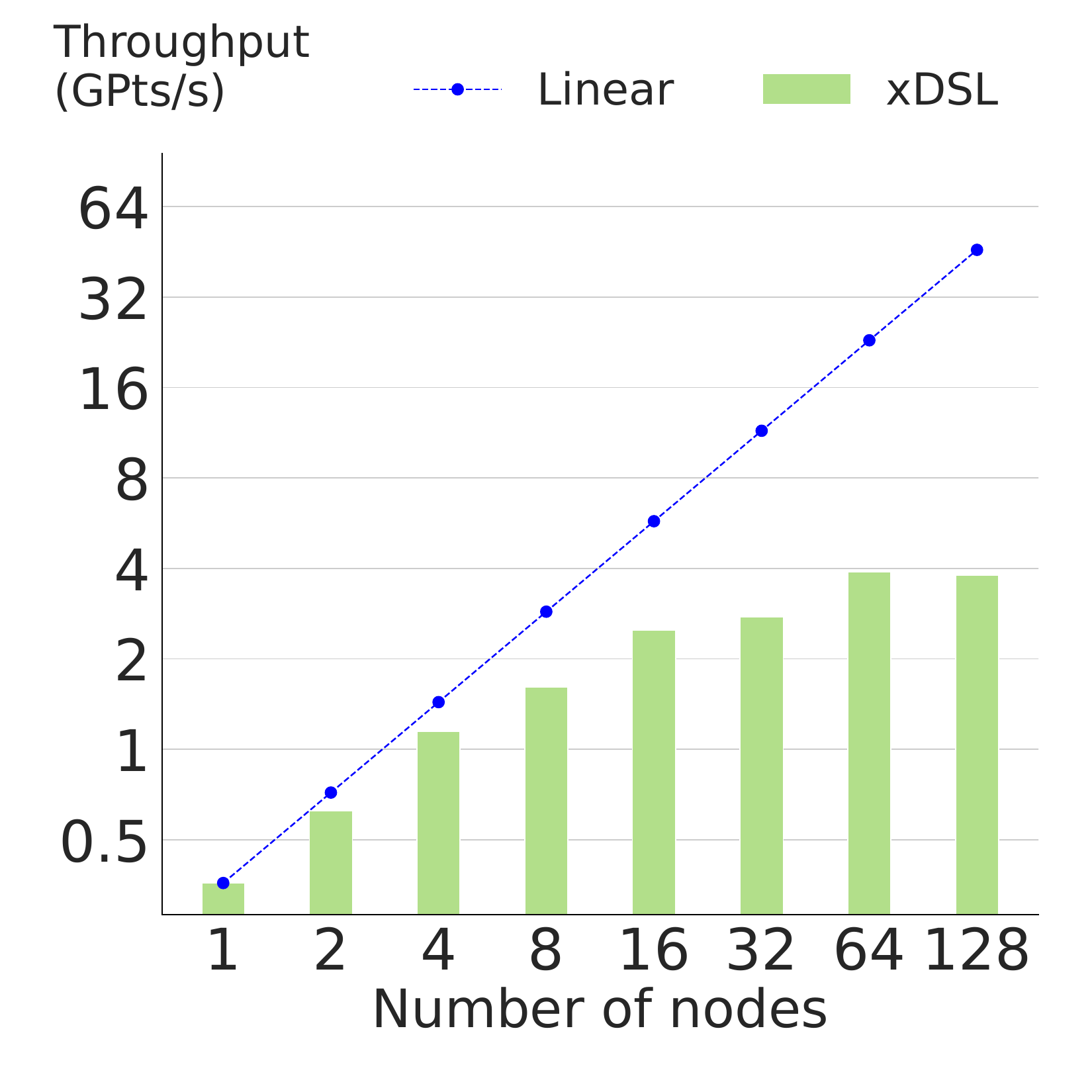}
    \caption{Tracer advection}\label{fig:traadv_dist_throughput}
  \end{subfigure}
  \caption{Multi-node strong scaling CPU throughput for xDSL-PSyclone for PW advection $[256,256,128]$ (a) 
   and tracer advection $[512,512,128]$ (b) on ARCHER2 (higher is better)}
\end{figure}
\Cref{fig:pwadvection_dist_throughput} reports the throughput for strong scaling of the
PW advection benchmark on ARCHER2 up to 128 nodes (16384 cores).
Unlike Devito, the PSyclone NEMO API we are using does not provide distributed memory capabilities, therefore, only xDSL-PSyclone results are presented here. The code was compiled with xDSL-PSyclone, and lowered using the \emph{stencil}, \emph{dmp} and \emph{mpi} dialects, leveraging a 2D \emph{dmp} \emph{decomposition strategy}, which is commonplace in these types of model due to tight coupling in the vertical dimension.\@ xDSL-PSyclone provides good strong scaling to eight nodes but then suffers from strong scaling effects due to the small global problem size. Likewise, \cref{fig:traadv_dist_throughput} reports throughput for the tracer advection benchmark where we observe similar behavior, with the 2D decomposition strategy limiting the strong scaling capability of xDSL-PSyclone.

\section{Related Work}\label{sec:related}
Numerous DSLs and compiler frameworks for HPC stencil computations
exist.
In contrast to building a siloed language and compiler infrastructure,
our work differs from many previous attempts: we leverage
the MLIR infrastructure to share maintenance support and
interoperability with other community-driven IRs.
High-level approaches to problem modeling using symbolic notation include the
Python-based OpenSBLI \citep{jacobs2017opensbli}, using Einstein
notation and the C/C++-based OPS library \citep{ops2018} underneath
for automated parallelism.
ExaStencils \citep{exaslangMPIKucuk:2016, exastencils_2020} has a
multi-layered high-level symbolic DSL for FDs, targetting multi-CPU
and multi-GPU.\@ STELLA \citep{Stella2015} (and susbequently GridTools \citep{gridtools2021}) uses a concise discretized mathematical
syntax and helped port the COSMO model \citep{Thaler2019} to NVIDIA GPUs.

Using a notation closer to modeling loops and arithmetic expressions,
notable code-generation frameworks include Patus \citep{Patus2011},
the Pochoir stencil compiler \citep{pochoir2011},
Physis \citep{physis2011} with C-based embedded DSL
and Mint \citep{mint2011unat}, a source-to-source translator for CUDA.\@
MSC \citep{li2021} facilitates halo exchanges for DMP
targeting many-core processor systems such as Sunway and Matrix.
SDSLc \citep{Rawat2015} targets CPUs and GPUs,
Shift Calculus \citep{Liao2015_shiftcalculus} differs by
using an ontology language. Lift \citep{LIFTSteuwer2015}, based upon
semantic-preserving rewrite rules to optimize stencils, StencilGen \citep{Rawat2018}
targets GPUs with more exotic optimizations, such as overlapped temporal blocking,
and Artemis \citep{rawat2019} primarily targets GPUs. Other efforts to benefit from MLIR
include domain-specific code generators using tensor-level abstractions \citep{essadki2023}.

Targetting diverse problem domains in science other than FD stencils,
for unstructured meshes, we encounter OP2 \citep{op2Reguly2016, mudalige2019},
Liszt \citep{Liszt2011}, Paralab/Finch \citep{heisler_finch_2023},
Firedrake \citep{FiredrakeUserManual}, UFL \citep{alnaes2012ufl}
and FEniCS \citep{logg2012automated}.
SPIRAL \citep{puschel2005spiral} targets Digital Signal Processing
and Tensor Contraction Engine \citep{auer2006automatic} for molecular physics.
On the medical applications side, MOD2IR \citep{mitenkov2023} focuses
on neuronal simulations, leveraging the LLVM toolchain to facilitate the optimization
of NMODL \citep{NMODL2000}.\@ limpetMLIR \citep{thangamani2023lifting} uses MLIR and
targets electrophysiology models, and Stride \citep{stride2022},
built on top of Devito, focuses on ultra-sound tomography imaging.
For tensor algebra, TACO \citep{kjolstad2017tensor} offers a high-level DSL
targeting a range of hardware accelerators.
On the image processing side, Halide \citep{halide2013, halide-dist2016} is a toolkit
for generating HPC codes from functional specifications in image processing
using programmer-specified scheduling transformations, the polyhedral-based 
PolyMage \citep{polymage2015} offers a high-level Python DSL and optimizing
code generator and Forma \citep{Forma2015} also targetting GPUs and claims to
provide an easier-to-handle DSL interface.

Focusing more on libraries and frameworks optimizing loop-dominated codes,
Python-based Loo.py \citep{klockner2014loopy}, generates code for OpenCL/CUDA,
Chill \citep{chen2008chill} guides high-level loop transformations,
YASK \citep{yask2016} specializing in data layout transformations to
optimize for high-bandwidth Intel architectures.
Numba \citep{lam2015numba} uses decorators for runtime-acceleration of Python code
using LLVM as a backend. Notable annotation-based approaches include
Tai-Chi \citep{taichi2019} and DaCe \citep{ziogas2021}.

Regarding frameworks to assist DSL developers in building DSLs for HPC code, 
AnyDSL \citep{anydsl2018} provides composable DSL development using partial
evaluation. Julia \citep{bezanson2017julia} combines dynamic types with
the ability to generate HPC code. Python-based projects like Codon \citep{codon:2023},
features static typing and compilation via LLVM, showcasing notable
speedups compared to native Python routines and Mojo \citep{mojo2023}
as a Pythonic frontend for MLIR, displays a significant performance boost
over state-of-the-art tools.
Finally, the MSF's \citep{msf_coullon2019extensibility}
approach differs in addressing the sharing of common abstractions between
stencil frameworks, providing a stencil-DSL operating in an ahead-of-time
non-embedded fashion.

Describing and optimizing communication patterns for DMP/MPI
scientific software is a focus of many libraries and compilers
\citep{zhao2018bricks, pencil2020sc, zhao2021ppopp, castillo2019}.
Halide \citep{halide-dist2016}, Exaslang \citep{exaslangMPIKucuk:2016} and
PETSc \citep{petscSFZhang:2022} facilitate large-scale
computational science with libraries supporting efficient execution at
all parallelism levels (SIMD, SMP, DMP).
Finally, \emph{ParallelStencil.jl} \citep{omlin2022distributed} is a high-level
Julia approach using bindings for MPI \emph{MPI.jl} \citep{Byrne2021}.

\section{Conclusions and Future Work}\label{sec:conclusions}

By merging the backend compilation stacks of two stencil-DSLs focusing on FD stencils, PSyclone and Devito, we presented a prototype to support these important domain-specific abstractions and their optimizations that can be shared across DSLs.
As a result, their developer communities can benefit from delivering peak performance for key applications reasoning about FD stencils, such as seismic and medical imaging and weather modeling, targeting CPUs, GPUs, and clusters thereof.
To enable broad sharing of abstractions, we expressed fundamental HPC abstractions such as
message passing and halo exchanges as SSA-based IRs for the MLIR compiler ecosystem.
We demonstrated that these abstractions enable the generation of code employing shared and distributed parallelism exhibiting competitive single-node and strong scaling performance to state-of-the-art
supercomputers. The MPI abstractions are generic and independent of the specific application domain, while the DMP abstractions are focused on halo exchanges specific to FD stencil computations.
Both enable the sharing of optimizations and code generation strategies for message passing across stencil computations that target supercomputers.
Our evaluation across stencil kernel workloads from Devito and PSyclone
shows competitive and even improved performance while broadening the
set of targets.

We hope that this work on HPC abstractions for MPI and distributed-memory code generation
transcends the specific application domains of our DSLs and motivates further research on exploiting
synergies enabled by collaborations across DSL and compiler frameworks in HPC.\@
We aim to establish a cohesive ecosystem of HPC-oriented IRs within
the compiler landscape, facilitating reasoning and HPC code generation.
These IRs promote modularity and interoperability by abstracting
away from domain-specific considerations, aiming to enable code generation
for efficient and scalable codes across DSLs targeting problems other than stencils.

A limitation identified by our benchmarking has been in the existing MLIR lowerings
from the \emph{scf} dialect to the OpenMP and GPU dialects. Further work optimizing these would be beneficial, potentially requiring developing bespoke lowerings which are able to consider the wider context that an \emph{scf.parallel} operation sits within. The ability to support one parallel region across all \emph{scf.parallel} loops in OpenMP and finer-grained control over the synchronization of GPU kernels are important aspects that would likely deliver significant performance improvements for codes like the tracer advection benchmark.
Further work includes enhancing the decomposition techniques to support advanced exchange schemes and DMP/MPI optimizations, such as diagonal communications in the cartesian grid topology and communication/computation overlap. The abstractions presented in this paper aim to be modular and extensible building blocks to facilitate these future developments.

\begin{acks}
This work was supported by the \grantsponsor{EPSRC}{Engineering and Physical Sciences Research Council (EPSRC)}{https://www.ukri.org/councils/epsrc/} grants \grantnum{EPSRC}{EP/W007789/1} and \grantnum{EPSRC}{EP/W007940/1}.
The authors would like to thank the shepherd and the reviewers for their comments and allocated time towards
improving this work.
The authors thank the xDSL, Devito, and PsyClone communities for their comments
and discussions.
\end{acks}

\ifarxiv

\fi

\begin{thebibliography}{82}


\ifx \showCODEN    \undefined \def \showCODEN     #1{\unskip}     \fi
\ifx \showDOI      \undefined \def \showDOI       #1{#1}\fi
\ifx \showISBNx    \undefined \def \showISBNx     #1{\unskip}     \fi
\ifx \showISBNxiii \undefined \def \showISBNxiii  #1{\unskip}     \fi
\ifx \showISSN     \undefined \def \showISSN      #1{\unskip}     \fi
\ifx \showLCCN     \undefined \def \showLCCN      #1{\unskip}     \fi
\ifx \shownote     \undefined \def \shownote      #1{#1}          \fi
\ifx \showarticletitle \undefined \def \showarticletitle #1{#1}   \fi
\ifx \showURL      \undefined \def \showURL       {\relax}        \fi
\providecommand\bibfield[2]{#2}
\providecommand\bibinfo[2]{#2}
\providecommand\natexlab[1]{#1}
\providecommand\showeprint[2][]{arXiv:#2}

\bibitem[Abadi et~al\mbox{.}(2015)]%
        {tensorflow2015-whitepaper}
\bibfield{author}{\bibinfo{person}{Mart\'{i}n Abadi}, \bibinfo{person}{Ashish
  Agarwal}, \bibinfo{person}{Paul Barham}, \bibinfo{person}{Eugene Brevdo},
  \bibinfo{person}{Zhifeng Chen}, \bibinfo{person}{Craig Citro},
  \bibinfo{person}{Greg~S. Corrado}, \bibinfo{person}{Andy Davis},
  \bibinfo{person}{Jeffrey Dean}, \bibinfo{person}{Matthieu Devin},
  \bibinfo{person}{Sanjay Ghemawat}, \bibinfo{person}{Ian Goodfellow},
  \bibinfo{person}{Andrew Harp}, \bibinfo{person}{Geoffrey Irving},
  \bibinfo{person}{Michael Isard}, \bibinfo{person}{Yangqing Jia},
  \bibinfo{person}{Rafal Jozefowicz}, \bibinfo{person}{Lukasz Kaiser},
  \bibinfo{person}{Manjunath Kudlur}, \bibinfo{person}{Josh Levenberg},
  \bibinfo{person}{Dandelion Man\'{e}}, \bibinfo{person}{Rajat Monga},
  \bibinfo{person}{Sherry Moore}, \bibinfo{person}{Derek Murray},
  \bibinfo{person}{Chris Olah}, \bibinfo{person}{Mike Schuster},
  \bibinfo{person}{Jonathon Shlens}, \bibinfo{person}{Benoit Steiner},
  \bibinfo{person}{Ilya Sutskever}, \bibinfo{person}{Kunal Talwar},
  \bibinfo{person}{Paul Tucker}, \bibinfo{person}{Vincent Vanhoucke},
  \bibinfo{person}{Vijay Vasudevan}, \bibinfo{person}{Fernanda Vi\'{e}gas},
  \bibinfo{person}{Oriol Vinyals}, \bibinfo{person}{Pete Warden},
  \bibinfo{person}{Martin Wattenberg}, \bibinfo{person}{Martin Wicke},
  \bibinfo{person}{Yuan Yu}, {and} \bibinfo{person}{Xiaoqiang Zheng}.}
  \bibinfo{year}{2015}\natexlab{}.
\newblock \bibinfo{title}{{TensorFlow}: Large-Scale Machine Learning on
  Heterogeneous Systems}.
\newblock
\newblock
\urldef\tempurl%
\url{https://www.tensorflow.org/}
\showURL{%
\tempurl}
\newblock
\shownote{Software available from tensorflow.org}.


\bibitem[Abadi et~al\mbox{.}(2016)]%
        {tensorflow}
\bibfield{author}{\bibinfo{person}{Mart\'{\i}n Abadi}, \bibinfo{person}{Paul
  Barham}, \bibinfo{person}{Jianmin Chen}, \bibinfo{person}{Zhifeng Chen},
  \bibinfo{person}{Andy Davis}, \bibinfo{person}{Jeffrey Dean},
  \bibinfo{person}{Matthieu Devin}, \bibinfo{person}{Sanjay Ghemawat},
  \bibinfo{person}{Geoffrey Irving}, \bibinfo{person}{Michael Isard},
  \bibinfo{person}{Manjunath Kudlur}, \bibinfo{person}{Josh Levenberg},
  \bibinfo{person}{Rajat Monga}, \bibinfo{person}{Sherry Moore},
  \bibinfo{person}{Derek~G. Murray}, \bibinfo{person}{Benoit Steiner},
  \bibinfo{person}{Paul Tucker}, \bibinfo{person}{Vijay Vasudevan},
  \bibinfo{person}{Pete Warden}, \bibinfo{person}{Martin Wicke},
  \bibinfo{person}{Yuan Yu}, {and} \bibinfo{person}{Xiaoqiang Zheng}.}
  \bibinfo{year}{2016}\natexlab{}.
\newblock \showarticletitle{TensorFlow: a system for large-scale machine
  learning}. In \bibinfo{booktitle}{\emph{Proceedings of the 12th USENIX
  Conference on Operating Systems Design and Implementation}} (Savannah, GA,
  USA) \emph{(\bibinfo{series}{OSDI'16})}. \bibinfo{publisher}{USENIX
  Association}, \bibinfo{address}{USA}, \bibinfo{pages}{265--283}.
\newblock
\showISBNx{9781931971331}
\urldef\tempurl%
\url{https://dl.acm.org/doi/10.5555/3026877.3026899}
\showURL{%
\tempurl}


\bibitem[Adams et~al\mbox{.}(2019)]%
        {adams2019lfric}
\bibfield{author}{\bibinfo{person}{S.~V. Adams}, \bibinfo{person}{R.~W. Ford},
  \bibinfo{person}{M. Hambley}, \bibinfo{person}{J.~M. Hobson},
  \bibinfo{person}{I. Kavčič}, \bibinfo{person}{C.~M. Maynard},
  \bibinfo{person}{T. Melvin}, \bibinfo{person}{E.~H. Müller},
  \bibinfo{person}{S. Mullerworth}, \bibinfo{person}{A.~R. Porter},
  \bibinfo{person}{M. Rezny}, \bibinfo{person}{B.~J. Shipway}, {and}
  \bibinfo{person}{R. Wong}.} \bibinfo{year}{2019}\natexlab{}.
\newblock \showarticletitle{{LFR}ic: Meeting the challenges of scalability and
  performance portability in Weather and Climate models}.
\newblock \bibinfo{journal}{\emph{J. Parallel and Distrib. Comput.}}
  \bibinfo{volume}{132} (\bibinfo{year}{2019}), \bibinfo{pages}{383--396}.
\newblock
\showISSN{0743-7315}
\urldef\tempurl%
\url{https://doi.org/10.1016/j.jpdc.2019.02.007}
\showDOI{\tempurl}


\bibitem[Afanasyev et~al\mbox{.}(2021)]%
        {gridtools2021}
\bibfield{author}{\bibinfo{person}{Anton Afanasyev}, \bibinfo{person}{Mauro
  Bianco}, \bibinfo{person}{Lukas Mosimann}, \bibinfo{person}{Carlos Osuna},
  \bibinfo{person}{Felix Thaler}, \bibinfo{person}{Hannes Vogt},
  \bibinfo{person}{Oliver Fuhrer}, \bibinfo{person}{Joost VandeVondele}, {and}
  \bibinfo{person}{Thomas~C. Schulthess}.} \bibinfo{year}{2021}\natexlab{}.
\newblock \showarticletitle{{GridTools}: a framework for portable weather and
  climate applications}.
\newblock \bibinfo{journal}{\emph{SoftwareX}}  \bibinfo{volume}{15}
  (\bibinfo{year}{2021}), \bibinfo{pages}{100707}.
\newblock
\urldef\tempurl%
\url{https://doi.org/10.1016/j.softx.2021.100707}
\showDOI{\tempurl}


\bibitem[Aln{\ae}s(2012)]%
        {alnaes2012ufl}
\bibfield{author}{\bibinfo{person}{Martin~Sandve Aln{\ae}s}.}
  \bibinfo{year}{2012}\natexlab{}.
\newblock \bibinfo{booktitle}{\emph{UFL: a finite element form language}}.
\newblock \bibinfo{publisher}{Springer Berlin Heidelberg},
  \bibinfo{address}{Berlin, Heidelberg}, \bibinfo{pages}{303--338}.
\newblock
\showISBNx{978-3-642-23099-8}
\urldef\tempurl%
\url{https://doi.org/10.1007/978-3-642-23099-8_17}
\showDOI{\tempurl}


\bibitem[Auer et~al\mbox{.}(2006)]%
        {auer2006automatic}
\bibfield{author}{\bibinfo{person}{Alexander~A. Auer}, \bibinfo{person}{Gerald
  Baumgartner}, \bibinfo{person}{David~E. Bernholdt}, \bibinfo{person}{Alina
  Bibireata}, \bibinfo{person}{Venkatesh Choppella}, \bibinfo{person}{Daniel
  Cociorva}, \bibinfo{person}{Xiaoyang Gao}, \bibinfo{person}{Robert Harrison},
  \bibinfo{person}{Sriram Krishnamoorthy}, \bibinfo{person}{Sandhya Krishnan},
  \bibinfo{person}{Chi-Chung Lam}, \bibinfo{person}{Qingda Lu},
  \bibinfo{person}{Marcel Nooijen}, \bibinfo{person}{Russell Pitzer},
  \bibinfo{person}{J. Ramanujam}, \bibinfo{person}{P. Sadayappan}, {and}
  \bibinfo{person}{Alexander Sibiryakov}.} \bibinfo{year}{2006}\natexlab{}.
\newblock \showarticletitle{Automatic code generation for many-body electronic
  structure methods: the tensor contraction engine}.
\newblock \bibinfo{journal}{\emph{Molecular Physics}} \bibinfo{volume}{104},
  \bibinfo{number}{2} (\bibinfo{year}{2006}), \bibinfo{pages}{211--228}.
\newblock
\urldef\tempurl%
\url{https://doi.org/10.1080/00268970500275780}
\showDOI{\tempurl}


\bibitem[Bezanson et~al\mbox{.}(2017)]%
        {bezanson2017julia}
\bibfield{author}{\bibinfo{person}{Jeff Bezanson}, \bibinfo{person}{Alan
  Edelman}, \bibinfo{person}{Stefan Karpinski}, {and} \bibinfo{person}{Viral~B.
  Shah}.} \bibinfo{year}{2017}\natexlab{}.
\newblock \showarticletitle{{Julia: A fresh approach to numerical computing}}.
\newblock \bibinfo{journal}{\emph{SIAM Rev.}} \bibinfo{volume}{59},
  \bibinfo{number}{1} (\bibinfo{date}{Sept.} \bibinfo{year}{2017}),
  \bibinfo{pages}{65--98}.
\newblock
\urldef\tempurl%
\url{https://doi.org/10.1137/141000671}
\showDOI{\tempurl}


\bibitem[Bisbas et~al\mbox{.}(2021)]%
        {bisbas2021}
\bibfield{author}{\bibinfo{person}{George Bisbas}, \bibinfo{person}{Fabio
  Luporini}, \bibinfo{person}{Mathias Louboutin}, \bibinfo{person}{Rhodri
  Nelson}, \bibinfo{person}{Gerard~J. Gorman}, {and} \bibinfo{person}{Paul
  H.~J. Kelly}.} \bibinfo{year}{2021}\natexlab{}.
\newblock \showarticletitle{Temporal blocking of finite-difference stencil
  operators with sparse ``off-the-grid'' sources}. In
  \bibinfo{booktitle}{\emph{2021 IEEE International Parallel and Distributed
  Processing Symposium (IPDPS)}}. \bibinfo{pages}{497--506}.
\newblock
\urldef\tempurl%
\url{https://doi.org/10.1109/IPDPS49936.2021.00058}
\showDOI{\tempurl}


\bibitem[Bisbas et~al\mbox{.}(2023)]%
        {bisbas2023automated}
\bibfield{author}{\bibinfo{person}{George Bisbas}, \bibinfo{person}{Rhodri
  Nelson}, \bibinfo{person}{Mathias Louboutin}, \bibinfo{person}{Paul H.~J.
  Kelly}, \bibinfo{person}{Fabio Luporini}, {and} \bibinfo{person}{Gerard
  Gorman}.} \bibinfo{year}{2023}\natexlab{}.
\newblock \bibinfo{title}{Automated {MPI} code generation for scalable
  finite-difference solvers}.
\newblock
\newblock
\showeprint[arxiv]{2312.13094}~[cs.DC]


\bibitem[Brown(2020)]%
        {brown2020exploring}
\bibfield{author}{\bibinfo{person}{Nick Brown}.}
  \bibinfo{year}{2020}\natexlab{}.
\newblock \showarticletitle{Exploring the acceleration of Nekbone on
  reconfigurable architectures}. In \bibinfo{booktitle}{\emph{2020 IEEE/ACM
  International Workshop on Heterogeneous High-performance Reconfigurable
  Computing (H2RC)}}. \bibinfo{pages}{19--28}.
\newblock
\urldef\tempurl%
\url{https://doi.org/10.1109/H2RC51942.2020.00008}
\showDOI{\tempurl}


\bibitem[Brown(2021)]%
        {brown2021accelerating}
\bibfield{author}{\bibinfo{person}{Nick Brown}.}
  \bibinfo{year}{2021}\natexlab{}.
\newblock \showarticletitle{Accelerating advection for atmospheric modelling on
  Xilinx and Intel FPGAs}. In \bibinfo{booktitle}{\emph{2021 IEEE International
  Conference on Cluster Computing (CLUSTER)}}. \bibinfo{pages}{767--774}.
\newblock
\urldef\tempurl%
\url{https://doi.org/10.1109/Cluster48925.2021.00113}
\showDOI{\tempurl}


\bibitem[Brown et~al\mbox{.}(2015)]%
        {brown2020highly}
\bibfield{author}{\bibinfo{person}{Nick Brown}, \bibinfo{person}{Michele
  Weiland}, \bibinfo{person}{Adrian Hill}, \bibinfo{person}{Ben Shipway},
  \bibinfo{person}{Chris Maynard}, \bibinfo{person}{Thomas Allen}, {and}
  \bibinfo{person}{Mike Rezny}.} \bibinfo{year}{2015}\natexlab{}.
\newblock \showarticletitle{A highly scalable Met Office NERC Cloud model}. In
  \bibinfo{booktitle}{\emph{Proceedings of the 3rd International Conference on
  Exascale Applications and Software}} (Edinburgh, UK)
  \emph{(\bibinfo{series}{EASC '15})}. \bibinfo{publisher}{University of
  Edinburgh}, \bibinfo{address}{GBR}, \bibinfo{pages}{132–137}.
\newblock
\showISBNx{9780992661519}


\bibitem[Byrne et~al\mbox{.}(2021)]%
        {Byrne2021}
\bibfield{author}{\bibinfo{person}{Simon Byrne}, \bibinfo{person}{Lucas~C.
  Wilcox}, {and} \bibinfo{person}{Valentin Churavy}.}
  \bibinfo{year}{2021}\natexlab{}.
\newblock \showarticletitle{MPI.jl: Julia bindings for the Message Passing
  Interface}.
\newblock \bibinfo{journal}{\emph{Proceedings of the JuliaCon Conferences}}
  \bibinfo{volume}{1}, \bibinfo{number}{1} (\bibinfo{year}{2021}),
  \bibinfo{pages}{68}.
\newblock
\urldef\tempurl%
\url{https://doi.org/10.21105/jcon.00068}
\showDOI{\tempurl}


\bibitem[Castillo et~al\mbox{.}(2019)]%
        {castillo2019}
\bibfield{author}{\bibinfo{person}{Emilio Castillo}, \bibinfo{person}{Nikhil
  Jain}, \bibinfo{person}{Marc Casas}, \bibinfo{person}{Miquel Moreto},
  \bibinfo{person}{Martin Schulz}, \bibinfo{person}{Ramon Beivide},
  \bibinfo{person}{Mateo Valero}, {and} \bibinfo{person}{Abhinav Bhatele}.}
  \bibinfo{year}{2019}\natexlab{}.
\newblock \showarticletitle{Optimizing Computation-Communication Overlap in
  Asynchronous Task-Based Programs}. In \bibinfo{booktitle}{\emph{Proceedings
  of the ACM International Conference on Supercomputing}} (Phoenix, Arizona)
  \emph{(\bibinfo{series}{ICS '19})}. \bibinfo{publisher}{Association for
  Computing Machinery}, \bibinfo{address}{New York, NY, USA},
  \bibinfo{pages}{380--391}.
\newblock
\showISBNx{9781450360791}
\urldef\tempurl%
\url{https://doi.org/10.1145/3330345.3330379}
\showDOI{\tempurl}


\bibitem[Chen et~al\mbox{.}(2008)]%
        {chen2008chill}
\bibfield{author}{\bibinfo{person}{Chun Chen}, \bibinfo{person}{Jacqueline
  Chame}, {and} \bibinfo{person}{Mary Hall}.} \bibinfo{year}{2008}\natexlab{}.
\newblock \bibinfo{booktitle}{\emph{CHiLL: A framework for composing high-level
  loop transformations}}.
\newblock \bibinfo{type}{{T}echnical {R}eport} 08-897.
  \bibinfo{institution}{University of Southern California}.
\newblock
\urldef\tempurl%
\url{https://citeseerx.ist.psu.edu/document?doi=6a4620589f63f3385707d2d590f7b7dc8ee4d74f}
\showURL{%
\tempurl}


\bibitem[Christen et~al\mbox{.}(2011)]%
        {Patus2011}
\bibfield{author}{\bibinfo{person}{Matthias Christen}, \bibinfo{person}{Olaf
  Schenk}, {and} \bibinfo{person}{Helmar Burkhart}.}
  \bibinfo{year}{2011}\natexlab{}.
\newblock \showarticletitle{{PATUS}: a Code Generation and Autotuning Framework
  for Parallel Iterative Stencil Computations on Modern Microarchitectures}. In
  \bibinfo{booktitle}{\emph{Proceedings of the 2011 IEEE International Parallel
  \& Distributed Processing Symposium}} \emph{(\bibinfo{series}{IPDPS '11})}.
  \bibinfo{publisher}{IEEE Computer Society}, \bibinfo{address}{Washington, DC,
  USA}, \bibinfo{pages}{676--687}.
\newblock
\showISBNx{978-0-7695-4385-7}
\urldef\tempurl%
\url{https://doi.org/10.1109/IPDPS.2011.70}
\showDOI{\tempurl}


\bibitem[Coullon et~al\mbox{.}(2019)]%
        {msf_coullon2019extensibility}
\bibfield{author}{\bibinfo{person}{H{\'e}l{\`e}ne Coullon},
  \bibinfo{person}{Julien Bigot}, {and} \bibinfo{person}{Christian P{\'e}rez}.}
  \bibinfo{year}{2019}\natexlab{}.
\newblock \showarticletitle{Extensibility and composability of a multi-stencil
  domain specific framework}.
\newblock \bibinfo{journal}{\emph{International Journal of Parallel
  Programming}}  \bibinfo{volume}{47} (\bibinfo{year}{2019}),
  \bibinfo{pages}{1046--1085}.
\newblock
\urldef\tempurl%
\url{https://doi.org/10.1007/s10766-017-0539-5}
\showDOI{\tempurl}


\bibitem[Cueto et~al\mbox{.}(2022)]%
        {stride2022}
\bibfield{author}{\bibinfo{person}{Carlos Cueto}, \bibinfo{person}{Oscar
  Bates}, \bibinfo{person}{George Strong}, \bibinfo{person}{Javier Cudeiro},
  \bibinfo{person}{Fabio Luporini}, \bibinfo{person}{Òscar {Calderón Agudo}},
  \bibinfo{person}{Gerard Gorman}, \bibinfo{person}{Lluis Guasch}, {and}
  \bibinfo{person}{Meng-Xing Tang}.} \bibinfo{year}{2022}\natexlab{}.
\newblock \showarticletitle{Stride: A flexible software platform for
  high-performance ultrasound computed tomography}.
\newblock \bibinfo{journal}{\emph{Computer Methods and Programs in
  Biomedicine}}  \bibinfo{volume}{221} (\bibinfo{year}{2022}),
  \bibinfo{pages}{106855}.
\newblock
\showISSN{0169-2607}
\urldef\tempurl%
\url{https://doi.org/10.1016/j.cmpb.2022.106855}
\showDOI{\tempurl}


\bibitem[Cytron et~al\mbox{.}(1991)]%
        {cytron1991efficiently}
\bibfield{author}{\bibinfo{person}{Ron Cytron}, \bibinfo{person}{Jeanne
  Ferrante}, \bibinfo{person}{Barry~K. Rosen}, \bibinfo{person}{Mark~N.
  Wegman}, {and} \bibinfo{person}{F.~Kenneth Zadeck}.}
  \bibinfo{year}{1991}\natexlab{}.
\newblock \showarticletitle{Efficiently computing static single assignment form
  and the control dependence graph}.
\newblock \bibinfo{journal}{\emph{ACM Trans. Program. Lang. Syst.}}
  \bibinfo{volume}{13}, \bibinfo{number}{4} (\bibinfo{date}{oct}
  \bibinfo{year}{1991}), \bibinfo{pages}{451--490}.
\newblock
\showISSN{0164-0925}
\urldef\tempurl%
\url{https://doi.org/10.1145/115372.115320}
\showDOI{\tempurl}


\bibitem[Denniston et~al\mbox{.}(2016)]%
        {halide-dist2016}
\bibfield{author}{\bibinfo{person}{Tyler Denniston}, \bibinfo{person}{Shoaib
  Kamil}, {and} \bibinfo{person}{Saman Amarasinghe}.}
  \bibinfo{year}{2016}\natexlab{}.
\newblock \showarticletitle{Distributed Halide}. In
  \bibinfo{booktitle}{\emph{Proceedings of the 21st ACM SIGPLAN Symposium on
  Principles and Practice of Parallel Programming}} (Barcelona, Spain)
  \emph{(\bibinfo{series}{PPoPP '16})}. \bibinfo{publisher}{Association for
  Computing Machinery}, \bibinfo{address}{New York, NY, USA}, Article
  \bibinfo{articleno}{5}, \bibinfo{numpages}{12}~pages.
\newblock
\showISBNx{9781450340922}
\urldef\tempurl%
\url{https://doi.org/10.1145/2851141.2851157}
\showDOI{\tempurl}


\bibitem[DeVito et~al\mbox{.}(2011)]%
        {Liszt2011}
\bibfield{author}{\bibinfo{person}{Zachary DeVito}, \bibinfo{person}{Niels
  Joubert}, \bibinfo{person}{Francisco Palacios}, \bibinfo{person}{Stephen
  Oakley}, \bibinfo{person}{Montserrat Medina}, \bibinfo{person}{Mike
  Barrientos}, \bibinfo{person}{Erich Elsen}, \bibinfo{person}{Frank Ham},
  \bibinfo{person}{Alex Aiken}, \bibinfo{person}{Karthik Duraisamy},
  \bibinfo{person}{Eric Darve}, \bibinfo{person}{Juan Alonso}, {and}
  \bibinfo{person}{Pat Hanrahan}.} \bibinfo{year}{2011}\natexlab{}.
\newblock \showarticletitle{Liszt: A domain specific language for building
  portable mesh-based PDE solvers}. In \bibinfo{booktitle}{\emph{SC '11:
  Proceedings of 2011 International Conference for High Performance Computing,
  Networking, Storage and Analysis}}. \bibinfo{pages}{1--12}.
\newblock
\urldef\tempurl%
\url{https://doi.org/10.1145/2063384.2063396}
\showDOI{\tempurl}


\bibitem[Eldridge et~al\mbox{.}(2021)]%
        {eldridge2021mlir}
\bibfield{author}{\bibinfo{person}{Schuyler Eldridge},
  \bibinfo{person}{Prithayan Barua}, \bibinfo{person}{Aliaksei Chapyzhenka},
  \bibinfo{person}{Adam Izraelevitz}, \bibinfo{person}{Jack Koenig},
  \bibinfo{person}{Chris Lattner}, \bibinfo{person}{Andrew Lenharth},
  \bibinfo{person}{George Leontiev}, \bibinfo{person}{Fabian Schuiki},
  \bibinfo{person}{Ram Sunder}, \bibinfo{person}{Andrew Young}, {and}
  \bibinfo{person}{Richard Xia}.} \bibinfo{year}{2021}\natexlab{}.
\newblock \showarticletitle{MLIR as hardware compiler infrastructure}. In
  \bibinfo{booktitle}{\emph{Workshop on Open-Source EDA Technology (WOSET)}}.
\newblock
\urldef\tempurl%
\url{https://woset-workshop.github.io/PDFs/2021/a06.pdf}
\showURL{%
\tempurl}


\bibitem[Essadki et~al\mbox{.}(2023)]%
        {essadki2023}
\bibfield{author}{\bibinfo{person}{Mohamed Essadki}, \bibinfo{person}{Bertrand
  Michel}, \bibinfo{person}{Bruno Maugars}, \bibinfo{person}{Oleksandr
  Zinenko}, \bibinfo{person}{Nicolas Vasilache}, {and} \bibinfo{person}{Albert
  Cohen}.} \bibinfo{year}{2023}\natexlab{}.
\newblock \showarticletitle{Code Generation for In-Place Stencils}. In
  \bibinfo{booktitle}{\emph{Proceedings of the 21st ACM/IEEE International
  Symposium on Code Generation and Optimization}} (Montr\'{e}al, QC, Canada)
  \emph{(\bibinfo{series}{CGO 2023})}. \bibinfo{publisher}{Association for
  Computing Machinery}, \bibinfo{address}{New York, NY, USA},
  \bibinfo{pages}{2--13}.
\newblock
\showISBNx{9798400701016}
\urldef\tempurl%
\url{https://doi.org/10.1145/3579990.3580006}
\showDOI{\tempurl}


\bibitem[Fehr et~al\mbox{.}(2023)]%
        {xdsl}
\bibfield{author}{\bibinfo{person}{Mathieu Fehr}, \bibinfo{person}{Michel
  Weber}, \bibinfo{person}{Christian Ulmann}, \bibinfo{person}{Alexandre
  Lopoukhine}, \bibinfo{person}{Martin Lücke}, \bibinfo{person}{Théo
  Degioanni}, \bibinfo{person}{Michel Steuwer}, {and} \bibinfo{person}{Tobias
  Grosser}.} \bibinfo{year}{2023}\natexlab{}.
\newblock \bibinfo{title}{Sidekick compilation with x{D}{S}{L}}.
\newblock
\newblock
\showeprint[arxiv]{2311.07422}~[cs.PL]


\bibitem[Govindarajan and Moses(2020)]%
        {SyFER2020}
\bibfield{author}{\bibinfo{person}{Sanath Govindarajan} {and}
  \bibinfo{person}{William~S. Moses}.} \bibinfo{year}{2020}\natexlab{}.
\newblock \bibinfo{title}{SyFER-MLIR: Integrating Fully Homomorphic Encryption
  Into the MLIR Compiler Framework}.
\newblock
\newblock
\urldef\tempurl%
\url{https://math.mit.edu/research/highschool/primes/materials/2020/Govindarajan-Moses.pdf}
\showURL{%
\tempurl}


\bibitem[Gysi et~al\mbox{.}(2021)]%
        {openearth2022}
\bibfield{author}{\bibinfo{person}{Tobias Gysi}, \bibinfo{person}{Christoph
  M{\"{u}}ller}, \bibinfo{person}{Oleksandr Zinenko}, \bibinfo{person}{Stephan
  Herhut}, \bibinfo{person}{Eddie Davis}, \bibinfo{person}{Tobias Wicky},
  \bibinfo{person}{Oliver Fuhrer}, \bibinfo{person}{Torsten Hoefler}, {and}
  \bibinfo{person}{Tobias Grosser}.} \bibinfo{year}{2021}\natexlab{}.
\newblock \showarticletitle{Domain-Specific Multi-Level {IR} Rewriting for
  {GPU:} The Open Earth Compiler for GPU-accelerated Climate Simulation}.
\newblock \bibinfo{journal}{\emph{{ACM} Trans. Archit. Code Optim.}}
  \bibinfo{volume}{18}, \bibinfo{number}{4} (\bibinfo{year}{2021}),
  \bibinfo{pages}{51:1--51:23}.
\newblock
\urldef\tempurl%
\url{https://doi.org/10.1145/3469030}
\showDOI{\tempurl}


\bibitem[Gysi et~al\mbox{.}(2015)]%
        {Stella2015}
\bibfield{author}{\bibinfo{person}{Tobias Gysi}, \bibinfo{person}{Carlos
  Osuna}, \bibinfo{person}{Oliver Fuhrer}, \bibinfo{person}{Mauro Bianco},
  {and} \bibinfo{person}{Thomas~C. Schulthess}.}
  \bibinfo{year}{2015}\natexlab{}.
\newblock \showarticletitle{{STELLA:} a domain-specific tool for structured
  grid methods in weather and climate models}. In
  \bibinfo{booktitle}{\emph{Proceedings of the International Conference for
  High Performance Computing, Networking, Storage and Analysis, {SC} 2015,
  Austin, TX, USA, November 15-20, 2015}},
  \bibfield{editor}{\bibinfo{person}{Jackie Kern} {and}
  \bibinfo{person}{Jeffrey~S. Vetter}} (Eds.). \bibinfo{publisher}{{ACM}},
  \bibinfo{address}{New York, NY, USA}, \bibinfo{pages}{41:1--41:12}.
\newblock
\urldef\tempurl%
\url{https://doi.org/10.1145/2807591.2807627}
\showDOI{\tempurl}


\bibitem[Ham et~al\mbox{.}(2023)]%
        {FiredrakeUserManual}
\bibfield{author}{\bibinfo{person}{David~A. Ham}, \bibinfo{person}{Paul H.~J.
  Kelly}, \bibinfo{person}{Lawrence Mitchell}, \bibinfo{person}{Colin~J.
  Cotter}, \bibinfo{person}{Robert~C. Kirby}, \bibinfo{person}{Koki Sagiyama},
  \bibinfo{person}{Nacime Bouziani}, \bibinfo{person}{Sophia Vorderwuelbecke},
  \bibinfo{person}{Thomas~J. Gregory}, \bibinfo{person}{Jack Betteridge},
  \bibinfo{person}{Daniel~R. Shapero}, \bibinfo{person}{Reuben~W. Nixon-Hill},
  \bibinfo{person}{Connor~J. Ward}, \bibinfo{person}{Patrick~E. Farrell},
  \bibinfo{person}{Pablo~D. Brubeck}, \bibinfo{person}{India Marsden},
  \bibinfo{person}{Thomas~H. Gibson}, \bibinfo{person}{Miklós Homolya},
  \bibinfo{person}{Tianjiao Sun}, \bibinfo{person}{Andrew T.~T. McRae},
  \bibinfo{person}{Fabio Luporini}, \bibinfo{person}{Alastair Gregory},
  \bibinfo{person}{Michael Lange}, \bibinfo{person}{Simon~W. Funke},
  \bibinfo{person}{Florian Rathgeber}, \bibinfo{person}{Gheorghe-Teodor
  Bercea}, {and} \bibinfo{person}{Graham~R. Markall}.}
  \bibinfo{year}{2023}\natexlab{}.
\newblock \bibinfo{booktitle}{\emph{Firedrake User Manual}
  (\bibinfo{edition}{first} ed.)}.
\newblock Imperial College London and University of Oxford and Baylor
  University and University of Washington.
\newblock
\urldef\tempurl%
\url{https://doi.org/10.25561/104839}
\showDOI{\tempurl}


\bibitem[Hammond et~al\mbox{.}(2023)]%
        {hammond2023mpi}
\bibfield{author}{\bibinfo{person}{Jeff Hammond}, \bibinfo{person}{Lisandro
  Dalcin}, \bibinfo{person}{Erik Schnetter}, \bibinfo{person}{Marc
  P\'{e}rache}, \bibinfo{person}{Jean-Baptiste Besnard}, \bibinfo{person}{Jed
  Brown}, \bibinfo{person}{Gonzalo~Brito Gadeschi}, \bibinfo{person}{Simon
  Byrne}, \bibinfo{person}{Joseph Schuchart}, {and} \bibinfo{person}{Hui
  Zhou}.} \bibinfo{year}{2023}\natexlab{}.
\newblock \showarticletitle{MPI Application Binary Interface Standardization}.
  In \bibinfo{booktitle}{\emph{Proceedings of the 30th European MPI Users'
  Group Meeting}} (<conf-loc>, <city>Bristol</city>, <country>United
  Kingdom</country>, </conf-loc>) \emph{(\bibinfo{series}{EuroMPI '23})}.
  \bibinfo{publisher}{Association for Computing Machinery},
  \bibinfo{address}{New York, NY, USA}, Article \bibinfo{articleno}{1},
  \bibinfo{numpages}{12}~pages.
\newblock
\showISBNx{9798400709135}
\urldef\tempurl%
\url{https://doi.org/10.1145/3615318.3615319}
\showDOI{\tempurl}


\bibitem[Heisler et~al\mbox{.}(2023)]%
        {heisler_finch_2023}
\bibfield{author}{\bibinfo{person}{Eric Heisler}, \bibinfo{person}{Aadesh
  Deshmukh}, \bibinfo{person}{Sandip Mazumder}, \bibinfo{person}{Ponnuswamy
  Sadayappan}, {and} \bibinfo{person}{Hari Sundar}.}
  \bibinfo{year}{2023}\natexlab{}.
\newblock \showarticletitle{Multi-discretization domain specific language and
  code generation for differential equations}.
\newblock \bibinfo{journal}{\emph{Journal of Computational Science}}
  \bibinfo{volume}{68} (\bibinfo{year}{2023}), \bibinfo{pages}{101981}.
\newblock
\showISSN{1877-7503}
\urldef\tempurl%
\url{https://doi.org/10.1016/j.jocs.2023.101981}
\showDOI{\tempurl}


\bibitem[Henretty et~al\mbox{.}(2011)]%
        {henretty2011}
\bibfield{author}{\bibinfo{person}{Tom Henretty}, \bibinfo{person}{Kevin
  Stock}, \bibinfo{person}{Louis-No{\"e}l Pouchet}, \bibinfo{person}{Franz
  Franchetti}, \bibinfo{person}{J. Ramanujam}, {and} \bibinfo{person}{P.
  Sadayappan}.} \bibinfo{year}{2011}\natexlab{}.
\newblock \showarticletitle{Data Layout Transformation for Stencil Computations
  on Short-Vector SIMD Architectures}. In \bibinfo{booktitle}{\emph{Compiler
  Construction}}, \bibfield{editor}{\bibinfo{person}{Jens Knoop}} (Ed.).
  \bibinfo{publisher}{Springer Berlin Heidelberg}, \bibinfo{address}{Berlin,
  Heidelberg}, \bibinfo{pages}{225--245}.
\newblock
\showISBNx{978-3-642-19861-8}
\urldef\tempurl%
\url{https://doi.org/10.1007/978-3-642-19861-8_13}
\showURL{%
\tempurl}


\bibitem[Hines and Carnevale(2000)]%
        {NMODL2000}
\bibfield{author}{\bibinfo{person}{M.~L. Hines} {and} \bibinfo{person}{N.~T.
  Carnevale}.} \bibinfo{year}{2000}\natexlab{}.
\newblock \showarticletitle{{Expanding NEURON's Repertoire of Mechanisms with
  NMODL}}.
\newblock \bibinfo{journal}{\emph{Neural Computation}} \bibinfo{volume}{12},
  \bibinfo{number}{5} (\bibinfo{date}{05} \bibinfo{year}{2000}),
  \bibinfo{pages}{995--1007}.
\newblock
\showISSN{0899-7667}
\urldef\tempurl%
\url{https://doi.org/10.1162/089976600300015475}
\showDOI{\tempurl}


\bibitem[Hu et~al\mbox{.}(2019)]%
        {taichi2019}
\bibfield{author}{\bibinfo{person}{Yuanming Hu}, \bibinfo{person}{Tzu-Mao Li},
  \bibinfo{person}{Luke Anderson}, \bibinfo{person}{Jonathan Ragan-Kelley},
  {and} \bibinfo{person}{Fr\'{e}do Durand}.} \bibinfo{year}{2019}\natexlab{}.
\newblock \showarticletitle{Taichi: A Language for High-Performance Computation
  on Spatially Sparse Data Structures}.
\newblock \bibinfo{journal}{\emph{ACM Trans. Graph.}} \bibinfo{volume}{38},
  \bibinfo{number}{6}, Article \bibinfo{articleno}{201} (\bibinfo{year}{2019}),
  \bibinfo{numpages}{16}~pages.
\newblock
\showISSN{0730-0301}
\urldef\tempurl%
\url{https://doi.org/10.1145/3355089.3356506}
\showDOI{\tempurl}


\bibitem[Jacobs et~al\mbox{.}(2017)]%
        {jacobs2017opensbli}
\bibfield{author}{\bibinfo{person}{Christian~T. Jacobs},
  \bibinfo{person}{Satya~P. Jammy}, {and} \bibinfo{person}{Neil~D. Sandham}.}
  \bibinfo{year}{2017}\natexlab{}.
\newblock \showarticletitle{OpenSBLI: A framework for the automated derivation
  and parallel execution of finite difference solvers on a range of computer
  architectures}.
\newblock \bibinfo{journal}{\emph{Journal of Computational Science}}
  \bibinfo{volume}{18} (\bibinfo{year}{2017}), \bibinfo{pages}{12--23}.
\newblock
\showISSN{1877-7503}
\urldef\tempurl%
\url{https://doi.org/10.1016/j.jocs.2016.11.001}
\showDOI{\tempurl}


\bibitem[Kjolstad et~al\mbox{.}(2017)]%
        {kjolstad2017tensor}
\bibfield{author}{\bibinfo{person}{Fredrik Kjolstad}, \bibinfo{person}{Shoaib
  Kamil}, \bibinfo{person}{Stephen Chou}, \bibinfo{person}{David Lugato}, {and}
  \bibinfo{person}{Saman Amarasinghe}.} \bibinfo{year}{2017}\natexlab{}.
\newblock \showarticletitle{The tensor algebra compiler}.
\newblock \bibinfo{journal}{\emph{Proc. ACM Program. Lang.}}
  \bibinfo{volume}{1}, \bibinfo{number}{OOPSLA}, Article
  \bibinfo{articleno}{77} (\bibinfo{year}{2017}), \bibinfo{numpages}{29}~pages.
\newblock
\urldef\tempurl%
\url{https://doi.org/10.1145/3133901}
\showDOI{\tempurl}


\bibitem[Kl\"{o}ckner(2014)]%
        {klockner2014loopy}
\bibfield{author}{\bibinfo{person}{Andreas Kl\"{o}ckner}.}
  \bibinfo{year}{2014}\natexlab{}.
\newblock \showarticletitle{Loo.py: transformation-based code generation for
  GPUs and CPUs}. In \bibinfo{booktitle}{\emph{Proceedings of ACM SIGPLAN
  International Workshop on Libraries, Languages, and Compilers for Array
  Programming}} (Edinburgh, United Kingdom)
  \emph{(\bibinfo{series}{ARRAY'14})}. \bibinfo{publisher}{Association for
  Computing Machinery}, \bibinfo{address}{New York, NY, USA},
  \bibinfo{pages}{82--87}.
\newblock
\showISBNx{9781450329378}
\urldef\tempurl%
\url{https://doi.org/10.1145/2627373.2627387}
\showDOI{\tempurl}


\bibitem[Kuckuk and Köstler(2016)]%
        {exaslangMPIKucuk:2016}
\bibfield{author}{\bibinfo{person}{Sebastian Kuckuk} {and}
  \bibinfo{person}{Harald Köstler}.} \bibinfo{year}{2016}\natexlab{}.
\newblock \showarticletitle{Automatic Generation of Massively Parallel Codes
  from ExaSlang}.
\newblock \bibinfo{journal}{\emph{Computation}} \bibinfo{volume}{4},
  \bibinfo{number}{3} (\bibinfo{year}{2016}).
\newblock
\showISSN{2079-3197}
\urldef\tempurl%
\url{https://doi.org/10.3390/computation4030027}
\showDOI{\tempurl}


\bibitem[Lam et~al\mbox{.}(2015)]%
        {lam2015numba}
\bibfield{author}{\bibinfo{person}{Siu~Kwan Lam}, \bibinfo{person}{Antoine
  Pitrou}, {and} \bibinfo{person}{Stanley Seibert}.}
  \bibinfo{year}{2015}\natexlab{}.
\newblock \showarticletitle{Numba: a LLVM-based Python JIT compiler}. In
  \bibinfo{booktitle}{\emph{Proceedings of the Second Workshop on the LLVM
  Compiler Infrastructure in HPC}} (Austin, Texas) \emph{(\bibinfo{series}{LLVM
  '15})}. \bibinfo{publisher}{Association for Computing Machinery},
  \bibinfo{address}{New York, NY, USA}, Article \bibinfo{articleno}{7},
  \bibinfo{numpages}{6}~pages.
\newblock
\showISBNx{9781450340052}
\urldef\tempurl%
\url{https://doi.org/10.1145/2833157.2833162}
\showDOI{\tempurl}


\bibitem[Lattner et~al\mbox{.}(2021)]%
        {mlir}
\bibfield{author}{\bibinfo{person}{Chris Lattner}, \bibinfo{person}{Mehdi
  Amini}, \bibinfo{person}{Uday Bondhugula}, \bibinfo{person}{Albert Cohen},
  \bibinfo{person}{Andy Davis}, \bibinfo{person}{Jacques Pienaar},
  \bibinfo{person}{River Riddle}, \bibinfo{person}{Tatiana Shpeisman},
  \bibinfo{person}{Nicolas Vasilache}, {and} \bibinfo{person}{Oleksandr
  Zinenko}.} \bibinfo{year}{2021}\natexlab{}.
\newblock \showarticletitle{{{MLIR}}: Scaling Compiler Infrastructure for
  Domain Specific Computation}. In \bibinfo{booktitle}{\emph{2021 {{IEEE/ACM}}
  International Symposium on Code Generation and Optimization (CGO)}}.
  \bibinfo{publisher}{IEEE}, \bibinfo{pages}{2--14}.
\newblock
\urldef\tempurl%
\url{https://doi.org/10.1109/CGO51591.2021.9370308}
\showDOI{\tempurl}


\bibitem[Lei{\ss}a et~al\mbox{.}(2018)]%
        {anydsl2018}
\bibfield{author}{\bibinfo{person}{Roland Lei{\ss}a}, \bibinfo{person}{Klaas
  Boesche}, \bibinfo{person}{Sebastian Hack}, \bibinfo{person}{Ars{\`{e}}ne
  P{\'{e}}rard{-}Gayot}, \bibinfo{person}{Richard Membarth},
  \bibinfo{person}{Philipp Slusallek}, \bibinfo{person}{Andr{\'{e}}
  M{\"{u}}ller}, {and} \bibinfo{person}{Bertil Schmidt}.}
  \bibinfo{year}{2018}\natexlab{}.
\newblock \showarticletitle{Any{DSL}: a partial evaluation framework for
  programming high-performance libraries}.
\newblock \bibinfo{journal}{\emph{Proc. {ACM} Program. Lang.}}
  \bibinfo{volume}{2}, \bibinfo{number}{{OOPSLA}} (\bibinfo{year}{2018}),
  \bibinfo{pages}{119:1--119:30}.
\newblock
\urldef\tempurl%
\url{https://doi.org/10.1145/3276489}
\showDOI{\tempurl}


\bibitem[Lengauer et~al\mbox{.}(2020)]%
        {exastencils_2020}
\bibfield{author}{\bibinfo{person}{Christian Lengauer}, \bibinfo{person}{Sven
  Apel}, \bibinfo{person}{Matthias Bolten}, \bibinfo{person}{Shigeru Chiba},
  \bibinfo{person}{Ulrich R{\"u}de}, \bibinfo{person}{J{\"u}rgen Teich},
  \bibinfo{person}{Armin Gr{\"o}{\ss}linger}, \bibinfo{person}{Frank Hannig},
  \bibinfo{person}{Harald K{\"o}stler}, \bibinfo{person}{Lisa Claus},
  \bibinfo{person}{Alexander Grebhahn}, \bibinfo{person}{Stefan Groth},
  \bibinfo{person}{Stefan Kronawitter}, \bibinfo{person}{Sebastian Kuckuk},
  \bibinfo{person}{Hannah Rittich}, \bibinfo{person}{Christian Schmitt}, {and}
  \bibinfo{person}{Jonas Schmitt}.} \bibinfo{year}{2020}\natexlab{}.
\newblock \showarticletitle{ExaStencils: Advanced Multigrid Solver Generation}.
  In \bibinfo{booktitle}{\emph{Software for Exascale Computing - SPPEXA
  2016-2019}}, \bibfield{editor}{\bibinfo{person}{Hans-Joachim Bungartz},
  \bibinfo{person}{Severin Reiz}, \bibinfo{person}{Benjamin Uekermann},
  \bibinfo{person}{Philipp Neumann}, {and} \bibinfo{person}{Wolfgang~E. Nagel}}
  (Eds.). \bibinfo{publisher}{Springer International Publishing},
  \bibinfo{address}{Cham}, \bibinfo{pages}{405--452}.
\newblock
\showISBNx{978-3-030-47956-5}
\urldef\tempurl%
\url{https://doi.org/10.1007/978-3-030-47956-5_14}
\showURL{%
\tempurl}


\bibitem[Li et~al\mbox{.}(2024)]%
        {li2024}
\bibfield{author}{\bibinfo{person}{Mingzhen Li}, \bibinfo{person}{Bangduo
  Chen}, \bibinfo{person}{Hailong Yang}, \bibinfo{person}{Zhongzhi Luan}, {and}
  \bibinfo{person}{Depei Qian}.} \bibinfo{year}{2024}\natexlab{}.
\newblock \showarticletitle{Building a domain-specific compiler for emerging
  processors with a reusable approach}. In \bibinfo{booktitle}{\emph{Science
  China Information Sciences}}, Vol.~\bibinfo{volume}{67}.
\newblock
\urldef\tempurl%
\url{https://doi.org/10.1007/s11432-022-3727-6}
\showDOI{\tempurl}


\bibitem[Li et~al\mbox{.}(2021)]%
        {li2021}
\bibfield{author}{\bibinfo{person}{Mingzhen Li}, \bibinfo{person}{Yi Liu},
  \bibinfo{person}{Hailong Yang}, \bibinfo{person}{Yongmin Hu},
  \bibinfo{person}{Qingxiao Sun}, \bibinfo{person}{Bangduo Chen},
  \bibinfo{person}{Xin You}, \bibinfo{person}{Xiaoyan Liu},
  \bibinfo{person}{Zhongzhi Luan}, {and} \bibinfo{person}{Depei Qian}.}
  \bibinfo{year}{2021}\natexlab{}.
\newblock \showarticletitle{Automatic Code Generation and Optimization of
  Large-Scale Stencil Computation on Many-Core Processors}. In
  \bibinfo{booktitle}{\emph{Proceedings of the 50th International Conference on
  Parallel Processing}} (Lemont, IL, USA) \emph{(\bibinfo{series}{ICPP '21})}.
  \bibinfo{publisher}{ACM}, \bibinfo{address}{New York, NY, USA}, Article
  \bibinfo{articleno}{34}, \bibinfo{numpages}{12}~pages.
\newblock
\showISBNx{9781450390682}
\urldef\tempurl%
\url{https://doi.org/10.1145/3472456.3473517}
\showDOI{\tempurl}


\bibitem[Liao et~al\mbox{.}(2015)]%
        {Liao2015_shiftcalculus}
\bibfield{author}{\bibinfo{person}{Chunhua Liao}, \bibinfo{person}{Pei-Hung
  Lin}, \bibinfo{person}{Daniel~J. Quinlan}, \bibinfo{person}{Yue Zhao}, {and}
  \bibinfo{person}{Xipeng Shen}.} \bibinfo{year}{2015}\natexlab{}.
\newblock \showarticletitle{Enhancing Domain Specific Language Implementations
  through Ontology}. In \bibinfo{booktitle}{\emph{Proceedings of the 5th
  International Workshop on Domain-Specific Languages and High-Level Frameworks
  for High Performance Computing}} (Austin, Texas)
  \emph{(\bibinfo{series}{WOLFHPC '15})}. \bibinfo{publisher}{ACM},
  \bibinfo{address}{New York, NY, USA}, Article \bibinfo{articleno}{3},
  \bibinfo{numpages}{9}~pages.
\newblock
\showISBNx{9781450340168}
\urldef\tempurl%
\url{https://doi.org/10.1145/2830018.2830022}
\showDOI{\tempurl}


\bibitem[Logg et~al\mbox{.}(2012)]%
        {logg2012automated}
\bibfield{editor}{\bibinfo{person}{A. Logg}, \bibinfo{person}{K.A. Mardal},
  {and} \bibinfo{person}{G. Wells}} (Eds.). \bibinfo{year}{2012}\natexlab{}.
\newblock \bibinfo{booktitle}{\emph{Automated Solution of Differential
  Equations by the Finite Element Method: The FEniCS Book}}.
\newblock \bibinfo{publisher}{Springer Berlin Heidelberg}.
\newblock
\showISBNx{9783642230998}
\urldef\tempurl%
\url{https://doi.org/10.1007/978-3-642-23099-8}
\showDOI{\tempurl}


\bibitem[Louboutin et~al\mbox{.}(2019)]%
        {devito2019}
\bibfield{author}{\bibinfo{person}{M. Louboutin}, \bibinfo{person}{M. Lange},
  \bibinfo{person}{F. Luporini}, \bibinfo{person}{N. Kukreja},
  \bibinfo{person}{P.~A. Witte}, \bibinfo{person}{F.~J. Herrmann},
  \bibinfo{person}{P. Velesko}, {and} \bibinfo{person}{G.~J. Gorman}.}
  \bibinfo{year}{2019}\natexlab{}.
\newblock \showarticletitle{Devito (v3.1.0): an embedded domain-specific
  language for finite differences and geophysical exploration}.
\newblock \bibinfo{journal}{\emph{Geoscientific Model Development}}
  \bibinfo{volume}{12}, \bibinfo{number}{3} (\bibinfo{year}{2019}),
  \bibinfo{pages}{1165--1187}.
\newblock
\urldef\tempurl%
\url{https://doi.org/10.5194/gmd-12-1165-2019}
\showDOI{\tempurl}


\bibitem[Luporini et~al\mbox{.}(2020)]%
        {devitoTOMS2020}
\bibfield{author}{\bibinfo{person}{Fabio Luporini}, \bibinfo{person}{Mathias
  Louboutin}, \bibinfo{person}{Michael Lange}, \bibinfo{person}{Navjot
  Kukreja}, \bibinfo{person}{Philipp Witte}, \bibinfo{person}{Jan
  H\"{u}ckelheim}, \bibinfo{person}{Charles Yount}, \bibinfo{person}{Paul H.~J.
  Kelly}, \bibinfo{person}{Felix~J. Herrmann}, {and} \bibinfo{person}{Gerard~J.
  Gorman}.} \bibinfo{year}{2020}\natexlab{}.
\newblock \showarticletitle{Architecture and Performance of Devito, a System
  for Automated Stencil Computation}.
\newblock \bibinfo{journal}{\emph{ACM Trans. Math. Softw.}}
  \bibinfo{volume}{46}, \bibinfo{number}{1}, Article \bibinfo{articleno}{6}
  (\bibinfo{date}{apr} \bibinfo{year}{2020}), \bibinfo{numpages}{28}~pages.
\newblock
\showISSN{0098-3500}
\urldef\tempurl%
\url{https://doi.org/10.1145/3374916}
\showDOI{\tempurl}


\bibitem[Majumder and Bondhugula(2024)]%
        {majumder2021hir}
\bibfield{author}{\bibinfo{person}{Kingshuk Majumder} {and}
  \bibinfo{person}{Uday Bondhugula}.} \bibinfo{year}{2024}\natexlab{}.
\newblock \bibinfo{title}{HIR: An MLIR-based Intermediate Representation for
  Hardware Accelerator Description}.
\newblock , \bibinfo{numpages}{13}~pages.
\newblock
\showISBNx{9798400703942}
\urldef\tempurl%
\url{https://doi.org/10.1145/3623278.3624767}
\showDOI{\tempurl}


\bibitem[Maruyama et~al\mbox{.}(2011)]%
        {physis2011}
\bibfield{author}{\bibinfo{person}{Naoya Maruyama}, \bibinfo{person}{Tatsuo
  Nomura}, \bibinfo{person}{Kento Sato}, {and} \bibinfo{person}{Satoshi
  Matsuoka}.} \bibinfo{year}{2011}\natexlab{}.
\newblock \showarticletitle{Physis: An Implicitly Parallel Programming Model
  for Stencil Computations on Large-Scale GPU-Accelerated Supercomputers}. In
  \bibinfo{booktitle}{\emph{Proceedings of 2011 International Conference for
  High Performance Computing, Networking, Storage and Analysis}} (Seattle,
  Washington) \emph{(\bibinfo{series}{SC '11})}. \bibinfo{publisher}{ACM},
  \bibinfo{address}{New York, NY, USA}, Article \bibinfo{articleno}{11},
  \bibinfo{numpages}{12}~pages.
\newblock
\showISBNx{9781450307710}
\urldef\tempurl%
\url{https://doi.org/10.1145/2063384.2063398}
\showDOI{\tempurl}


\bibitem[Melvin et~al\mbox{.}(2017)]%
        {melvin2017lfric}
\bibfield{author}{\bibinfo{person}{Thomas Melvin}, \bibinfo{person}{Steve
  Mullerworth}, \bibinfo{person}{Rupert Ford}, \bibinfo{person}{Chris Maynard},
  {and} \bibinfo{person}{Mike Hobson}.} \bibinfo{year}{2017}\natexlab{}.
\newblock \showarticletitle{LFRic: Building a new Unified Model}. In
  \bibinfo{booktitle}{\emph{EGU General Assembly Conference Abstracts}}.
  \bibinfo{pages}{13021}.
\newblock


\bibitem[Meurer et~al\mbox{.}(2017)]%
        {SymPy}
\bibfield{author}{\bibinfo{person}{Aaron Meurer},
  \bibinfo{person}{Christopher~P. Smith}, \bibinfo{person}{Mateusz Paprocki},
  \bibinfo{person}{Ond\v{r}ej \v{C}ert\'{i}k}, \bibinfo{person}{Sergey~B.
  Kirpichev}, \bibinfo{person}{Matthew Rocklin}, \bibinfo{person}{Amit Kumar},
  \bibinfo{person}{Sergiu Ivanov}, \bibinfo{person}{Jason~K. Moore},
  \bibinfo{person}{Sartaj Singh}, \bibinfo{person}{Thilina Rathnayake},
  \bibinfo{person}{Sean Vig}, \bibinfo{person}{Brian~E. Granger},
  \bibinfo{person}{Richard~P. Muller}, \bibinfo{person}{Francesco Bonazzi},
  \bibinfo{person}{Harsh Gupta}, \bibinfo{person}{Shivam Vats},
  \bibinfo{person}{Fredrik Johansson}, \bibinfo{person}{Fabian Pedregosa},
  \bibinfo{person}{Matthew~J. Curry}, \bibinfo{person}{Andy~R. Terrel},
  \bibinfo{person}{\v{S}t\v{e}p\'{a}n Rou\v{c}ka}, \bibinfo{person}{Ashutosh
  Saboo}, \bibinfo{person}{Isuru Fernando}, \bibinfo{person}{Sumith Kulal},
  \bibinfo{person}{Robert Cimrman}, {and} \bibinfo{person}{Anthony Scopatz}.}
  \bibinfo{year}{2017}\natexlab{}.
\newblock \showarticletitle{SymPy: symbolic computing in Python}.
\newblock \bibinfo{journal}{\emph{PeerJ Computer Science}}  \bibinfo{volume}{3}
  (\bibinfo{date}{Jan.} \bibinfo{year}{2017}), \bibinfo{pages}{e103}.
\newblock
\showISSN{2376-5992}
\urldef\tempurl%
\url{https://doi.org/10.7717/peerj-cs.103}
\showDOI{\tempurl}


\bibitem[Mitenkov et~al\mbox{.}(2023)]%
        {mitenkov2023}
\bibfield{author}{\bibinfo{person}{George Mitenkov}, \bibinfo{person}{Ioannis
  Magkanaris}, \bibinfo{person}{Omar Awile}, \bibinfo{person}{Pramod Kumbhar},
  \bibinfo{person}{Felix Sch\"{u}rmann}, {and} \bibinfo{person}{Alastair~F.
  Donaldson}.} \bibinfo{year}{2023}\natexlab{}.
\newblock \showarticletitle{{MOD2IR}: High-Performance Code Generation for a
  Biophysically Detailed Neuronal Simulation {DSL}}. In
  \bibinfo{booktitle}{\emph{Proceedings of the 32nd ACM SIGPLAN International
  Conference on Compiler Construction}} (Montr\'{e}al, QC, Canada)
  \emph{(\bibinfo{series}{CC 2023})}. \bibinfo{publisher}{ACM},
  \bibinfo{address}{New York, NY, USA}, \bibinfo{pages}{203--215}.
\newblock
\showISBNx{9798400700880}
\urldef\tempurl%
\url{https://doi.org/10.1145/3578360.3580268}
\showDOI{\tempurl}


\bibitem[Modular({[n.\,d.]})]%
        {mojo2023}
\bibfield{author}{\bibinfo{person}{Modular}.}
  \bibinfo{year}{[n.\,d.]}\natexlab{}.
\newblock \bibinfo{title}{Mojo: Programming language for all of AI}.
\newblock
\newblock
\urldef\tempurl%
\url{https://www.modular.com/mojo}
\showURL{%
\tempurl}


\bibitem[Mudalige et~al\mbox{.}(2019)]%
        {mudalige2019}
\bibfield{author}{\bibinfo{person}{G.~R. Mudalige}, \bibinfo{person}{I.~Z.
  Reguly}, \bibinfo{person}{S.~P. Jammy}, \bibinfo{person}{C.~T. Jacobs},
  \bibinfo{person}{M.~B. Giles}, {and} \bibinfo{person}{N.~D. Sandham}.}
  \bibinfo{year}{2019}\natexlab{}.
\newblock \showarticletitle{Large-scale performance of a DSL-based multi-block
  structured-mesh application for Direct Numerical Simulation}.
\newblock \bibinfo{journal}{\emph{J. Parallel and Distrib. Comput.}}
  \bibinfo{volume}{131} (\bibinfo{year}{2019}), \bibinfo{pages}{130--146}.
\newblock
\showISSN{0743-7315}
\urldef\tempurl%
\url{https://doi.org/10.1016/j.jpdc.2019.04.019}
\showDOI{\tempurl}


\bibitem[Mullapudi et~al\mbox{.}(2015)]%
        {polymage2015}
\bibfield{author}{\bibinfo{person}{Ravi~Teja Mullapudi}, \bibinfo{person}{Vinay
  Vasista}, {and} \bibinfo{person}{Uday Bondhugula}.}
  \bibinfo{year}{2015}\natexlab{}.
\newblock \showarticletitle{PolyMage: Automatic Optimization for Image
  Processing Pipelines}. In \bibinfo{booktitle}{\emph{Proceedings of the
  Twentieth International Conference on Architectural Support for Programming
  Languages and Operating Systems}} (Istanbul, Turkey)
  \emph{(\bibinfo{series}{ASPLOS '15})}. \bibinfo{publisher}{Association for
  Computing Machinery}, \bibinfo{address}{New York, NY, USA},
  \bibinfo{pages}{429--443}.
\newblock
\showISBNx{9781450328357}
\urldef\tempurl%
\url{https://doi.org/10.1145/2694344.2694364}
\showDOI{\tempurl}


\bibitem[Omlin et~al\mbox{.}(2022)]%
        {omlin2022distributed}
\bibfield{author}{\bibinfo{person}{Samuel Omlin}, \bibinfo{person}{Ludovic
  R{\"a}ss}, {and} \bibinfo{person}{Ivan Utkin}.}
  \bibinfo{year}{2022}\natexlab{}.
\newblock \bibinfo{title}{Distributed Parallelization of x{PU} Stencil
  Computations in {J}ulia}.
\newblock
\newblock
\showeprint[arxiv]{2211.15716}~[cs.DC]


\bibitem[Paszke et~al\mbox{.}(2019)]%
        {pytorch2019nips}
\bibfield{author}{\bibinfo{person}{Adam Paszke}, \bibinfo{person}{Sam Gross},
  \bibinfo{person}{Francisco Massa}, \bibinfo{person}{Adam Lerer},
  \bibinfo{person}{James Bradbury}, \bibinfo{person}{Gregory Chanan},
  \bibinfo{person}{Trevor Killeen}, \bibinfo{person}{Zeming Lin},
  \bibinfo{person}{Natalia Gimelshein}, \bibinfo{person}{Luca Antiga},
  \bibinfo{person}{Alban Desmaison}, \bibinfo{person}{Andreas K\"{o}pf},
  \bibinfo{person}{Edward Yang}, \bibinfo{person}{Zach DeVito},
  \bibinfo{person}{Martin Raison}, \bibinfo{person}{Alykhan Tejani},
  \bibinfo{person}{Sasank Chilamkurthy}, \bibinfo{person}{Benoit Steiner},
  \bibinfo{person}{Lu Fang}, \bibinfo{person}{Junjie Bai}, {and}
  \bibinfo{person}{Soumith Chintala}.} \bibinfo{year}{2019}\natexlab{}.
\newblock \showarticletitle{PyTorch: an imperative style, high-performance deep
  learning library}.
\newblock In \bibinfo{booktitle}{\emph{Proceedings of the 33rd International
  Conference on Neural Information Processing Systems}}.
  \bibinfo{publisher}{Curran Associates Inc.}, \bibinfo{address}{Red Hook, NY,
  USA}.
\newblock
\urldef\tempurl%
\url{https://dl.acm.org/doi/10.5555/3454287.3455008}
\showURL{%
\tempurl}


\bibitem[Piacsek and Williams(1970)]%
        {piacsek1970conservation}
\bibfield{author}{\bibinfo{person}{Steve~A Piacsek} {and}
  \bibinfo{person}{Gareth~P Williams}.} \bibinfo{year}{1970}\natexlab{}.
\newblock \showarticletitle{Conservation properties of convection difference
  schemes}.
\newblock \bibinfo{journal}{\emph{J. Comput. Phys.}} \bibinfo{volume}{6},
  \bibinfo{number}{3} (\bibinfo{year}{1970}), \bibinfo{pages}{392--405}.
\newblock
\showISSN{0021-9991}
\urldef\tempurl%
\url{https://doi.org/10.1016/0021-9991(70)90038-0}
\showDOI{\tempurl}


\bibitem[Porter et~al\mbox{.}(2018)]%
        {porter2018portable}
\bibfield{author}{\bibinfo{person}{A.~R. Porter}, \bibinfo{person}{J.
  Appleyard}, \bibinfo{person}{M. Ashworth}, \bibinfo{person}{R.~W. Ford},
  \bibinfo{person}{J. Holt}, \bibinfo{person}{H. Liu}, {and}
  \bibinfo{person}{G.~D. Riley}.} \bibinfo{year}{2018}\natexlab{}.
\newblock \showarticletitle{Portable multi- and many-core performance for
  finite-difference or finite-element codes -- application to the free-surface
  component of NEMO (NEMOLite2D 1.0)}.
\newblock \bibinfo{journal}{\emph{Geoscientific Model Development}}
  \bibinfo{volume}{11}, \bibinfo{number}{8} (\bibinfo{year}{2018}),
  \bibinfo{pages}{3447--3464}.
\newblock
\urldef\tempurl%
\url{https://doi.org/10.5194/gmd-11-3447-2018}
\showDOI{\tempurl}


\bibitem[Puschel et~al\mbox{.}(2005)]%
        {puschel2005spiral}
\bibfield{author}{\bibinfo{person}{M. Puschel}, \bibinfo{person}{J.~M.~F.
  Moura}, \bibinfo{person}{J.~R. Johnson}, \bibinfo{person}{D. Padua},
  \bibinfo{person}{M.~M. Veloso}, \bibinfo{person}{B.~W. Singer},
  \bibinfo{person}{Jianxin Xiong}, \bibinfo{person}{F. Franchetti},
  \bibinfo{person}{A. Gacic}, \bibinfo{person}{Y. Voronenko},
  \bibinfo{person}{K. Chen}, \bibinfo{person}{R.~W. Johnson}, {and}
  \bibinfo{person}{N. Rizzolo}.} \bibinfo{year}{2005}\natexlab{}.
\newblock \showarticletitle{SPIRAL: Code Generation for DSP Transforms}.
\newblock \bibinfo{journal}{\emph{Proc. IEEE}} \bibinfo{volume}{93},
  \bibinfo{number}{2} (\bibinfo{year}{2005}), \bibinfo{pages}{232--275}.
\newblock
\urldef\tempurl%
\url{https://doi.org/10.1109/JPROC.2004.840306}
\showDOI{\tempurl}


\bibitem[Ragan{-}Kelley et~al\mbox{.}(2013)]%
        {halide2013}
\bibfield{author}{\bibinfo{person}{Jonathan Ragan{-}Kelley},
  \bibinfo{person}{Connelly Barnes}, \bibinfo{person}{Andrew Adams},
  \bibinfo{person}{Sylvain Paris}, \bibinfo{person}{Fr{\'{e}}do Durand}, {and}
  \bibinfo{person}{Saman~P. Amarasinghe}.} \bibinfo{year}{2013}\natexlab{}.
\newblock \showarticletitle{Halide: a language and compiler for optimizing
  parallelism, locality, and recomputation in image processing pipelines}. In
  \bibinfo{booktitle}{\emph{{ACM} {SIGPLAN} Conference on Programming Language
  Design and Implementation, {PLDI} '13, Seattle, WA, USA, June 16-19, 2013}},
  \bibfield{editor}{\bibinfo{person}{Hans{-}Juergen Boehm} {and}
  \bibinfo{person}{Cormac Flanagan}} (Eds.). \bibinfo{publisher}{{ACM}},
  \bibinfo{address}{New York, NY, USA}, \bibinfo{pages}{519--530}.
\newblock
\urldef\tempurl%
\url{https://doi.org/10.1145/2491956.2462176}
\showDOI{\tempurl}


\bibitem[Ravishankar et~al\mbox{.}(2015)]%
        {Forma2015}
\bibfield{author}{\bibinfo{person}{Mahesh Ravishankar}, \bibinfo{person}{Justin
  Holewinski}, {and} \bibinfo{person}{Vinod Grover}.}
  \bibinfo{year}{2015}\natexlab{}.
\newblock \showarticletitle{Forma: A DSL for Image Processing Applications to
  Target GPUs and Multi-Core CPUs}. In \bibinfo{booktitle}{\emph{Proceedings of
  the 8th Workshop on General Purpose Processing Using GPUs}} (San Francisco,
  CA, USA) \emph{(\bibinfo{series}{GPGPU-8})}. \bibinfo{publisher}{Association
  for Computing Machinery}, \bibinfo{address}{New York, NY, USA},
  \bibinfo{pages}{109--120}.
\newblock
\showISBNx{9781450334075}
\urldef\tempurl%
\url{https://doi.org/10.1145/2716282.2716290}
\showDOI{\tempurl}


\bibitem[Rawat et~al\mbox{.}(2015)]%
        {Rawat2015}
\bibfield{author}{\bibinfo{person}{Prashant Rawat}, \bibinfo{person}{Martin
  Kong}, \bibinfo{person}{Tom Henretty}, \bibinfo{person}{Justin Holewinski},
  \bibinfo{person}{Kevin Stock}, \bibinfo{person}{Louis-No\"{e}l Pouchet},
  \bibinfo{person}{J. Ramanujam}, \bibinfo{person}{Atanas Rountev}, {and}
  \bibinfo{person}{P. Sadayappan}.} \bibinfo{year}{2015}\natexlab{}.
\newblock \showarticletitle{SDSLc: A Multi-Target Domain-Specific Compiler for
  Stencil Computations}. In \bibinfo{booktitle}{\emph{Proceedings of the 5th
  International Workshop on Domain-Specific Languages and High-Level Frameworks
  for High Performance Computing}} (Austin, Texas)
  \emph{(\bibinfo{series}{WOLFHPC '15})}. \bibinfo{publisher}{ACM},
  \bibinfo{address}{New York, NY, USA}, Article \bibinfo{articleno}{6},
  \bibinfo{numpages}{10}~pages.
\newblock
\showISBNx{9781450340168}
\urldef\tempurl%
\url{https://doi.org/10.1145/2830018.2830025}
\showDOI{\tempurl}


\bibitem[Rawat et~al\mbox{.}(2018)]%
        {Rawat2018}
\bibfield{author}{\bibinfo{person}{Prashant~Singh Rawat},
  \bibinfo{person}{Miheer Vaidya}, \bibinfo{person}{Aravind Sukumaran-Rajam},
  \bibinfo{person}{Mahesh Ravishankar}, \bibinfo{person}{Vinod Grover},
  \bibinfo{person}{Atanas Rountev}, \bibinfo{person}{Louis~Noel Pouchet}, {and}
  \bibinfo{person}{P. Sadayappan}.} \bibinfo{year}{2018}\natexlab{}.
\newblock \showarticletitle{Domain-Specific Optimization and Generation of
  High-Performance {GPU} Code for Stencil Computations}.
\newblock \bibinfo{journal}{\emph{Proc. IEEE}} \bibinfo{volume}{106},
  \bibinfo{number}{11} (\bibinfo{year}{2018}), \bibinfo{pages}{1902--1920}.
\newblock
\showISSN{00189219}
\urldef\tempurl%
\url{https://doi.org/10.1109/JPROC.2018.2862896}
\showDOI{\tempurl}


\bibitem[Rawat et~al\mbox{.}(2019)]%
        {rawat2019}
\bibfield{author}{\bibinfo{person}{Prashant~Singh Rawat},
  \bibinfo{person}{Miheer Vaidya}, \bibinfo{person}{Aravind Sukumaran-Rajam},
  \bibinfo{person}{Atanas Rountev}, \bibinfo{person}{Louis-No\"{e}l Pouchet},
  {and} \bibinfo{person}{P. Sadayappan}.} \bibinfo{year}{2019}\natexlab{}.
\newblock \showarticletitle{On Optimizing Complex Stencils on {GPU}s}. In
  \bibinfo{booktitle}{\emph{2019 IEEE International Parallel and Distributed
  Processing Symposium (IPDPS)}}. \bibinfo{publisher}{IEEE},
  \bibinfo{pages}{641--652}.
\newblock
\urldef\tempurl%
\url{https://doi.org/10.1109/IPDPS.2019.00073}
\showDOI{\tempurl}


\bibitem[Reguly et~al\mbox{.}(2016)]%
        {op2Reguly2016}
\bibfield{author}{\bibinfo{person}{István~Z. Reguly},
  \bibinfo{person}{Gihan~R. Mudalige}, \bibinfo{person}{Carlo Bertolli},
  \bibinfo{person}{Michael~B. Giles}, \bibinfo{person}{Adam Betts},
  \bibinfo{person}{Paul H.~J. Kelly}, {and} \bibinfo{person}{David Radford}.}
  \bibinfo{year}{2016}\natexlab{}.
\newblock \showarticletitle{Acceleration of a Full-Scale Industrial CFD
  Application with OP2}.
\newblock \bibinfo{journal}{\emph{IEEE Transactions on Parallel and Distributed
  Systems}} \bibinfo{volume}{27}, \bibinfo{number}{5} (\bibinfo{year}{2016}),
  \bibinfo{pages}{1265--1278}.
\newblock
\urldef\tempurl%
\url{https://doi.org/10.1109/TPDS.2015.2453972}
\showDOI{\tempurl}


\bibitem[Reguly et~al\mbox{.}(2018)]%
        {ops2018}
\bibfield{author}{\bibinfo{person}{Istv{\'{a}}n~Z. Reguly},
  \bibinfo{person}{Gihan~R. Mudalige}, {and} \bibinfo{person}{Michael~B.
  Giles}.} \bibinfo{year}{2018}\natexlab{}.
\newblock \showarticletitle{Loop Tiling in Large-Scale Stencil Codes at
  Run-Time with {OPS}}.
\newblock \bibinfo{journal}{\emph{{IEEE} Trans. Parallel Distributed Syst.}}
  \bibinfo{volume}{29}, \bibinfo{number}{4} (\bibinfo{year}{2018}),
  \bibinfo{pages}{873--886}.
\newblock
\urldef\tempurl%
\url{https://doi.org/10.1109/TPDS.2017.2778161}
\showDOI{\tempurl}


\bibitem[Rodriguez-Canal et~al\mbox{.}(2023a)]%
        {fortran-fpga}
\bibfield{author}{\bibinfo{person}{G. Rodriguez-Canal}, \bibinfo{person}{N.
  Brown}, \bibinfo{person}{T. Dykes}, \bibinfo{person}{J. Jones}, {and}
  \bibinfo{person}{U. Haus}.} \bibinfo{year}{2023}\natexlab{a}.
\newblock \showarticletitle{Fortran High-Level Synthesis: Reducing the Barriers
  to Accelerating HPC Codes on FPGAs}. In \bibinfo{booktitle}{\emph{2023 33rd
  International Conference on Field-Programmable Logic and Applications
  (FPL)}}. \bibinfo{publisher}{IEEE Computer Society}, \bibinfo{address}{Los
  Alamitos, CA, USA}, \bibinfo{pages}{10--18}.
\newblock
\urldef\tempurl%
\url{https://doi.org/10.1109/FPL60245.2023.00010}
\showDOI{\tempurl}


\bibitem[Rodriguez-Canal et~al\mbox{.}(2023b)]%
        {rodriguez2023stencil}
\bibfield{author}{\bibinfo{person}{Gabriel Rodriguez-Canal},
  \bibinfo{person}{Nick Brown}, \bibinfo{person}{Maurice Jamieson},
  \bibinfo{person}{Emilien Bauer}, \bibinfo{person}{Anton Lydike}, {and}
  \bibinfo{person}{Tobias Grosser}.} \bibinfo{year}{2023}\natexlab{b}.
\newblock \showarticletitle{Stencil-HMLS: A multi-layered approach to the
  automatic optimisation of stencil codes on FPGA}. In
  \bibinfo{booktitle}{\emph{Proceedings of the SC '23 Workshops of The
  International Conference on High Performance Computing, Network, Storage, and
  Analysis}} (<conf-loc>, <city>Denver</city>, <state>CO</state>,
  <country>USA</country>, </conf-loc>) \emph{(\bibinfo{series}{SC-W '23})}.
  \bibinfo{publisher}{Association for Computing Machinery},
  \bibinfo{address}{New York, NY, USA}, \bibinfo{pages}{556--565}.
\newblock
\showISBNx{9798400707858}
\urldef\tempurl%
\url{https://doi.org/10.1145/3624062.3624543}
\showDOI{\tempurl}


\bibitem[{Science and Technology Facilities Council (STFC)}(2021)]%
        {psyclone_bench}
\bibfield{author}{\bibinfo{person}{{Science and Technology Facilities Council
  (STFC)}}.} \bibinfo{year}{2021}\natexlab{}.
\newblock \bibinfo{title}{{PS}yclone{B}ench: small benchmarks used to inform
  the development of the {PS}yclone Domain-Specific Compiler}.
\newblock
\newblock
\urldef\tempurl%
\url{https://github.com/stfc/PSycloneBench}
\showURL{%
\tempurl}
\newblock
\shownote{Accessed 2023-04-24}.


\bibitem[Shajii et~al\mbox{.}(2023)]%
        {codon:2023}
\bibfield{author}{\bibinfo{person}{Ariya Shajii}, \bibinfo{person}{Gabriel
  Ramirez}, \bibinfo{person}{Haris Smajlovi\'{c}}, \bibinfo{person}{Jessica
  Ray}, \bibinfo{person}{Bonnie Berger}, \bibinfo{person}{Saman Amarasinghe},
  {and} \bibinfo{person}{Ibrahim Numanagi\'{c}}.}
  \bibinfo{year}{2023}\natexlab{}.
\newblock \showarticletitle{Codon: A Compiler for High-Performance Pythonic
  Applications and {DSL}s}. In \bibinfo{booktitle}{\emph{Proceedings of the
  32nd ACM SIGPLAN International Conference on Compiler Construction}}
  (Montr\'{e}al, QC, Canada) \emph{(\bibinfo{series}{CC 2023})}.
  \bibinfo{publisher}{ACM}, \bibinfo{address}{New York, NY, USA},
  \bibinfo{pages}{191--202}.
\newblock
\showISBNx{9798400700880}
\urldef\tempurl%
\url{https://doi.org/10.1145/3578360.3580275}
\showDOI{\tempurl}


\bibitem[Steuwer et~al\mbox{.}(2015)]%
        {LIFTSteuwer2015}
\bibfield{author}{\bibinfo{person}{Michel Steuwer}, \bibinfo{person}{Christian
  Fensch}, \bibinfo{person}{Sam Lindley}, {and} \bibinfo{person}{Christophe
  Dubach}.} \bibinfo{year}{2015}\natexlab{}.
\newblock \showarticletitle{Generating Performance Portable Code Using Rewrite
  Rules: From High-Level Functional Expressions to High-Performance Open{CL}
  Code}.
\newblock \bibinfo{journal}{\emph{SIGPLAN Not.}} \bibinfo{volume}{50},
  \bibinfo{number}{9} (\bibinfo{date}{aug} \bibinfo{year}{2015}),
  \bibinfo{pages}{205--217}.
\newblock
\showISSN{0362-1340}
\urldef\tempurl%
\url{https://doi.org/10.1145/2858949.2784754}
\showDOI{\tempurl}


\bibitem[Tang et~al\mbox{.}(2011)]%
        {pochoir2011}
\bibfield{author}{\bibinfo{person}{Yuan Tang}, \bibinfo{person}{Rezaul~Alam
  Chowdhury}, \bibinfo{person}{Bradley~C. Kuszmaul}, \bibinfo{person}{Chi-Keung
  Luk}, {and} \bibinfo{person}{Charles~E. Leiserson}.}
  \bibinfo{year}{2011}\natexlab{}.
\newblock \showarticletitle{The Pochoir Stencil Compiler}. In
  \bibinfo{booktitle}{\emph{Proceedings of the Twenty-third Annual ACM
  Symposium on Parallelism in Algorithms and Architectures}} (San Jose,
  California, USA) \emph{(\bibinfo{series}{SPAA '11})}.
  \bibinfo{publisher}{ACM}, \bibinfo{address}{New York, NY, USA},
  \bibinfo{pages}{117--128}.
\newblock
\showISBNx{978-1-4503-0743-7}
\urldef\tempurl%
\url{https://doi.org/10.1145/1989493.1989508}
\showDOI{\tempurl}


\bibitem[Thaler et~al\mbox{.}(2019)]%
        {Thaler2019}
\bibfield{author}{\bibinfo{person}{Felix Thaler}, \bibinfo{person}{Stefan
  Moosbrugger}, \bibinfo{person}{Carlos Osuna}, \bibinfo{person}{Mauro Bianco},
  \bibinfo{person}{Hannes Vogt}, \bibinfo{person}{Anton Afanasyev},
  \bibinfo{person}{Lukas Mosimann}, \bibinfo{person}{Oliver Fuhrer},
  \bibinfo{person}{Thomas~C. Schulthess}, {and} \bibinfo{person}{Torsten
  Hoefler}.} \bibinfo{year}{2019}\natexlab{}.
\newblock \showarticletitle{Porting the COSMO Weather Model to Manycore CPUs}.
  In \bibinfo{booktitle}{\emph{Proceedings of the Platform for Advanced
  Scientific Computing Conference}} (Zurich, Switzerland)
  \emph{(\bibinfo{series}{PASC '19})}. \bibinfo{publisher}{Association for
  Computing Machinery}, \bibinfo{address}{New York, NY, USA}, Article
  \bibinfo{articleno}{13}, \bibinfo{numpages}{11}~pages.
\newblock
\showISBNx{9781450367707}
\urldef\tempurl%
\url{https://doi.org/10.1145/3324989.3325723}
\showDOI{\tempurl}


\bibitem[Thangamani et~al\mbox{.}(2023)]%
        {thangamani2023lifting}
\bibfield{author}{\bibinfo{person}{Arun Thangamani}, \bibinfo{person}{Tiago
  Trevisan}, \bibinfo{person}{Vincent Loechner}, \bibinfo{person}{St{\'e}phane
  Genaud}, {and} \bibinfo{person}{B{\'e}renger Bramas}.}
  \bibinfo{year}{2023}\natexlab{}.
\newblock \showarticletitle{Lifting Code Generation of Cardiac Physiology
  Simulation to Novel Compiler Technology}. In \bibinfo{booktitle}{\emph{21st
  ACM/IEEE International Symposium on Code Generation and Optimization
  (CGO'23)}}. ACM, \bibinfo{publisher}{{ACM}}, \bibinfo{address}{New York, NY,
  USA}, \bibinfo{pages}{68--80}.
\newblock
\urldef\tempurl%
\url{https://doi.org/10.1145/3579990.3580008}
\showDOI{\tempurl}


\bibitem[Unat et~al\mbox{.}(2011)]%
        {mint2011unat}
\bibfield{author}{\bibinfo{person}{Didem Unat}, \bibinfo{person}{Xing Cai},
  {and} \bibinfo{person}{Scott~B. Baden}.} \bibinfo{year}{2011}\natexlab{}.
\newblock \showarticletitle{Mint: Realizing {CUDA} Performance in {3D} Stencil
  Methods with Annotated {C}}. In \bibinfo{booktitle}{\emph{Proceedings of the
  International Conference on Supercomputing}} (Tucson, Arizona, USA)
  \emph{(\bibinfo{series}{ICS '11})}. \bibinfo{publisher}{ACM},
  \bibinfo{address}{New York, NY, USA}, \bibinfo{pages}{214--224}.
\newblock
\showISBNx{9781450301022}
\urldef\tempurl%
\url{https://doi.org/10.1145/1995896.1995932}
\showDOI{\tempurl}


\bibitem[Wang and Chandramowlishwaran(2020)]%
        {pencil2020sc}
\bibfield{author}{\bibinfo{person}{Hengjie Wang} {and} \bibinfo{person}{Aparna
  Chandramowlishwaran}.} \bibinfo{year}{2020}\natexlab{}.
\newblock \showarticletitle{Pencil: A Pipelined Algorithm for Distributed
  Stencils}. In \bibinfo{booktitle}{\emph{SC20: International Conference for
  High Performance Computing, Networking, Storage and Analysis}}.
  \bibinfo{pages}{1--16}.
\newblock
\urldef\tempurl%
\url{https://doi.org/10.1109/SC41405.2020.00089}
\showDOI{\tempurl}


\bibitem[Yount et~al\mbox{.}(2016)]%
        {yask2016}
\bibfield{author}{\bibinfo{person}{Charles Yount}, \bibinfo{person}{Josh
  Tobin}, \bibinfo{person}{Alexander Breuer}, {and} \bibinfo{person}{Alejandro
  Duran}.} \bibinfo{year}{2016}\natexlab{}.
\newblock \showarticletitle{{YASK} - Yet Another Stencil Kernel: {A} Framework
  for {HPC} Stencil Code-Generation and Tuning}. In
  \bibinfo{booktitle}{\emph{Sixth International Workshop on Domain-Specific
  Languages and High-Level Frameworks for High Performance Computing,
  WOLFHPC@SC 2016, Salt Lake, UT, USA, November 14, 2016}}.
  \bibinfo{publisher}{{IEEE} Computer Society}, \bibinfo{pages}{30--39}.
\newblock
\urldef\tempurl%
\url{https://doi.org/10.1109/WOLFHPC.2016.08}
\showDOI{\tempurl}


\bibitem[Zhang et~al\mbox{.}(2022)]%
        {petscSFZhang:2022}
\bibfield{author}{\bibinfo{person}{Junchao Zhang}, \bibinfo{person}{Jed Brown},
  \bibinfo{person}{Satish Balay}, \bibinfo{person}{Jacob Faibussowitsch},
  \bibinfo{person}{Matthew Knepley}, \bibinfo{person}{Oana Marin},
  \bibinfo{person}{Richard~Tran Mills}, \bibinfo{person}{Todd Munson},
  \bibinfo{person}{Barry~F. Smith}, {and} \bibinfo{person}{Stefano Zampini}.}
  \bibinfo{year}{2022}\natexlab{}.
\newblock \showarticletitle{The PetscSF Scalable Communication Layer}.
\newblock \bibinfo{journal}{\emph{IEEE Transactions on Parallel and Distributed
  Systems}} \bibinfo{volume}{33}, \bibinfo{number}{4} (\bibinfo{year}{2022}),
  \bibinfo{pages}{842--853}.
\newblock
\urldef\tempurl%
\url{https://doi.org/10.1109/TPDS.2021.3084070}
\showDOI{\tempurl}


\bibitem[Zhao et~al\mbox{.}(2021)]%
        {zhao2021ppopp}
\bibfield{author}{\bibinfo{person}{Tuowen Zhao}, \bibinfo{person}{Mary Hall},
  \bibinfo{person}{Hans Johansen}, {and} \bibinfo{person}{Samuel Williams}.}
  \bibinfo{year}{2021}\natexlab{}.
\newblock \showarticletitle{Improving Communication by Optimizing On-Node Data
  Movement with Data Layout}. In \bibinfo{booktitle}{\emph{Proceedings of the
  26th ACM SIGPLAN Symposium on Principles and Practice of Parallel
  Programming}} (Virtual Event, Republic of Korea)
  \emph{(\bibinfo{series}{PPoPP '21})}. \bibinfo{publisher}{Association for
  Computing Machinery}, \bibinfo{address}{New York, NY, USA},
  \bibinfo{pages}{304--317}.
\newblock
\showISBNx{9781450382946}
\urldef\tempurl%
\url{https://doi.org/10.1145/3437801.3441598}
\showDOI{\tempurl}


\bibitem[Zhao et~al\mbox{.}(2018)]%
        {zhao2018bricks}
\bibfield{author}{\bibinfo{person}{Tuowen Zhao}, \bibinfo{person}{Samuel
  Williams}, \bibinfo{person}{Mary Hall}, {and} \bibinfo{person}{Hans
  Johansen}.} \bibinfo{year}{2018}\natexlab{}.
\newblock \showarticletitle{Delivering Performance-Portable Stencil
  Computations on CPUs and GPUs Using Bricks}. In
  \bibinfo{booktitle}{\emph{2018 IEEE/ACM International Workshop on
  Performance, Portability and Productivity in HPC (P3HPC)}}.
  \bibinfo{pages}{59--70}.
\newblock
\urldef\tempurl%
\url{https://doi.org/10.1109/P3HPC.2018.00009}
\showDOI{\tempurl}


\bibitem[Ziogas et~al\mbox{.}(2021)]%
        {ziogas2021}
\bibfield{author}{\bibinfo{person}{Alexandros~Nikolaos Ziogas},
  \bibinfo{person}{Timo Schneider}, \bibinfo{person}{Tal Ben-Nun},
  \bibinfo{person}{Alexandru Calotoiu}, \bibinfo{person}{Tiziano De~Matteis},
  \bibinfo{person}{Johannes de Fine~Licht}, \bibinfo{person}{Luca Lavarini},
  {and} \bibinfo{person}{Torsten Hoefler}.} \bibinfo{year}{2021}\natexlab{}.
\newblock \showarticletitle{Productivity, Portability, Performance:
  Data-Centric Python}. In \bibinfo{booktitle}{\emph{Proceedings of the
  International Conference for High Performance Computing, Networking, Storage
  and Analysis}} (St. Louis, Missouri) \emph{(\bibinfo{series}{SC '21})}.
  \bibinfo{publisher}{Association for Computing Machinery},
  \bibinfo{address}{New York, NY, USA}, Article \bibinfo{articleno}{95},
  \bibinfo{numpages}{13}~pages.
\newblock
\showISBNx{9781450384421}
\urldef\tempurl%
\url{https://doi.org/10.1145/3458817.3476176}
\showDOI{\tempurl}


\end{thebibliography}
\end{document}